\documentclass[prd,aps,showpacs,10pt]{revtex4}
\usepackage[english]{babel}
\usepackage{amsmath}
\usepackage{amssymb}
\usepackage{longtable}

\usepackage[pdftitle={Physical Perturbations in Cosmology},
pdfauthor={P. G. Miedema / W. A. van Leeuwen}, pdfsubject={Gauge
Invariant Cosmological Density Perturbations},
pdfstartview={FitH}, pdfkeywords={Structure Formation, Cosmic
Background Fluctuations}, bookmarks, bookmarksnumbered,
colorlinks, linkcolor={blue}, citecolor={blue}]{hyperref}

\renewcommand{\vec}[1]{\mbox{\boldmath$#1$}}

\newcommand{\vecs}[1]{\mbox{\scriptsize\boldmath$#1$}}

\newcommand{\nul}{{\mbox{$\scriptscriptstyle(0)$}}}
\newcommand{\een}{{\mbox{$\scriptscriptstyle(1)$}}}
\newcommand{\twee}{{\mbox{$\scriptscriptstyle(2)$}}}
\newcommand{\gi}{{\scriptstyle\mathrm{gi}}}
\newcommand{\dif}{\mathrm{d}}
\newcommand{\me}{\mathrm{e}}
\newcommand{\mi}{\mathrm{i}}

\begin{document}

\title{Density Perturbations in the Early Universe}

\author{P.\ G.\ Miedema}
\email{p.g.miedema@hccnet.nl}
\affiliation{Royal Netherlands Military Academy\\
   Hogeschoollaan 2, NL--4818 CR  Breda, The Netherlands}

\author{W.\ A.\ van Leeuwen}
\affiliation{University of
Amsterdam, Institute for Theoretical Physics,\\
   Valckenierstraat 65, NL--1018 XE Amsterdam, The Netherlands}

\date{November 7, 2005}

\begin{abstract}
We propose a way to construct manifestly gauge \emph{independent}
quantities out of the gauge \emph{dependent} quantities occurring in
the linearized Einstein equations. Thereupon, we show that these
gauge-invariant combinations can be identified with measurable
perturbations to the particle and energy densities. In order to
follow the evolution of these quantities in time, we rewrite the
full set of linearized Einstein equations as a set of four first
order, ordinary differential equations for the four physical
quantities in question, namely 1.\ the perturbations to the energy
density, 2.\ the particle number density, 3.\ the divergence of the
spatial fluid velocity and 4.\ the local spatial curvature due to a
density perturbation.

Next, we recombine the linearized perturbation equations, which
still contain gauge functions, in such a way that the resulting
equations contain only gauge-invariant quantities; as it should be,
all gauge functions have disappeared. In this way, we have arrived
at a formulation of the equations for cosmological perturbations
that is manifestly gauge-invariant. This set of linearized and
manifestly gauge-invariant Einstein equations for the
gauge-invariant combinations constitute the main result of this
paper.

The set of equations for the new gauge-invariant quantities, however
complicated it may seem, is more tractable than those of earlier
treatments. This lucky circumstance entails that, unlike the
situation encountered in former treatments, we do not need to have
recourse to any approximation. Therefore, we obtain results that,
sometimes, differ slightly from those found so far. As an
illustration, we consider the equations for two cases that can be
treated analytically, namely the radiation-dominated and
matter-dominated eras of a flat Friedmann-Lema\^\i
tre-Robertson-Walker universe.

In the radiation-dominated era we find, for small-scale
perturbations, acoustic waves with an \emph{increasing} amplitude,
while standard treatments predict acoustic waves with a
\emph{decaying} amplitude. For large-scale perturbations we find
exactly the same growth rates as in the standard literature.

Finally we mention that, when considering the non-relativistic limit
of the linearized Einstein equations we find ---not surprisingly---
the Poisson equation. Unlike what has been done sofar, however, we
did not need to limit the discussion to a static universe, i.e., we
found the Poisson equation to be valid not only in a static, but
also in an expanding universe.

\end{abstract}


\maketitle


\section{Introduction}

In order to link the density perturbations at the end of the period
of inflation and the observed temperature perturbations in the
cosmic $2.7\,\mathrm{K}$ microwave background radiation, one needs
the linearized Einstein equations. These equations, which determine
the growth of densities in the expanding universe after the era of
inflation, are the main subject of this article. Our final results
are contained in the Eqs.~(\ref{eq:epsilongi})--(\ref{eq:T-pert})
with (\ref{subeq:coeff}) and~(\ref{eq:sigmagi}).

Mathematically, the problem is very simple, in principle. First,
all quantities relevant to the problem are split up into two
parts: `a background part' and `a perturbation part'. The
background parts are chosen to satisfy the Einstein equations for
an isotropic universe, i.e., one chooses for the background
quantities the Friedmann-Lema\^\i tre-Robertson-Walker-solution.
Because of the isotropy, the background quantities depend on the
time coordinate~$t$ only. The perturbation parts are supposed to
be small compared to their background counterparts, and to depend
on the space-time coordinate $x=(ct,\vec{x})$. The background and
perturbations are often referred to as `zero order' and `first
order' quantities, respectively, and we will use this terminology
also in this article. After substituting the sum of the zero order
and first order parts of all relevant quantities into the Einstein
equations, all terms that are products of two or more quantities
of first order are neglected. This procedure leads, by
construction, to a set of \emph{linear} differential equations for
the quantities of first order. The solution of this set of linear
differential equations is then reduced to a standard problem of
the theory of ordinary, linear differential equations.

\subsection{Historical embedding}

The first systematic and extensive study of cosmological
perturbations is due to Lifshitz~\cite{lifshitz1946,c15,I.12}.
Almost half a century later, Mukhanov, Feldman and Brandenberger, in
their 1992 review article~\cite{mfb1992} entitled `Theory of
Cosmological Perturbations' mention or discuss more than 60 articles
on the subject, and, thereupon, suggest their own approach to the
problem. An article of Chung-Pei~Ma and Bertschinger
\cite{chung-pei1995}, entitled `Cosmological Perturbation Theory in
the Synchronous and Conformal Newtonian Gauges' critically reviews
the existing literature on the subject, including the work of
Mukhanov~\emph{et~al.}

For a solid overview of the literature we refer to the references
by the aforementioned authors~\cite{mfb1992,chung-pei1995}. For a
nice historical overview of gauge theories we refer to
O'Raifeartaigh and Straumann~\cite{ORN2000}. An actual linking of
inflationary perturbations and observable radiation background
anisotropies has been performed by van Tent~\cite{tent2002}.

The fact that so many studies are devoted to a problem that is
nothing but obtaining the solution of a set of ordinary, linear
differential equations is due to the fact that there are several
complicating factors, not regarding the mathematics involved, but
with respect to the physical interpretation of the solutions. In
this article we will try and explain what has been done so far, and
we will define, at the same time, \textit{gauge-invariant
expressions for the physical perturbations to the energy and
particle number densities}, which differ from
gauge-invariant combinations encountered in the literature. In the
limit of low velocities, \emph{but not necessarily in the limit of
slow expansion of the universe}, our first order equations reduce,
not surprisingly, to one single equation, the well-known Poisson
equation. We thus find that the Poisson equation is valid in an
expanding universe. This result, which has not been found earlier,
is an immediate result from our choice of gauge-invariant
quantities. See Sec.~\ref{nrl}.

Let us now try and explain in some detail what exactly is the
complicating factor in any treatment of the linearized Einstein
equations.

\subsection{Origin of the problem}\label{intro-noot}

At the very moment that one has split up a physical quantity into
a zero order and a first order part, one introduces an ambiguity.
To be explicit, let us consider the two quantities that play the
leading roles in the theory of cosmological perturbations. They
are: the energy density of the universe, $\varepsilon(x)$, and the
particle number density of the universe, $n(x)$. The linearized
Einstein equations contain as \emph{known} functions the zero
order functions $\varepsilon_\nul(t)$ and $n_\nul(t)$, which
describe the evolution of the background, i.e., they describe the
evolution of the unperturbed universe and they obey the
unperturbed Einstein equations, and as \emph{unknown} functions
the perturbations $\varepsilon_\een(x)$ and $n_\een(x)$. The
latter are the solutions to be obtained from the linearized
Einstein equations. The subindexes~$0$ and~$1$, which indicate the
order, have been put between round brackets, in order to
distinguish them from tensor indices. In all calculations,
products of a zero order and a first order quantity are considered
to be of first order, and are retained, whereas products of first
order quantities are neglected. Having said all this, we can say
where and how the ambiguity mentioned above arises.

The linearized Einstein equations, which, as stated above already,
are the equations which determine the first order quantities
$\varepsilon_\een(x)$ and $n_\een(x)$, do not fix these quantities
uniquely. In fact, it turns out that next to any solution for
$\varepsilon_\een$ and $n_\een$ of the linearized Einstein
equations, there exist solutions of the form
\begin{subequations}
\label{subeq:split-e-n}
\begin{eqnarray}
  \hat{\varepsilon}_\een(x) & = & \varepsilon_\een(x)+\psi(x)\partial_0\varepsilon_\nul(t),
     \label{e-ijk} \\
   \hat{n}_{\een} (x) & = & n_{\een}(x)+\psi(x) \partial_0 n_{\nul}(t),
         \label{n-ijk}
\end{eqnarray}
\end{subequations}
which also satisfy the linearized Einstein equations. Here the
symbol $\partial_0$ stands for the derivative with respect to
$x^0=ct$. In the equations~(\ref{subeq:split-e-n}) the function
$\psi(x)$ is an arbitrary but `small' function of the space-time
coordinate $x=(ct,\vec{x})$, i.e., we consider~$\psi(x)$ to be of
first order. We will derive the equations (\ref{subeq:split-e-n})
at a later point in this article; here it is sufficient to note
that, as (\ref{e-ijk}) and (\ref{n-ijk}) show, the perturbations
$\varepsilon_\een$ and $n_\een$ are fixed by the linearized
Einstein equations up to terms that are proportional to an
arbitrary, small function $\psi(x)$, usually called a gauge
function in this context. Since a physical quantity, i.e., a
directly measurable property of a system, may not depend on an
arbitrary function, the quantities $\varepsilon_\een$ and $n_\een$
cannot be interpreted as the real, physical, values of the
perturbations in the energy density or the particle number
density. But, if $\varepsilon_\een$ and $n_\een$ are not the
physical perturbations, what \emph{are} the real perturbations?
This is the notorious `gauge problem' encountered in any treatment
of cosmological perturbations. Many different answers to this
question can be found in the literature, none of which is
completely satisfactory, a fact which explains the ongoing
discussion on this subject. It is our hope that the answer given
in the present article will turn out to be the definitive one.

\subsection{Gauge-invariant combinations}  \label{intro-mies}

What we will do in this article is leave $\psi(x)$ undetermined,
i.e., we do \emph{not} choose any particular gauge, but we eliminate
the gauge function $\psi(x)$ from the theory altogether. This is not
a new idea: many articles have been written in which gauge-invariant
quantities, i.e., quantities independent of a gauge function are
used. What one essentially does in any gauge-invariant approach is
to try and construct expressions for the perturbations [like we do
in Eqs.~(\ref{subeq:gi-en})], which are such that $\psi(x)$
disappears from the defining expressions of the physical quantities.
These quantities, the perturbations in energy density and particle
number density in our case, are then shown to obey a set of linear
equations, not containing the gauge function $\psi(x)$ anymore [see
Eqs.~(\ref{eq:epsilongi}) and~(\ref{eq:ngi}) with
(\ref{subeq:coeff}) and~(\ref{eq:sigmagi})]. These equations follow,
by elimination of the gauge dependent quantities in favor of the
gauge-invariant ones, in a straightforward way from the usual
linearized Einstein equations, which did contain $\psi(x)$. In this
way, the theory is no longer plagued by the gauge freedom that is
inherent to the original equations and their solutions: $\psi(x)$
has disappeared altogether, as it should, not with brute force, but
as a natural consequence of the definitions~(\ref{subeq:gi-en}) of
the perturbations to the energy and the particle number densities.

As noted above, our approach also belongs to the class of
gauge-invariant cosmological theories, where terms are added to the
perturbations in such a way that they become independent of the
particular choice of coordinates. However, this can be done, in
principle, in infinitely many ways, since any linear combination of
gauge-invariant quantities is gauge-invariant also. Our treatment
distinguishes itself from earlier treatments by the fact that our
set of first order equations reduces to the usual non-relativistic
theory in the limit that the three-part of the cosmological fluid
velocity four-vector $U^\mu$ is small compared to the velocity of
light. Consequently, our splitting up of the energy density and the
particle number density in a zero order and a first order part is
such that the first order part reduces to the non-relativistic
expression. In other words, our treatment of perturbations is
`around' the classical, Newtonian theory of gravity. By the way,
this is not a \emph{conditio sine qua non} for a treatment of
cosmological perturbations to be true, valid or even useful for some
particular purpose, but it is a very desirable property.

We will come back to this point in Sec.~\ref{nrl}.

\subsection{Structure of the article}\label{intro-wim}

This article being quite extended, it might be useful to outline
shortly its structure.

In Sec.~\ref{frommat-mies} we define, in Eqs.~(\ref{subeq:gi-en}),
the physical perturbations to the energy density,
$\varepsilon_\een^\gi$ and the particle number density $n_\een^\gi$
that we propose in this article. In Sec.~\ref{frommat-zus}, we
explain how the gauge problem actually arises in cosmology once one
chooses to solve the Einstein equations perturbatively, i.e., by
means of a series expansion. Furthermore, we show in this section
that the quantities $\varepsilon_\een^\gi$ and $n_\een^\gi$, defined
in Sec.~\ref{frommat-mies}, are gauge-invariant.

In Sec.~\ref{frommat-non} we further motivate our choice of these
physical perturbations $\varepsilon^\gi_\een$ and $n^\gi_\een$ by
noting that they reduce to the perturbations as they occur in
Newton's theory of gravitation, contrarily to what is the case for
the perturbations defined in earlier treatments. The proof of this
statement is postponed to Sec.~\ref{nrl}. In Sec.~\ref{sec:an-exam}
we apply our perturbation theory to the cases of a
radiation-dominated and the matter-dominated universe and we show
that our approach yields in the radiation-dominated era small-scale
\emph{growing} density perturbations, in contrast to the
\emph{decaying} density perturbations found in earlier literature.

In Sec.~\ref{synchronous} we introduce a particular system of
coordinates, the well-known synchronous coordinates, and rewrite the
Einstein equations with respect to this particular coordinates. In
Sec.~\ref{backpert} we restrict the problem of obtaining a solution
of the perturbed Einstein equations to universes for which the zero
order solution is homogeneous and isotropic, the so-called
Friedmann-Robertson-Walker or
Friedmann-Lema\^{\i}tre-Robertson-Walker universes. This section is
rather technical: we treat zero order and first order quantities in
Secs.~\ref{backpert-zero} and \ref{backpert-first}, zero order
Einstein equations and conservation laws in
Secs.~\ref{nulde-einstein} and~\ref{nulde-conservation}, and,
finally, first order equations and conservation laws in
Secs.~\ref{eerste-einstein} and \ref{eerste-conservation}.

In Sec.~\ref{klasse} the first order equations are split up
according to their tensorial character, i.e., whether they are a
scalar, a vector or a second rank tensor. The tensorial, the
vectorial and the scalar parts of the perturbations, and the
equations they should obey, are derived in Secs.~\ref{tensor},
\ref{vector} and~\ref{scalar}, respectively. Since, in this article,
we are only interested in the energy density and the particle number
density, which both are scalar quantities, we only need the scalar
equations. In Sec.~\ref{evo-scal}, the linearized Einstein equations
for scalar perturbations found in Sec.~\ref{scalar} are rewritten
and simplified: in this section we show that the set of conservation
laws and constraint equations can be rewritten as a set of four
first order ordinary differential equations and one algebraic
equation. It is found that only three scalars play a role in the
theory of density perturbations, implying that there is very little
choice for the construction of gauge-invariant quantities.

A summary of all relevant equations is given in Sec.~\ref{resume}\@.
Sec.~\ref{sec:pertub-therm} is devoted to the perturbations in some
thermodynamical quantities related to the energy density and
particle number density perturbations. In Sec.~\ref{thirdstep} the
perturbation equations derived in Sec.~\ref{evo-scal} are rewritten
in a manifestly gauge-invariant form. In Sec.~\ref{nrl} we consider
the Newtonian limit and in Sec.~\ref{sec:an-exam} we apply the
evolution equations~(\ref{subeq:final}) to the two main eras of a
flat \textsc{flrw} universe. Finally, in Sec.~\ref{sec:stan-th} we
relate our results to results of earlier work.

In Appendix~\ref{sec:eq-state} some useful thermodynamic relations
are collected. Appendix~\ref{giofoe} is devoted to gauge
transformations. In Appendix~\ref{sec:gauge-newton} we show that in
the non-relativistic limit there remains some gauge freedom, left
over from the general theory relativity. Details to the derivations
in Sec.~\ref{thirdstep} are given in Appendix~\ref{sec:deriv-egi}.
In Appendix~\ref{app:standard-equation} we derive the standard
equations using the perturbation equations~(\ref{subeq:pertub-gi}).
Finally, Appendix~\ref{notations} gives a list of the symbols and
notations used in the main text.

\subsection{Gauge-invariant quantities}\label{frommat-mies}

In the existing literature on cosmological perturbations, one has
attempted to solve the problem that corresponds to the
non-uniqueness of the perturbations $\varepsilon_\een$ and $n_\een$
(\ref{subeq:split-e-n}) in two, essentially different, ways. The
first way is to impose an extra condition on the gauge field
$\psi(x)$~\cite{hwang-noh-1997,noh-hwang-2004a,Hwang-Noh-2005,
noh-hwang-2005a,Bertschinger1996}. Another way to get rid of the
gauge field $\psi(x)$ is to take linear combinations of the matter
variables and \emph{components of the perturbed metric tensor} to
construct gauge-invariant quantities (Bardeen~\cite{c13}, Mukhanov,
\emph{et~al}.~\cite{mfb1992}). The latter approach is generally
considered better than the one where one fixes a gauge, because it
not only leads to quantities that are independent of an undetermined
function, as should be the case for a physical quantity, but it also
does not rely on any particular choice for the gauge function, and,
therefore, has less arbitrariness.

In this article we approach the gauge problem in a Bardeen- or
Mukhanov-like way. However, instead of adding terms containing
perturbations of the metric tensor field $g_{\mu\nu}$, we add a term
that is proportional to (minus) the divergence of the (normalized)
cosmological four velocity $U^\mu$:
\begin{subequations}
\label{subeq:gi-en}
\begin{eqnarray}
 \varepsilon^\gi_\een & := & \varepsilon_\een -
   \frac{\partial_0\varepsilon_\nul}{\partial_0
   \theta_\nul}\theta_\een,       \label{gien} \\
 n^\gi_\een & := & n_\een - \frac{\partial_0
     n_\nul}{\partial_0 \theta_\nul}
      \theta_\een,       \label{gidi}
\end{eqnarray}
\end{subequations}
where $\theta_\nul$ and $\theta_\een$ are the background and
perturbation part of the covariant four-divergence
$c^{-1}U^\mu{}_{;\mu}$ of the cosmological fluid velocity field
$U^{\mu}(x)$. At first sight, these additional terms look quite
artificial. However, in Sec.~\ref{frommat-zus} we will show how they
arise very naturally.

In Sec.~\ref{frommat-zus}, we will show that the quantities
$\varepsilon^\gi_\een$ and $n^\gi_\een$ do \emph{not} change if we
switch from the old coordinates $x^\mu$ to new coordinates
$\hat{x}^\mu$ according to
\begin{equation}
   \hat{x}^\mu = x^\mu - \xi^\mu(x), \label{func}
\end{equation}
where the $\xi^\mu (x)$ ($\mu=0,\ldots,3$) are four arbitrary
functions, considered to be of first order, of the old coordinates
$x^\mu$. In other words, we will first show that
\begin{subequations}
\begin{eqnarray}
   \hat{\varepsilon}^\gi_\een (x) & = & \varepsilon^\gi_\een (x), \\
   \hat{n}^\gi_\een (x) & = & n^\gi_\een (x),
\end{eqnarray}
\end{subequations}
i.e., the perturbations (\ref{subeq:gi-en}) are independent of
$\xi^\mu(x)$, i.e., gauge-invariant. The combinations
(\ref{subeq:gi-en}) are not completely obvious indeed, but the proof
in Sec.~\ref{frommat-zus} that they will be gauge-invariant will
take away the mystery completely: the above gauge-invariant
combinations (\ref{subeq:gi-en}) will then even turn out the
`obvious' gauge-invariant quantities to study. Before showing this,
we round off the present reasoning.

If we transform the linearized Einstein equations with
$\varepsilon_\een (x)$ and $n_\een (x)$ to the new coordinates
$\hat{x}$, (\ref{func}), it will turn out that we find equations in
which only the zero component of the gauge functions $\xi^\mu(x)$,
($\mu=0,1,2,3$), occurs. We will call it $\psi(x)$:
\begin{equation}
\psi(x)\equiv \xi^0(x). \label{defpsi}
\end{equation}
In cosmology, the gauge function $\psi(x)$ is to be treated as a
first order quantity, i.e., as a small (or `infinitesimal') change
of the coordinates. If we eliminate
---with the help of (\ref{subeq:gi-en})--- the quantities
$\varepsilon_\een(x)$ and $n_\een(x)$ from the linearized Einstein
equations in favor of the gauge-invariant quantities
$\varepsilon^\gi_\een (x)$ and $n^\gi_\een (x)$, we obtain equations
in which the gauge function $\psi(x)$ is absent altogether. The
disappearance of the gauge function $\psi(x)$ is due entirely to the
introduction of the gauge-invariant quantities
$\varepsilon^\gi_\een(x)$ and $n^\gi_\een(x)$ rather than
$\varepsilon_\een(x)$ and $n_\een(x)$.

The gauge-invariant quantities $\varepsilon^\gi_\een$ and
$n^\gi_\een$ \emph{could be} the physical perturbations of the
energy density and the particle number density we are after, since
they are independent of the gauge field $\psi(x)$, but this is not
true a priori, since any linear combination of gauge-invariant
quantities is gauge-invariant. Hence, it is not evident that the
particular combinations proposed in the equations
(\ref{subeq:gi-en})  correspond to physical, i.e., measurable
quantities.   However, as we will show in section \ref{nrl}, their
non-relativistic limit is just what one would expect.

\subsection{Final results} \label{sec:ofr}

The key to the solution of the gauge problem of cosmology are the
equations (\ref{subeq:pertub-flrw})--(\ref{con-sp-1}), which are
derived in Sec.~\ref{evo-scal}\@. We obtain from these equations
(see Sec.~\ref{thirdstep}) a new set of evolution equations for the
gauge-invariant perturbations $\varepsilon^\gi_\een$ and
$n^\gi_\een$. The equation for~$\varepsilon^\gi_\een$ is a simple
second order differential equation of the form:
\begin{equation}\label{eq:epsilongi}
    \ddot{\varepsilon}^\gi_\een+a_1\dot{\varepsilon}^\gi_\een+
  a_2\varepsilon^\gi_\een = a_3 \sigma^\gi_\een,
\end{equation}
where~$\sigma^\gi_\een$ is given by the function~(\ref{eq:sigmagi}).
A dot denotes the derivative with respect to $x^0=ct$. The
coefficients~$a_1$, $a_2$ and~$a_3$ are complicated combinations
of~$\varepsilon_\nul$, $n_\nul$, the spatial curvature
$\mbox{$^3\!R_\nul$}$, the Hubble function~$H$ and the equation of
state $p_\nul=p(\varepsilon_\nul,n_\nul)$ and its partial
derivatives [see Eq.~(\ref{subeq:coeff})]. After one has solved this
equation for~$\varepsilon^\gi_\een$, one can calculate the
gauge-invariant perturbation to the particle number density with the
help of the formula [see Eq.~(\ref{eq:entropy-gi})]
\begin{equation}\label{eq:ngi}
  n_\een^\gi = \sigma_\een^\gi +
  \frac{n_\nul}{\varepsilon_\nul(1+w)}\varepsilon_\een^\gi,
\end{equation}
where $w:=p_\nul/\varepsilon_\nul$. The second order differential
equation~(\ref{eq:epsilongi}) for the first order energy density
perturbation~$\varepsilon^\gi_\een$ and the
expression~(\ref{eq:ngi}) for the first order particle density
perturbation ~$n^\gi_\een$ are the main results of this paper. They
are obtained without any approximation.

With their help one can relate the (measurable) fluctuation in the
cosmic background temperature [see Eq.~(\ref{eq:de-dp-dT-gi})]
\begin{equation}\label{eq:T-pert}
  T^\gi_\een=\dfrac{\varepsilon^\gi_\een-
    \left(\dfrac{\partial \varepsilon}{\partial n}\right)_{\!T}n^\gi_\een}
    {\left(\dfrac{\partial \varepsilon}{\partial T}\right)_{\!n}},
\end{equation}
to fluctuations $T^\gi_\een(t_0,\vec{x})$ at a time~$t_0$ at the
end of the era of inflation.

In order to get some insight into what these equations imply, we
consider two particular cases: the flat \textsc{flrw} universe in
the radiation-dominated and the matter-dominated eras. As is common
usage, we consider the so-called `contrast functions'
$\delta_\varepsilon:={\varepsilon^\gi_\een}/{\varepsilon_\nul}$ and
$\delta_n:=n^\gi_\een/n_\nul$ [see Eqs.~(\ref{subeq:final})] rather
than the fluctuations $\varepsilon^\gi_\een$ and $n^\gi_\een$
themselves. In the radiation-dominated era we find for these
contrast functions the system of equations~(\ref{subeq:final-rad})
with the exact solutions~(\ref{eq:coupled-rad}) and~(\ref{nu13}).
For large-scale perturbations these solutions are in complete
agreement with the solutions found in the standard theory. For
small-scale perturbations, however, Eqs.~(\ref{subeq:final-rad})
yield solutions different from those found in the standard theory:
we find an oscillating solution with an \emph{increasing} amplitude,
whereas the standard result is an oscillating solution with a
\emph{decaying} amplitude.

In the matter-dominated era we find the
equation~(\ref{eq:delta-dust}), which is slightly different from the
standard equation~(\ref{eq:delta-dust-standard}). For the density
contrast function, $\delta_\varepsilon$, we find the solution
(\ref{eq:matter-physical}) rather than the solution
(\ref{eq:matter-non-physical}) found in the standard literature.

The  purpose of this article is to present a logical and
straightforward derivation of the linearized Einstein equations,
culminating in the equations~(\ref{eq:epsilongi}), (\ref{eq:ngi})
and~(\ref{eq:T-pert}). It is not our purpose to solve and discuss
these equations in detail in the various stages of the evolution of
the universe, and analyze the possible cosmological consequences
that might be contained in our equations.

\subsection{Gauge-invariant first order perturbations}
\label{frommat-zus}

We now proceed with the proof that $\varepsilon^\gi_\een$ and
$n^\gi_\een$ are gauge-invariant. To that end, we start by recalling
the defining expression for the Lie derivative of an arbitrary
tensor field $A^{\alpha\cdots\beta}{}_{\mu\cdots\nu}$ with respect
to a vector field $\xi^{\tau}(x)$. It reads
\begin{eqnarray}
\lefteqn{\left(\mathcal{L}_\xi
A\right)^{\alpha\cdots\beta}{}_{\mu\cdots\nu}=
  A^{\alpha\cdots\beta}{}_{\mu\cdots\nu;\tau} \xi^{\tau} } \nonumber \\
 && -\,A^{\tau\cdots\beta}{}_{\mu\cdots\nu}\xi^{\alpha}{}_{;\tau}-\cdots-
    A^{\alpha\cdots\tau}{}_{\mu\cdots\nu}\xi^\beta{}_{;\tau}   \nonumber \\
 && +\,A^{\alpha\cdots\beta}{}_{\tau\cdots\nu}\xi^{\tau}{}_{;\mu}+\cdots+
    A^{\alpha\cdots\beta}{}_{\mu\cdots\tau}\xi^{\tau}{}_{;\nu}    ,   \label{lie1}
\end{eqnarray}
where the semi-colon denotes the covariant derivative. At the
right-hand side, there is a term with a plus sign for each lower
index and a term with a minus sign for each upper index. Recall
also, that the covariant derivative in the expression for the Lie
derivative may be replaced by an ordinary derivative, since the
Lie derivative is, by definition, independent of the connection.
This fact simplifies some of the calculations below.

Now, let $\{ x^{\mu} \}$ and $\{ \hat{x}{}^{\mu} = x^{\mu} -
\xi^{\mu} (x) \}$ be two sets of coordinate systems, where
$\xi^{\mu} (x)$ is an arbitrary ---but infinitesimal, i.e., in
this article, of first order--- vector field. Then the components
$\hat{A}^{\alpha\cdots\beta}{}_{\mu\cdots\nu}(x)$ of the tensor
$A$ with respect to the new coordinates $\hat{x}{}^{\mu}$ can be
related to the components of the tensor
$A^{\alpha\cdots\beta}{}_{\mu\cdots\nu}(x)$, defined with respect
to the old coordinates $\{ x^{\mu} \}$ with the help of the Lie
derivative. Up to and including terms containing first order
derivatives one has
\begin{equation}
  \hat{A}^{\alpha\cdots\beta}{}_{\mu\cdots\nu}(x) =
  A^{\alpha\cdots\beta}{}_{\mu\cdots\nu}(x)+
  \left({\mathcal{L}_{\xi}}A\right)^{\alpha\cdots\beta}{}_{\mu\cdots\nu}(x)+\cdots.
     \label{lie2}
\end{equation}
For a derivation of this equation, see Weinberg~\cite{c8},
Chap.~10, Sec.~9.

Note that $x$ in the left-hand side corresponds to a point, $P$
say, of space-time with coordinates $x^\mu$ in the coordinate
frame \{$x$\}, while in the right-hand side $x$ corresponds to
another point, $Q$ say, with exactly the same coordinates $x^\mu$,
but now with respect to the coordinate frame \{$\hat{x}$\}. Thus,
equation (\ref{lie2}) is an expression that relates one tensor
field $A$ at two different points of space-time, points that are
related via the relation (\ref{func}).

The following observation is crucial. Because of the general
covariance of the Einstein equations, they are invariant under
general coordinate transformations $x\rightarrow\hat{x}$ and, in
particular, under coordinate transformations given
by~(\ref{func}). Hence, if some tensorial quantity $A(x)$ of
rank~$n$ ($n=0,1,\ldots$) satisfies the Einstein equations with as
source term the energy-momentum tensor $T$, the quantity
$\hat{A}(x)=A(x)+\mathcal{L}_\xi A(x)$ satisfies the Einstein
equations with source term $\hat{T}(x)=T(x)+\mathcal{L}_\xi T(x)$,
for a universe \emph{with precisely the same physical content}.
Because of the linearity of the linearized Einstein equations, a
linear combination of any two solutions is also a solution. In
particular, $\mathcal{L}_\xi A$, being the difference of ${A}$ and
$\hat{A}$, is a solution of the linearized Einstein equations with
source term $\mathcal{L}_\xi T$. In first order, $\mathcal{L}_\xi
A(x)$ may be replaced by $\mathcal{L}_\xi A_\nul(t)$, where
$A_\nul(t)$ is the solution for $A(t)$ of the zero order Einstein
equations. The freedom to add a term of the form $\mathcal{L}_\xi
A_\nul(t)$, with $\xi^\mu$ ($\mu=0,1,2,3$) four arbitrary
functions of first order, to any solution of the Einstein
equations of the first order, is the reason that none of the first
order solutions is uniquely defined, and, hence, does not
correspond in a unique way to a measurable property of the
universe. This is the notorious gauge problem referred to in the
introduction of this article. The additional terms
$\mathcal{L}_\xi A_\nul(t)$ are called `gauge modes'.

Combining (\ref{lie1}) and (\ref{lie2}) we have
\begin{eqnarray}
\lefteqn{\hat{A}^{\alpha\cdots\beta}{}_{\mu\cdots\nu}(x) =
 A^{\alpha\cdots\beta}{}_{\mu\cdots\nu}(x)+
  A^{\alpha\cdots\beta}{}_{\mu\cdots\nu;\tau}\xi^{\tau} } \nonumber \\
 && -\, A^{\tau\cdots\beta}{}_{\mu\cdots\nu}\xi^{\alpha}{}_{;\tau}-\cdots-
     A^{\alpha\cdots\tau}{}_{\mu\cdots\nu}\xi^{\beta}{}_{;\tau}   \nonumber \\
 &&  +\,A^{\alpha\cdots\beta}{}_{\tau\cdots\nu} \xi^{\tau}{}_{;\mu}+\cdots+
     A^{\alpha\cdots\beta}{}_{\mu\cdots\tau} \xi^{\tau}{}_{;\nu}.   \label{dakjeserop}
\end{eqnarray}
We now apply the equation (\ref{dakjeserop}) to the case that $A$
is a scalar $\sigma$, a four-vector $V^\mu$ and a tensor
$A_\mu{}_\nu$, respectively,
\begin{subequations}
\label{sca-vec-ten}
\begin{eqnarray}
  \hat{\sigma}(x)&=&\sigma(x)+\xi^\tau(x) \partial_\tau\sigma(x), \label{sigma}\\
  \hat{V}^\mu&=&V^\mu+V^\mu{}_{;\tau}\xi^\tau-V^\tau\xi^\mu{}_{;\tau},
     \label{lievec}\\
\hat{A}_{\mu\nu}&=& A_{\mu\nu}+
   A_{\mu\nu;\tau}\xi^\tau+A_{\tau\nu}\xi^\tau{}_{;\mu}+
   A_{\mu\tau}\xi^\tau{}_{;\nu}.
\label{lieder}
\end{eqnarray}
\end{subequations}
For the metric tensor, $g_{\mu\nu}$ we find in particular, from
Eq.~(\ref{lieder}),
\begin{equation}
   \hat{g}_{\mu\nu} = g_{\mu\nu} + \xi_{\mu;\nu} + \xi_{\nu;\mu},
           \label{killing}
\end{equation}
where we have used that the covariant derivative of the metric
vanishes.

Our construction of gauge-invariant perturbations totally rest upon
these equations for hatted quantities. In case $\sigma(x)$ is some
scalar quantity obeying the Einstein equations, $\sigma(x)$ can be
split up in the usual way in a zero order and a first order part:
\begin{equation}
    \sigma(x) := \sigma_\nul(t) + \sigma_\een(x), \label{sigma1}
\end{equation}
where $\sigma_\nul (t) $ is some background quantity, and hence,
not dependent on the spatial coordinates. Then (\ref{sigma})
becomes
\begin{equation}
   \hat{\sigma}(x)=\sigma_\nul(t) + \sigma_\een(x)+
   \xi^0 (x)\partial_0 \sigma_\nul(t) + \xi^\mu(x) \partial_\mu
  \sigma_\een(x).   \label{sigmahat1}
\end{equation}
The last term, being a product of the first order quantity
$\xi^{\mu} (x)$ and the first order quantity $\partial_{\mu}
\sigma_\een$, will be neglected. We thus find
\begin{equation}
  \hat{\sigma} (x) = \sigma_\nul (t) + \hat{\sigma}_\een(x),
     \label{sigmahat2}
\end{equation}
with
\begin{equation}
  \hat{\sigma}_\een(x) := \sigma_\een(x) +
    \psi(x) \partial_0\sigma_\nul(t),   \label{sigmahat3}
\end{equation}
where we used (\ref{defpsi}). Similarly, we find from
(\ref{lievec}) and (\ref{killing})
\begin{equation}
  \hat{V}^\mu_\een(x)=V^\mu_\een+
      V_{\nul;\tau}^\mu\xi^\tau-V_\nul^\tau\xi^\mu{}_{;\tau},
 \label{transvec}
\end{equation}
and
\begin{equation}
  \hat{g}_{\een\mu\nu}(x)=g_{\een\mu\nu}(x)+\xi_{\mu;\nu}+\xi_{\nu;\mu}.
    \label{transmetric}
\end{equation}
The latter two equations will be used later.

We are now in a position that we can conclude the proof of the
statement that $\varepsilon^\gi_\een$ and $n^\gi_\een$ are
gauge-invariant. To that end, we now write down the equation
(\ref{sigmahat3}) once again, for another arbitrary scalar quantity
$\omega(x)$ obeying the Einstein equations. We then find the
analogue of Eq.~(\ref{sigmahat3})
\begin{equation}
   \hat{\omega}_\een(x) = \omega_\een(x) + \psi(x)\partial_0
     \omega_\nul(t). \label{omegahat3}
\end{equation}
The left-hand sides of (\ref{sigmahat3}) and (\ref{omegahat3})
give the value of the perturbation at the point with coordinates
$x$ with respect to the old coordinate system $\{x\}$; the
right-hand sides of (\ref{sigmahat3}) and (\ref{omegahat3})
contains quantities with the same values of the coordinates, $x$,
but now with respect to the new coordinate system $\{\hat{x}\}$.
Eliminating the function~$\psi(x)$ from Eqs.~(\ref{sigmahat3})
and~(\ref{omegahat3}) yields
\begin{equation}
\hat{\sigma}_\een(x) - \frac{\partial_0 \sigma_\nul(t)}{\partial_0
\omega_\nul(t)} \hat{\omega}_\een(x) = \sigma_\een(x) -
\frac{\partial_0\sigma_\nul(t)}{\partial_0 \omega_\nul(t)}
\omega_\een(x). \label{invariance1}
\end{equation}
In other words, the particular linear combination occurring in the
right-hand side of (\ref{invariance1}) of any two quantities
$\omega$ and $\sigma$ transforming as scalars under an arbitrary
space-time transformation  ---of first order,
cf.~Eq.~(\ref{func})--- is gauge-invariant, and, hence, a possible
candidate for a physical quantity.

The equation (\ref{invariance1}) is the key equation of this article
as far as the scalar quantities $\varepsilon_\een$ and $n_\een$ are
concerned. It tells us how to combine the scalar quantities
occurring in the linearized Einstein equations in such a way that
they become gauge independent. The equation (\ref{invariance1}) can
be used to immediately derive the expressions (\ref{subeq:gi-en})
for the gauge-invariant energy and particle number densities.

In fact, let $U^{\mu} (x)$ be the four-velocity of the
cosmological fluid. In the theory of cosmological perturbations
[see Eqs.~(\ref{subeq:einstein-flrw})--(\ref{FRW3})
and~(\ref{subeq:pertub-flrw})--(\ref{con-sp-1})] there will turn
out to be only three scalars, namely
\begin{subequations}
\label{subeq:ent}
\begin{eqnarray}
    \varepsilon(x) & = & c^{-2} T^{\mu\nu}(x) U_\mu(x) U_\nu(x), \label{eps1} \\
    n(x) & = & c^{-2} N^\mu(x) U_\mu(x), \label{en1} \\
    \theta(x) & = & c^{-1} U^\mu{}_{;\mu}(x), \label{exp1}
\end{eqnarray}
\end{subequations}
where
\begin{equation}\label{eq:current}
    N^\mu:=nU^\mu,
\end{equation}
is the cosmological particle current four-vector normalized
according to $U^\mu U_\mu=c^2$. These scalars are split up
according to
\begin{subequations}
\label{subeq:ent-split}
\begin{eqnarray}
   \varepsilon (x) & = & \varepsilon_\nul (t) + \varepsilon_\een (x), \label{eps2} \\
   n(x) & = & n_\nul(t) + n_\een(x), \label{en2} \\
   \theta(x) & = & \theta_\nul(t) + \theta_\een(x), \label{exp2}
\end{eqnarray}
\end{subequations}
where the background quantities $\varepsilon_\nul(t)$, $n_\nul(t)$
and $\theta_\nul(t)$ are solutions of the unperturbed Einstein
equations. They depend on the time coordinate $t$ only. The relation
(\ref{invariance1}) inspires us to consider the gauge-invariant
combinations
\begin{subequations}
\label{subeq:ent-gi}
\begin{eqnarray}
   \varepsilon^\gi_\een(x) & := & \varepsilon_\een(x) - \frac{\partial_0
     \varepsilon_\nul(t)}{\partial_0 \omega_\nul(t)} \omega_\een(x),  \label{eps3} \\
   n^\gi_\een(x) & := & n_\een(x) - \frac{\partial_0 n_\nul(t)}
     {\partial_0 \omega_\nul(t)} \omega_\een(x), \label{en3} \\
   \theta^\gi_\een(x) & := & \theta_\een(x) - \frac{\partial_0
     \theta_\nul(t)}{\partial_0 \omega_\nul(t)} \omega_\een(x).  \label{exp3}
\end{eqnarray}
\end{subequations}
The question remains what to choose for $\omega$ in these three
cases. In principle, for $\omega$ we could choose any of the
following three scalar functions available in the theory, i.e., we
could choose $\varepsilon$, $n$ or $\theta$. As follows
from~(\ref{eps3}) and~(\ref{en3}), the choices $\omega =
\varepsilon$ and $\omega = n$ would lead to
$\varepsilon^\gi_\een(x)=0$ or $n^\gi_\een(x)=0$, respectively.
This would mean that perturbations in the energy or the particle
number density would show up only in the second order. Apparently,
these choices would not be suitable since the perturbations of
precisely these quantities are the object of study. We therefore
are left with the choice
\begin{equation}
  \omega = \theta, \label{omega1}
\end{equation}
which implies the equations~(\ref{gien}) and (\ref{gidi}) for the
energy and particle number density perturbations, as was to be
shown, and, moreover,
\begin{equation}
   \theta^\gi_\een=0. \label{thetagi}
\end{equation}
The latter equation implies that perturbations in the divergence of
the cosmological velocity field~$\theta(x)$ will only show up in
second order. This constitutes no problem, since we are not
interested in this quantity. Inserting (\ref{omega1}) into
(\ref{eps3}) and (\ref{en3}) we obtain the expressions
(\ref{subeq:gi-en}). Hence, it now has been shown that
$\varepsilon^\gi_\een$ and $n^\gi_\een$ are indeed invariant under
the infinitesimal coordinate transformation (\ref{func}), i.e., that
they are gauge-invariant.

The linearized Einstein equations we are about to study contain the
gauge function $\psi(x)$. If we eliminate $\varepsilon_\een(x)$ and
$n_\een(x)$ from these equations in favor of
$\varepsilon^\gi_\een(x)$ and $n^\gi_\een(x)$, with the help of the
expressions (\ref{subeq:gi-en}), we obtain equations, which do not
depend on $\psi(x)$ at all: all terms with $\psi(x)$ cancel
automatically. This could be expected, because the equations should
have solutions for $\varepsilon^\gi_\een(x)$ and $n^\gi_\een(x)$
which are, by construction, independent of the
diffeomorphism~(\ref{func}). Stated differently, the fact that
$\psi(x)$ disappears from the linearized perturbation theory is in
line with the earlier observation that~(\ref{subeq:gi-en}) are gauge
independent combinations for arbitrary diffeomorphisms~(\ref{func}).
In Sec.~\ref{thirdstep} we will show in detail how this will happen.

\subsection{Non-relativistic limit of the gauge-invariant combinations}
\label{frommat-non}

The following question may arise. It is clear now that
$\varepsilon_\een(x)$ and $n_\een(x)$ are not gauge-invariant, but
that the combinations $\varepsilon^\gi_\een(x)$ and $n^\gi_\een(x)$
are so. Hence, $\varepsilon_\een(x)$ and $n_\een(x)$ cannot be
physical perturbations, while $\varepsilon^\gi_\een (x)$ and
$n^\gi_\een(x)$ could be physical. But how can one be sure that the
particular combinations (\ref{subeq:gi-en}) are indeed the real
physical perturbations? The fact that they are gauge-invariant is a
necessary, but not a sufficient reason. Since, moreover, any linear
combination of gauge-invariant quantities is a new gauge-invariant
quantity, it is not clear at all, yet, that the particular
combinations~(\ref{subeq:gi-en}) are the right ones.

This issue can be settled by considering the non-relativistic limit,
i.e., the limit of low spatial velocities with respect to the speed
of light. In this particular case the linearized Einstein equations
(\ref{subeq:pertub-gi})--(\ref{subeq:pertub-gi-e-n}) imply, see
Eqs.~(\ref{Egi-poisson-present}) and (\ref{eq:ident}),
\begin{equation}\label{poisson}
    \nabla^2\varphi(\vec{x})=
    4\pi G\dfrac{\varepsilon^\gi_\een(t_\mathrm{p},\vec{x})}{c^2},
\end{equation}
where $t_\mathrm{p}$ indicates the present time, $c$ is the speed of
light, $G$ Newton's gravitational constant and $\nabla^2$ is the
usual Laplace operator. This is the well-known Poisson equation of
the Newtonian theory of gravity. Hence, in the non-relativistic
regime, the mathematical combination $\varepsilon^\gi_\een$
(\ref{gien}), divided by $c^2$, is to be interpreted as the normal,
or `right' physical perturbation of the mass density $\varrho_\een$,
implying that $\varepsilon^\gi_\een$ may be interpreted as the
`right' or physical perturbation in the energy density indeed. Since
the non-relativistic limit of another Einstein equation,
Eq.~(\ref{nu2-flat}), will turn out to imply
$n^\gi_\een(t_\mathrm{p},\vec{x})=
\varepsilon^\gi_\een(t_\mathrm{p},\vec{x})/(mc^2)$,
Eq.~(\ref{nu2-flat-newt}), we see that also
$n^\gi_\een$~(\ref{gidi}) is the right physical perturbation. This
answers the question raised at the beginning of this section with
respect to $\varepsilon^\gi_\een$ and $n^\gi_\een$: these quantities
are, indeed, the physical perturbations we were looking for.

\section{The Einstein equations rewritten in synchronous coordinates}
   \label{synchronous}

As most authors do, we choose a synchronous system of reference. A
synchronous system of reference is a system in which the line
element for the metric has the form:
\begin{equation}
  \dif s^2=c^2\dif t^2-g_{ij}(t,\vec{x})\dif x^i \dif x^j.
       \label{line-element}
\end{equation}
The name synchronous stems from the fact that surfaces with
$t=\mathrm{constant}$ are surfaces of simultaneity for observers
at rest with respect to the synchronous coordinates, i.e.,
observers for which the three coordinates $x^i$ ($i=1,2,3$) remain
constant. A synchronous system can be used for an arbitrary
space-time manifold, not necessarily a homogeneous or homogeneous
and isotropic one. In a synchronous system, the coordinate~$t$
measures the proper time along lines of constant $x^i$. From
(\ref{line-element}) we can read off that ($x^0=ct$):
\begin{equation}
    g_{00}(t,\vec{x})=1, \quad g_{0i}(t,\vec{x})=0.    \label{sync-cond}
\end{equation}
From the form of the line element in four-space,
Eq.~(\ref{line-element}), it follows that minus
$g_{ij}(t,\vec{x})$, ($i=1,2,3$), is the metric of a
three-di\-men\-sional subspaces with constant~$t$. Because of
(\ref{sync-cond}), knowing the three-geometry in all
hypersurfaces, is equivalent to knowing the geometry of
space-time. The following abbreviations will prove useful when we
rewrite the Einstein equations with respect to synchronous
coordinates:
\begin{equation}
\label{def-gammas}
    \varkappa_{ij} := -\tfrac{1}{2}\dot{g}_{ij}, \quad
    \varkappa^i{}_j := g^{ik}\varkappa_{kj}, \quad
    \varkappa^{ij}:=+\tfrac{1}{2}\dot{g}^{ij},
\end{equation}
where a dot denotes differentiation with respect to~$x^0=ct$. From
Eqs.~(\ref{sync-cond})--(\ref{def-gammas}) it follows that the
connection coefficients of (four-di\-men\-sional) space-time
\begin{equation}
   \Gamma^\lambda{}_{\mu\nu}=\tfrac{1}{2} g^{\lambda\rho}
   \left( g_{\rho\mu,\nu}+g_{\rho\nu,\mu}-g_{\mu\nu,\rho} \right),
        \label{concoef1}
\end{equation}
in synchronous coordinates are given by
\begin{subequations}
\label{subeq:con}
\begin{eqnarray}
   && \Gamma^0{}_{00}=\Gamma^i{}_{00}=\Gamma^0{}_{i0}=\Gamma^0{}_{0i}=0,
           \label{con1}  \\
   && \Gamma^0{}_{ij}=\varkappa_{ij},   \quad
       \Gamma^i{}_{0j}=\Gamma^i{}_{j0}=-\varkappa^i{}_j,     \label{con2}  \\
   && \Gamma^k{}_{ij}=\tfrac{1}{2} g^{kl}
   \left( g_{li,j}+g_{lj,i}-g_{ij,l} \right).   \label{con3}
\end{eqnarray}
\end{subequations}
From Eq.~(\ref{con3}) it follows that the~$\Gamma^k{}_{ij}$ are
also the connection coefficients of (three-dimensional) subspaces
of constant time.

The Ricci tensor $R_{\mu\nu}:={R^{\lambda}}_{\mu\lambda\nu}$ is,
in terms of the connection coefficients, given by
\begin{equation}
  R_{\mu\nu}=
   \Gamma^\lambda{}_{\mu\nu,\lambda}-
   \Gamma^\lambda{}_{\mu\lambda,\nu}+
   \Gamma^\sigma{}_{\mu\nu}\Gamma^\lambda{}_{\lambda\sigma}-
   \Gamma^\sigma{}_{\mu\lambda} \Gamma^\lambda{}_{\nu\sigma}.  \label{Ricci1}
\end{equation}
Upon substituting Eqs.~(\ref{subeq:con}) into Eq.~(\ref{Ricci1})
one finds for the components of the Ricci tensor
\begin{subequations}
\label{subeq:ricci}
\begin{eqnarray}
   R_{00} & = & \dot{\varkappa}^k{}_k - \varkappa^l{}_k \varkappa^k{}_l, \label{R00} \\
   R_{0i} & = & \varkappa^k{}_{k|i}-\varkappa^k{}_{i|k}, \label{R0i} \\
   R_{ij} & = & \dot{\varkappa}_{ij} - \varkappa_{ij} \varkappa^k{}_k +
        2\varkappa_{ik}\varkappa^k{}_j + \mbox{$^3\!R_{ij}$},  \label{Rij}
\end{eqnarray}
\end{subequations}
where the vertical bar in Eq.~(\ref{R0i}) denotes covariant
differentiation with respect to the metric~$g_{ij}$ of a
three-dimensional subspace:
\begin{equation}
   \varkappa^i{}_{j|k} := \varkappa^i{}_{j,k} + \Gamma^i{}_{lk} \varkappa^l{}_j -
        \Gamma^l{}_{jk} \varkappa^i{}_l.   \label{threecov}
\end{equation}
The quantities $\mbox{$^3\!R_{ij}$}$ in Eq.~(\ref{Rij}) are found
to be given by
\begin{equation}
  \mbox{$^3\!R_{ij}$} = \Gamma^k{}_{ij,k}-\Gamma^k{}_{ik,j}+
   \Gamma^l{}_{ij}\Gamma^k{}_{kl}-\Gamma^l{}_{ik} \Gamma^k{}_{jl}.
         \label{Ricci-drie}
\end{equation}
Hence, $\mbox{$^3\!R_{ij}$}$ is the Ricci tensor of the
three-dimensional subspaces of constant time. For the components
$R^\mu{}_\nu=g^{\mu\tau}R_{\tau\nu}$ of the Ricci
tensor~(\ref{subeq:ricci}), we get
\begin{subequations}
\label{subeq:Rmixed}
\begin{eqnarray}
   R^0{}_0 & = & \dot{\varkappa}^k{}_k - \varkappa^l{}_k \varkappa^k{}_l,
           \label{Rm00} \\
   R^0{}_i & = & \varkappa^k{}_{k|i}-\varkappa^k{}_{i|k}, \label{Rm0i} \\
   R^i{}_j & = & \dot{\varkappa}^i{}_j - \varkappa^i{}_j \varkappa^k{}_k +
        \mbox{$^3\!R^i{}_j$},   \label{Rmij}
\end{eqnarray}
\end{subequations}
where we have used Eqs.~(\ref{sync-cond})--(\ref{def-gammas}).

The Einstein equations read
\begin{equation}
   G^{\mu\nu} - \Lambda g^{\mu\nu}=\kappa T^{\mu\nu}, \label{ein-verg}
\end{equation}
where $G^{\mu\nu}$, the Einstein tensor, is given by
\begin{equation}
  G^{\mu\nu} = R^{\mu\nu} - \tfrac{1}{2}R^\alpha{}_\alpha g^{\mu\nu}.
     \label{ein-ten}
\end{equation}
In (\ref{ein-verg}) $\Lambda$ is a positive constant, the
well-known cosmological constant. The constant~$\kappa$ is
given~by
\begin{equation}
   \kappa := \frac{8\pi G}{c^4},   \label{kappa}
\end{equation}
with $G$ Newton's gravitational constant and~$c$ the speed of
light. In view of the Bianchi identities one has
\begin{equation}
  G^{\mu\nu}{}_{;\nu} = 0,    \label{bianchi}
\end{equation}
hence, since $g^{\mu\nu}{}_{;\nu}=0$, the source term $T^{\mu\nu}$
of the Einstein equations must fulfill the properties
\begin{equation}
    T^{\mu\nu}{}_{;\nu} = 0.    \label{behoud}
\end{equation}
These equations are the energy-momentum conservation laws. An
alternative way to write the Einstein equations~(\ref{ein-verg})
reads
\begin{equation}
   R^\mu{}_\nu=\kappa(T^\mu{}_\nu-
          \tfrac{1}{2}\delta^\mu{}_\nu T^\alpha{}_\alpha) -
          \Lambda\delta^\mu{}_\nu.     \label{einstein}
\end{equation}
Upon substituting the components (\ref{subeq:Rmixed}) into the
Einstein equations~(\ref{einstein}), and eliminating the time
derivative of $\varkappa^k{}_k$ from the $R^0{}_0$-equation with
the help of the $R^i{}_j$-equations, the Einstein equations can be
cast in the form
\begin{subequations}
\label{subeq:Ein-syn}
\begin{eqnarray}
    (\varkappa^k{}_k)^2 - \mbox{$^3\!R$} -
       \varkappa^k{}_l\varkappa^l{}_k & \!\! = \!\! &
        2(\kappa T^0{}_0 + \Lambda),  \label{Ein-syn1}   \\
    \varkappa^k{}_{k|i}-\varkappa^k{}_{i|k} & \!\! = \!\! &
           \kappa T^0{}_i, \label{Ein-syn2} \\
    \dot{\varkappa}^i{}_j - \varkappa^i{}_j \varkappa^k{}_k +
        \mbox{$^3\!R^i{}_{j}$} & \!\! = \!\! &
    \kappa(T^i{}_j - \tfrac{1}{2}\delta^i{}_j T^\mu{}_\mu)-\Lambda\delta^i{}_j,
            \label{Ein-syn3}
\end{eqnarray}
\end{subequations}
where
\begin{equation}
  \mbox{$^3\!R$}:=g^{ij}\,\mbox{$^3\!R_{ij}$}=\mbox{$^3\!R^k{}_k$},
     \label{drieR}
\end{equation}
is the curvature scalar of the three-dimensional subspaces of
constant time. The (differential) equations (\ref{Ein-syn3}) are
the so-called dynamical Einstein equations: they define the
evolution (of the time derivative) of the (spatial part of the)
metric. The (algebraic) equations (\ref{Ein-syn1})
and~(\ref{Ein-syn2}) are constraint equations: they relate the
initial conditions, and, once these are satisfied at one time,
they are satisfied automatically at all times.

The right-hand side of Eqs.~(\ref{subeq:Ein-syn}) contain the
components of the energy momentum tensor $T_{\mu\nu}$, which, for
a perfect fluid, are given by
\begin{equation}\label{Tmunu}
   T^\mu{}_\nu = (\varepsilon+p)u^\mu u_\nu - p \delta^\mu{}_\nu,
\end{equation}
where~$u^\mu(x)=c^{-1}U^\mu(x)$ is the hydrodynamic fluid
four-velocity normalized to unity: ($u^\mu u_\mu=1$),
$\varepsilon(x)$ the energy density and~$p(x)$ the pressure at a
point~$x$ in space-time. In this expression we neglect terms
containing the shear and volume viscosity, and other terms related
to irreversible processes. The equation of state for the pressure
\begin{equation}
     p=p(n,\varepsilon),     \label{toestand}
\end{equation}
where $n(x)$ is the particle number density at a point~$x$ in
space-time, is supposed to be a given function of $n$ and
$\varepsilon$ (see also Appendix~\ref{sec:eq-state} for equations
of state in alternative forms).

As stated above already, the Einstein equations (\ref{Ein-syn1})
and (\ref{Ein-syn2}) are constraint equations to the Einstein
equations (\ref{Ein-syn3}) only: they tell us what relations
should exist between the initial values of the various unknown
functions, in order that the Einstein equations be solvable. In
the following, we shall suppose that these conditions are
satisfied. Thus we are left with the nine equations
(\ref{Ein-syn3}), of which, because of the symmetry of $g_{ij}$,
only six are independent. These six equations, together with the
four equations (\ref{behoud}) constitute a set of ten equations
for the eleven $(6+3+1+1)$ independent quantities $g_{ij}$, $u^i$,
$\varepsilon$ and $n$. The eleventh equation needed to close the
system of equations is the particle number conservation law
$N^\mu{}_{;\mu}=0$, i.e.,
\begin{equation}
     (nu^\mu)_{;\mu}=0,     \label{pncl}
\end{equation}
[see Eq.~(\ref{eq:current})] where a semicolon denotes covariant
differentiation with respect to the metric tensor $g_{\mu\nu}$.
This equation can be rewritten in terms of the fluid expansion
scalar defined by Eq.~(\ref{exp1}). Using Eqs.~(\ref{subeq:con}),
we can rewrite the four-divergence (\ref{exp1}) in the form
\begin{equation}
  \theta = \dot{u}^0 - \varkappa^k{}_k u^0 + \vartheta,  \label{vierdiv}
\end{equation}
where the three-divergence $\vartheta$ is given by
\begin{equation}
    \vartheta := u^k{}_{|k}.    \label{driediv}
\end{equation}
Using now Eqs.~(\ref{exp1}), (\ref{subeq:con}), (\ref{threecov})
and~(\ref{vierdiv}), the four energy-momentum conservation
laws~(\ref{behoud}) and the particle number conservation
law~(\ref{pncl}) can be rewritten as
\begin{subequations}
\label{subeq:cons-laws}
\begin{eqnarray}
  \dot{T}^{00}+T^{0k}{}_{|k}+\varkappa^k{}_l T^l{}_k -
       \varkappa^k{}_k T^{00} & = & 0,     \label{Tnulnu}  \\
  \dot{T}^{i0}+T^{ik}{}_{|k}-2\varkappa^i{}_k T^{k0} -
               \varkappa^k{}_k T^{i0} & = & 0,     \label{Tinu}
\end{eqnarray}
\end{subequations}
and
\begin{equation}\label{deeltjes}
    \dot{n}u^0 + n_{,k}u^k + n\theta = 0,
\end{equation}
respectively. Since $T^{0i}$ is a vector and $T^{ij}$ is a tensor
with respect to coordinate transformations in a subspace of
constant time, and, hence, are tensorial quantities in this
three-dimensional subspace, we could use in
(\ref{subeq:cons-laws}) a bar to denote covariant differentiation
with respect to the metric $g_{ij}(t,\vec{x})$ of such a subspace
of constant time~$t$.

The Einstein equations (\ref{subeq:Ein-syn}) and conservation laws
(\ref{subeq:cons-laws}) and~(\ref{deeltjes}) describe a universe
filled with a perfect fluid and with a positive cosmological
constant. The  fluid pressure $p$ is described by an equation of
state of the form~(\ref{toestand}): in this stage we only need
that it is some function of the particle number density $n$ and
the energy density~$\varepsilon$.

It is now time to actually derive the zero  and first order
Einstein equations. To that end, we expand all quantities in the
form of series, and derive, recursively, equations for the
successive terms of these series. Furthermore, we will now limit
the discussion to a particular class of universes, namely the
collection of universes that, apart from a small, local
perturbation in space-time, are homogeneous and isotropic, the
so-called Friedmann-Lema\^\i tre-Robertson-Walker (\textsc{flrw})
universes.

\section{Zero and first order equations for the FLRW universe}   \label{backpert}

Consider space-times which are a continuous collection (foliation)
of three-dimensional, space-like slices of space-time, each of
which is maximally symmetric. Expressed in synchronous
coordinates, this statement means that we consider spaces with
metric $g_{\mu\nu}$ given by
\begin{subequations}
\label{subeq:metric}
\begin{eqnarray}
  && g_{00}(t,\vec{x})=1, \quad g_{0i}(t,\vec{x})=0, \label{m1}   \\
  && g_{ij}(t,\vec{x})=-a^2(t)\tilde{g}_{ij}(\vec{x}),   \label{m2}
\end{eqnarray}
\end{subequations}
where $\tilde{g}_{ij}(\vec{x})$ is the metric of such a
three-dimensional maximally symmetric subspace. The minus sign
in~(\ref{m2}) has been introduced in order to switch from the
conventional four-dimensional space-time with signature
$(+,-,-,-)$ to the conventional three-dimensional spatial metric
with signature $(+,+,+)$. We write $a^2(t)$ rather than $a(t)$ for
the time dependent proportionality factor in~(\ref{m2}), the
so-called scale factor or expansion factor of the universe, for
reasons that will become clear later: the scale factor $a(t)$ will
turn out to be identifiable in some cases, as the `radius of the
universe', i.e., the slices of the foliation are three-dimensional
surfaces in four-space, with radius~$a(t)$: see
Eq.~(\ref{spRicci}). Essentially, however, we only suppose that
the three-part of the metric can be factorized according
to~(\ref{m2}), also in case the slices are no surfaces of
hyperspheres.

We now expand all quantities concerned in series. We will
distinguish the successive terms of a series by a subindex between
brackets.
\begin{subequations}
\label{subeq:exp-scalar}
\begin{eqnarray}
  \varepsilon & = & \varepsilon_\nul +
     \eta \varepsilon_\een +
  \eta^2 \varepsilon_\twee+\cdots, \label{F1} \\
  n & = & n_\nul + \eta n_\een +
     \eta^2 n_\twee+\cdots,      \label{F2} \\
  p & = & p_\nul + \eta p_\een +
    \eta^2 p_\twee+\cdots,         \label{F3} \\
  \theta & = & \theta_\nul +
     \eta \theta_\een +
    \eta^2 \theta_\twee+\cdots,         \label{F3a} \\
  \vartheta & = & \vartheta_\nul +
     \eta \vartheta_\een +
    \eta^2 \vartheta_\twee+\cdots,         \label{F3b} \\
 \mbox{$^3\!R$} & = & \mbox{$^3\!R_\nul$} +
   \eta\, \mbox{$^3\!R_\een$} +
    \eta^2\, \mbox{$^3\!R_\twee$} + \cdots, \label{F9}
\end{eqnarray}
\end{subequations}
where the subindex zero refers to quantities of the unperturbed,
homogeneous and isotropic \textsc{flrw} universe. In order to
derive the background and first order Einstein equations, we need
ancillary quantities, which will also be expanded in series:
\begin{subequations}
\label{subeq:exp-vec-tens}
\begin{eqnarray}
  u^\mu & = & u^\mu_\nul +
   \eta u^\mu_\een +
     \eta^2 u^\mu_\twee + \cdots, \label{F4} \\
  g_{ij} & = & g_{\nul ij} +
      \eta g_{\een ij} +
     \eta^2 g_{\twee ij} + \cdots, \label{F5} \\
  \varkappa_{ij} & = & \varkappa_{\nul ij} +
     \eta \varkappa_{\een ij} +
     \eta^2 \varkappa_{\twee ij} + \cdots, \label{F6} \\
  T^\mu{}_\nu & = & T^\mu_{\nul\nu} +
      \eta T^\mu_{\een\nu} +
    \eta^2 T^\mu_{\twee\nu} + \cdots, \label{F4a} \\
\mbox{$^3\!R_{ij}$} & = &
      \mbox{$^3\!R_{\nul ij}$} +
      \eta\, \mbox{$^3\!R_{\een ij}$} +
  \eta^2\,\mbox{$^3\!R_{\twee ij}$} + \cdots, \label{F8} \\
 \Gamma^k{}_{ij} & = & \Gamma^k_{\nul ij} +
    \eta \Gamma^k_{\een ij} +
    \eta^2 \Gamma^k_{\twee ij} + \cdots. \label{F7}
\end{eqnarray}
\end{subequations}
In Eqs.~(\ref{subeq:exp-scalar}) and~(\ref{subeq:exp-vec-tens})
$\eta$ ($\eta=1$) is a bookkeeping parameter, the function of
which is to enable us in actual calculations to easily distinguish
between the terms of different orders.

\subsection{Zero order  quantities} \label{backpert-zero}

This section is concerned with the background or zero order
quantities occurring in the Einstein equations. All results of this
section are totally standard, and given here only to fix the
notation unambiguously.

A tensor in a maximally symmetric space is called maximally
form-invariant if its Lie-derivatives with respect to all Killing
vectors of the maximally symmetric space vanish. Essentially, this
means that the tensor is `the same' in all directions and at all
places, just as the metric of a maximally symmetric space is `the
same' in all directions and all places. It can be proved that the
only maximally form-invariant scalar function which exists in a
maximally symmetric space is a constant, that the only maximally
form-invariant vector in a maximally symmetric space is the null
vector, and that the only maximally form-invariant second rank
tensor in a maximally symmetric space of dimension $N\ge3$ is
proportional to the metric tensor (see Weinberg~\cite{c8},
Chap.~13). In our case, of a four-dimensional space-time which is a
foliation of three-dimensional maximally symmetric spaces, this
leads to the following, general, conclusions. A scalar can only be a
function of time and the spatial part $A^i$ of a vector $A^\mu$ is
the null vector. Furthermore, the spatial components $F^{ij}$ of a
tensor $F^{\mu\nu}$ must be proportional to the metric
$\tilde{g}^{ij}$ and the components $F^{0j}$ are identically zero.

In particular, the background energy density
$\varepsilon_\nul(t,\vec{x})$, Eq.~(\ref{F1}), a space-time scalar,
is independent of the coordinates $\vec{x}=(x^1, x^2, x^3)$ of the
maximally symmetric subspace. Similarly, the background particle
number density~$n_\nul$, Eq.~(\ref{F2}), the background
pressure~$p_\nul$, Eq.~(\ref{F3}), and the fluid expansion scalar
$\theta_\nul$, Eq.~(\ref{F3a}), all space-time scalars, are
functions of time only. Thus we have, with respect to our particular
coordinates,
\begin{equation}\label{scalar4}
   \varepsilon_\nul(t,\vec{x})=\varepsilon_\nul(t), \quad
     n_\nul(t,\vec{x})=n_\nul(t), \quad
    p_\nul(t,\vec{x})=p_\nul(t), \quad
    \theta_\nul(t,\vec{x})=\theta_\nul(t).
\end{equation}
The three-divergence $\vartheta_\nul$, Eq.~(\ref{F3b}), and the
curvature of the three-dimensional subspaces~$\mbox{$^3\!R_\nul$}$,
Eq.~(\ref{F9}), which are scalars only with respect to spatial
coordinate transformations, are also functions of time only, i.e.,
\begin{equation}\label{scalar3}
  \vartheta_\nul(t,\vec{x})=\vartheta_\nul(t), \quad
      \mbox{$^3\!R_\nul$}(t,\vec{x})=\mbox{$^3\!R_\nul$}(t).
\end{equation}
Furthermore, the component of the four-vector~$u^\mu$,
Eq.~(\ref{F4}), tangent to a maximally symmetric space is a
three-vector in that maximally symmetric space. The only maximally
form-invariant vector in a maximally symmetric space is the null
vector. Hence,
\begin{equation}
   u^i_\nul = 0.   \label{u0i}
\end{equation}
Consequently, $u^\mu$ is proportional to $\delta^\mu{}_0$.
Moreover, since $u^\mu$ is a unit vector, we have
\begin{equation}
    u^\mu_\nul=\delta^\mu{}_0.  \label{u0}
\end{equation}
The zero order background metric tensor, occurring in
Eq.~(\ref{F5}), has been supposed to be that of a maximally
symmetric space, i.e.,
\begin{equation}
   g_{\nul ij}(t,\vec{x})=-a^2(t)\tilde{g}_{ij}(\vec{x}),
       \label{metricmaxsym}
\end{equation}
compare Eq.~(\ref{m2}). The time derivative of the three-part of
the metric $g_{\nul ij}$, $\varkappa_{\nul ij}$, Eq.~(\ref{F6}),
may be expressed in the usual Hubble function
$\mathcal{H}(t):=(\dif a/\dif t)/a(t)$. We prefer to use a
function $H(t)=c^{-1}\mathcal{H}(t)$, which we will call Hubble
function also. Recalling that a dot denotes differentiation with
respect to $ct$, we have
\begin{equation}
        H:=\frac{\dot{a}}{a}. \label{Hubble}
\end{equation}
Substituting the expansion (\ref{F5}) into the definitions
(\ref{def-gammas}), we obtain
\begin{equation}\label{metricFRW}
  \varkappa_{\nul ij} = -Hg_{\nul ij}, \!\!\quad \!
  \varkappa^i_{\nul j} = -H\delta^i{}_j, \!\!\quad \!
  \varkappa_\nul^{ij} = -Hg_\nul^{ij},
\end{equation}
where we considered only terms up to the zero order in the
bookkeeping parameter $\eta$.

Similarly, with Eqs.~(\ref{vierdiv}), (\ref{driediv}),
(\ref{F3a}), (\ref{F3b}), (\ref{u0}) and~(\ref{metricFRW}) we find
for the background fluid expansion scalar, $\theta_\nul$, and the
three-divergence, $\vartheta_\nul$,
\begin{equation}
    \theta_\nul = 3 H, \quad
       \vartheta_\nul = 0.   \label{fes2}
\end{equation}

Using Eqs.~(\ref{Tmunu}), (\ref{F1}), (\ref{F3}), (\ref{F5}),
(\ref{u0}) and~(\ref{metricmaxsym}) we find for the components of
the energy momentum tensor, Eq.~(\ref{F4a}),
\begin{equation}\label{emt}
  T^0_{\nul 0} = \varepsilon_\nul, \quad
     T^i_{\nul 0} =  0, \quad
      T^i_{\nul j} = -p_\nul\delta^i{}_j,
\end{equation}
where the background pressure $p_\nul$ is given by the equation of
state~(\ref{toestand}), which, for the background pressure, is
defined by
\begin{equation}
   p_\nul = p(n_\nul,\varepsilon_\nul).
       \label{toestandback}
\end{equation}
i.e., the subindex zero in $p_\nul$ refers to the zero order
quantities it depends on; it is not a different function of its
arguments.

Finally, the Ricci tensor of the three-dimensional maximally
symmetric subspaces is proportional to the metric tensor of that
subspace, i.e.,
\begin{equation}
    \mbox{$^3\!R_{\nul ij}$} = K \tilde{g}_{ij},
           \label{Riccimaxsym}
\end{equation}
(see Weinberg~\cite{c8}, Chap.~15, Sec.~1). The quantity~$K$ is
time independent, as may be seen as follows. As follows from
Eqs.~(\ref{F8}) and~(\ref{F7}), the background three-dimensional
Ricci tensor, (\ref{Ricci-drie}), is given by
\begin{equation}
  \mbox{$^3\!R_{\nul ij}$} =
   \Gamma^k_{\nul ij,k}-\Gamma^k_{\nul ik,j}+
   \Gamma^l_{\nul ij} \Gamma^k_{\nul kl}-
   \Gamma^l_{\nul ik}
   \Gamma^k_{\nul jl},      \label{Ricci-drie-back}
\end{equation}
where the connection coefficients $\Gamma^k_{\nul ij}$ are given
by
\begin{equation}
   \Gamma^k_{\nul ij}=\tfrac{1}{2} g^{kl}_\nul
   \left( g_{\nul li,j}+g_{\nul lj,i}-g_{\nul ij,l} \right),  \label{cc0}
\end{equation}
where $g^{ij}_\nul$ and $g_{\nul ij}$ depend on time. From
Eq.~(\ref{metricmaxsym}), we have
\begin{equation}
   g_\nul^{ij}(t,\vec{x}) = -\frac{1}{a^2(t)}
      \tilde{g}^{ij}(\vec{x}).   \label{A36}
\end{equation}
Hence, the connection coefficients $\Gamma^k_{\nul ij}$ are equal
to the connection coefficients $\tilde{\Gamma}^k{}_{ij}$ of the
metric~$\tilde{g}_{ij}$:
\begin{equation}
    \Gamma^k_{\nul ij} = \tilde{\Gamma}^k{}_{ij}:=
      \tfrac{1}{2} \tilde{g}^{kl}
   \left( \tilde{g}_{li,j}+\tilde{g}_{lj,i}-\tilde{g}_{ij,l}\right).
      \label{con3FRW}
\end{equation}
Therefore, they do not depend on time. As a consequence,
$\mbox{$^3\!R_{\nul ij}$}$ is time independent, implying that $K$ is
a constant. Maximally symmetric spaces may have curvature~$K$ that
is positive, negative or zero. In the latter case the space is
called \emph{flat}.

We now choose coordinates $(r,\theta,\phi)$ such that the metric
coefficients $g_{\nul ij}$ get the well-known Robertson-Walker
form
\begin{equation}
  g_{\nul ij} = -a^2(t)\left(
    \begin{array}{ccc}
      \dfrac{1}{1-kr^2}    & 0   & 0                  \\
            0              & r^2 & 0                  \\
            0              &   0 &   r^2\sin^2\theta
    \end{array}\right),  \label{gFLRW}
\end{equation}
where $k=0,\pm1$. Comparing (\ref{metricmaxsym}) and (\ref{gFLRW})
we see that for this choice of coordinates we have
\begin{equation}
   \tilde{g}_{ij}=\mathrm{diag}
      \left(\frac{1}{1-kr^2}, r^2, r^2\sin^2\theta \right).
\label{tildegij}
\end{equation}
Substituting~(\ref{gFLRW}) into~(\ref{Ricci-drie-back}), combined
with~(\ref{con3FRW}), we find
\begin{equation}
    \mbox{$^3\!R_{\nul ij}$} = 2k \tilde{g}_{ij}.   \label{RFLRW}
\end{equation}
We thus find, for the chosen \textsc{rw} coordinates
\begin{equation}
    K = 2k.  \label{K2k}
\end{equation}
From Eqs.~(\ref{RFLRW}) and~(\ref{K2k}) we have
\begin{equation}
   \mbox{$^3\!R^i_{\nul j}$}(t) = -\frac{2k}{a^2(t)}\delta^i{}_j,
       \label{Rijmixed}
\end{equation}
implying that the zero order curvature scalar
$\mbox{$^3\!R_\nul$}=g^{ij}_\nul\,\mbox{$^3\!R_{\nul ij}$}$ is
given by
\begin{equation}
    \mbox{$^3\!R_\nul$}(t) = -\frac{6k}{a^2(t)},    \label{spRicci}
\end{equation}
where we have used Eqs.~(\ref{A36}) and~(\ref{RFLRW}). Note, that
in view of our choice of the metric $(+,-,-,-)$, spaces of
positive curvature~$k$ (such as spheres) have a negative curvature
scalar~$\mbox{$^3\!R_\nul$}$.

The results~(\ref{metricmaxsym}), (\ref{metricFRW}), (\ref{emt})
and~(\ref{RFLRW}) are in agreement with the general observation
that maximally form-invariant tensors of the second rank are
proportional to the metric tensor of the space concerned.

Thus we have found all background quantities: they are either
space independent or proportional to $g_{\nul ij}(t,\vec{x})$,
Eq.~(\ref{metricmaxsym}). The latter is proportional to
$\tilde{g}_{ij}(\vec{x})$, the metric characteristic for spatial
sections of constant time.

\subsection{First order quantities}\label{backpert-first}

In this quite technical section we express all quantities occurring
in the Einstein equations in terms of zero and first order
quantities. The equations of state for the energy and pressure,
$\varepsilon(n,T)$ and $p(n,T)$, are not specified yet.

Upon substituting the series~(\ref{F4}) into the normalization
condition $u^\mu u_\mu=1$, one finds, equating equal powers of the
bookkeeping parameter~$\eta$,
\begin{equation}
   u^0_\een=0,  \label{du0}
\end{equation}
for the first order perturbation to the four-velocity. Writing the
inverse of~(\ref{F5}) as
\begin{equation}
   g^{kl}=g_\nul^{kl} +
      \eta g_\een^{kl} + \cdots, \label{gup}
\end{equation}
where $g_\nul^{kl}$ is the inverse, (\ref{A36}), of $g_{\nul kl}$,
(\ref{gFLRW}), we find
\begin{equation}
   g_\een^{kl} = -g_\nul^{ki}
         g_\nul^{lj}
         g_{\een ij},      \label{g1up}
\end{equation}
and
\begin{equation}
   g^k_{\een i} =
     -g^{kl}_\nul
     g_{\een li}.  \label{mixedg}
\end{equation}
It is convenient to introduce
\begin{equation}
   h_{ij} := - g_{\een ij}, \label{hij}
\end{equation}
so that
\begin{equation}
  h^{ij}=g_\een^{ij}, \quad
      h^i{}_j=g_\nul^{ik} h_{kj}.   \label{hijgij}
\end{equation}
For the time derivative of the first order perturbations to the
metric, $\varkappa_{\een ij}$, Eq.~(\ref{def-gammas}), we get
\begin{equation}
  \varkappa_{\een ij}=\tfrac{1}{2}\dot{h}_{ij},  \quad
  \varkappa^i_{\een j}= \tfrac{1}{2}\dot{h}^i{}_j,  \quad
  \varkappa^{ij}_\een=\tfrac{1}{2}\dot{h}^{ij}.
\label{dgam}
\end{equation}

The first order perturbation~$\theta_\een$ to the fluid expansion
scalar~$\theta$, Eq.~(\ref{vierdiv}), can be found in the same
way. Using~(\ref{F3a}) and (\ref{u0}) one arrives~at
\begin{equation}
  \theta_\een=
  \vartheta_\een-\tfrac{1}{2}\dot{h}^k{}_k, \label{fes5}
\end{equation}
where we used Eqs.~(\ref{du0}) and~(\ref{dgam}). The first order
perturbation $\vartheta_\een$ to the three-divergence $\vartheta$,
Eq.~(\ref{driediv}), is
\begin{equation}
   \vartheta_\een = u^k_{\een|k},      \label{den1a}
\end{equation}
where we have used that
\begin{equation}
   (u^k{}_{|k})_\een= u^k_{\een|k},  \label{div-1}
\end{equation}
which is a consequence of
\begin{equation}
   \Gamma^k_{\een lk} u_\nul^l = 0.
\end{equation}
The latter equality follows from Eq.~(\ref{u0i}).

Upon substituting the series expansion (\ref{F1}), and
(\ref{F3})--(\ref{F5}) into Eq.~(\ref{Tmunu}) and equating equal
powers of $\eta$, one finds for the first order perturbation to
the energy-momentum tensor
\begin{eqnarray}
    T^0_{\een 0} & = & \varepsilon_\een, \nonumber \\
    T^i_{\een 0} & = & (\varepsilon_\nul+p_\nul) u^i_\een,   \label{pertemt} \\
    T^i_{\een j} & = & -p_\een \delta^i{}_j, \nonumber
\end{eqnarray}
where we have used Eqs.~(\ref{u0}) and~(\ref{du0}). The first
order perturbation to the pressure is related to
$\varepsilon_\een$ and $n_\een$ by the first order perturbation to
the equation of state~(\ref{toestand}). We have
\begin{equation}
    p_\een=p_n n_\een +
    p_\varepsilon \varepsilon_\een,   \label{perttoes}
\end{equation}
where~$p_n$ and~$p_\varepsilon$ are the partial derivatives
of~$p(n,\varepsilon)$ with respect to~$n$ and~$\varepsilon$,
\begin{equation}
    p_n := \left( \frac{\partial p}{\partial n} \right)_{\!\varepsilon}, \quad
  p_\varepsilon :=
        \left( \frac{\partial p}{\partial \varepsilon} \right)_{\!n}.
         \label{perttoes1}
\end{equation}
Since we consider only first order quantities, the partial
derivatives are functions of the background quantities only, i.e.,
\begin{equation}\label{eq:pn-pe-back}
    p_n=p_n(n_\nul,\varepsilon_\nul), \quad
    p_\varepsilon=p_\varepsilon(n_\nul,\varepsilon_\nul).
\end{equation}

Using Eq.~(\ref{con3}) and the expansions~(\ref{F5})
and~(\ref{F7}), we find for the first order perturbations of the
connection coefficients
\begin{equation}
  \Gamma^k_{\een ij}=-g^{kl}_\nul g_{\een lm}\Gamma^m_{\nul ij}+
    \tfrac{1}{2} g^{kl}_\nul \left( g_{\een li,j}+ g_{\een lj,i}-
       g_{\een ij,l} \right).  \label{cc1}
\end{equation}
The first order perturbation $\Gamma^k_{\een ij}$,
Eq.~(\ref{cc1}), occurring in the non-tensor $\Gamma^k{}_{ij}$,
Eq.~(\ref{F7}), happen to be expressible as a tensor. Indeed,
using Eq.~(\ref{hij}), one can rewrite Eq.~(\ref{cc1}) in the form
\begin{equation}
    \Gamma^k_{\een ij}=
   -\tfrac{1}{2} g^{kl}_\nul
     (h_{li|j}+h_{lj|i}-h_{ij|l}). \label{con3pert}
\end{equation}
Using the expansion for $\mbox{$^3\!R_{ij}$}$, (\ref{F8}), and
$\Gamma^k{}_{ij}$, (\ref{F7}), one finds for the first order
perturbation to the Ricci tensor~(\ref{Ricci-drie})
\begin{equation}
  \mbox{$^3\!R_{\een ij}$} =
\Gamma^k_{\een ij,k}-\Gamma^k_{\een ik,j}+
   \Gamma^l_{\nul ij} \Gamma^k_{\een kl}+
   \Gamma^l_{\een ij} \Gamma^k_{\nul kl}-
   \Gamma^l_{\nul ik} \Gamma^k_{\een jl}-
   \Gamma^l_{\een ik}\Gamma^k_{\nul jl}, \label{Ricci-drie-pert}
\end{equation}
which can be rewritten in the compact form
\begin{equation}
   \mbox{$^3\!R_{\een ij}$} =
\Gamma^k_{\een ij|k}-\Gamma^k_{\een ik|j}.
    \label{palatini}
\end{equation}
By substituting Eq.~(\ref{con3pert}) into Eq.~(\ref{palatini}),
one can express the first order perturbation to the Ricci tensor
of the three-dimensional subspace in terms of the perturbation to
the metric and its covariant derivatives
\begin{equation}
  \mbox{$^3\!R_{\een ij}$} =
   -\tfrac{1}{2} g^{kl}_\nul
     (h_{li|j|k}+h_{lj|i|k}-h_{ij|l|k}-h_{lk|i|j}). \label{deltaR3}
\end{equation}
The perturbation $\mbox{$^3\!R^i_{\een j}$}$ is given by
\begin{equation}
   \mbox{$^3\!R^i_{\een j}$}:=(g^{ip}\,\mbox{$^3\!R_{pj}$})_\een=
      g^{ip}_\nul\,\mbox{$^3\!R_{\een pj}$} +
      \tfrac{1}{3}\,\mbox{$^3\!R_\nul$} h^i{}_j,   \label{mixedRij}
\end{equation}
where we have used Eqs.~(\ref{metricmaxsym}), (\ref{RFLRW}),
(\ref{spRicci}) and~(\ref{hijgij}). Upon substituting
Eq.~(\ref{deltaR3}) into Eq.~(\ref{mixedRij}) we get
\begin{equation}
  \mbox{$^3\!R^i_{\een j}$} =
      -\tfrac{1}{2}g^{ip}_\nul(h^k{}_{p|j|k}+h^k{}_{j|p|k}-h^k{}_{k|p|j})+
      \tfrac{1}{2}g^{kl}_\nul h^i{}_{j|k|l} +
      \tfrac{1}{3}\,\mbox{$^3\!R_\nul$} h^i{}_j.   \label{R1mixed}
\end{equation}
Taking $i=j$ in Eq.~(\ref{R1mixed}) and summing over the repeated
index, we find for the first order perturbation to the curvature
scalar of the three-dimensional spaces
\begin{equation}
\mbox{$^3\!R_\een$} =
   g_\nul^{ij} (h^k{}_{k|i|j}-h^k{}_{i|j|k}) +
     \tfrac{1}{3}\,\mbox{$^3\!R_\nul$} h^k{}_k.   \label{driekrom}
\end{equation}
We thus have expressed all quantities occurring in the relevant
dynamical equations, i.e., the system of equations formed by the
Einstein equations combined with the conservation laws, in terms
of zero and first order quantities to be solved from these
equations. In the Secs.~\ref{nulde} and \ref{eerste} below we
derive the background and first order evolution equations,
respectively. To that end we substitute the
series~(\ref{subeq:exp-scalar}) and~(\ref{subeq:exp-vec-tens})
into the Einstein equations (\ref{subeq:Ein-syn}) and conservation
laws (\ref{subeq:cons-laws}) and~(\ref{deeltjes}). By equating the
powers of $\eta^0$, $\eta^1$, \ldots, we obtain the zero order,
the first order and higher order dynamical equations, constraint
equations and conservation laws. We will carry out this scheme for
the zero  and first order equations only.

\subsection{Zero order equations}  \label{nulde}

With the help of Sec.~\ref{backpert-zero} and the
expansions~(\ref{subeq:exp-scalar}) and~(\ref{subeq:exp-vec-tens})
we now can find from the Einstein equations~(\ref{subeq:Ein-syn})
and conservation laws~(\ref{subeq:cons-laws}) and~(\ref{deeltjes})
the zero order Einstein equations and the conservation laws.
Furthermore, in view of the symmetry induced by the isotropy, it
is possible to switch from the six quantities $g_{ij}$ and the six
quantities $\varkappa_{ij}$ to the curvature
$\mbox{$^3\!R_\nul$}(t)$ and the Hubble function $H(t)$ only.

\subsubsection{Einstein equations}\label{nulde-einstein}

Upon substituting Eqs.~(\ref{metricFRW}) and~(\ref{emt}) into the
$(0,0)$-component of the Einstein equations, Eq.~(\ref{Ein-syn1}),
one finds
\begin{equation}
   3H^2-\tfrac{1}{2}\,\mbox{$^3\!R_\nul$}=
     \kappa\varepsilon_\nul + \Lambda.        \label{endencon0}
\end{equation}
The $(0,i)$-components of the Einstein equations,
Eqs.~(\ref{Ein-syn2}), are identically fulfilled, as follows from
Eqs.~(\ref{metricFRW}) and~(\ref{emt}). We thus are left with the
six $(i,j)$-components of the Einstein equations,
Eqs.~(\ref{Ein-syn3}). In view of Eqs.~(\ref{metricFRW}),
(\ref{emt}) and~(\ref{Rijmixed}) we find that $\varkappa^1_{\nul
1}=\varkappa^2_{\nul 2}=\varkappa^3_{\nul 3}$, $T^1_{\nul
1}=T^2_{\nul 2}=T^3_{\nul 3}$ and $\mbox{$^3\!R^1_{\nul
1}$}=\mbox{$^3\!R^2_{\nul 2}$}= \mbox{$^3\!R^3_{\nul 3}$}$,
whereas for $i\neq j$ these quantities vanish. Hence, the six
$(i,j)$-components reduce to one equation,
\begin{equation}
    \dot{H} = -3H^2+\tfrac{1}{3}\,\mbox{$^3\!R_\nul$}
   +\tfrac{1}{2}\kappa(\varepsilon_\nul - p_\nul) + \Lambda.    \label{dyn0}
\end{equation}
In Eqs.~(\ref{endencon0})~and~(\ref{dyn0}) the background
curvature~$\mbox{$^3\!R_\nul$}$ is given by Eq.~(\ref{spRicci}).
It is, however, of convenience to determine this quantity from a
differential equation. Eliminating $a(t)$ from Eqs.~(\ref{Hubble})
and~(\ref{spRicci}) we obtain
\begin{equation}
  \mbox{$^3\!\dot{R}_\nul$} +
    2H\,\mbox{$^3\!R_\nul$} = 0,   \label{momback}
\end{equation}
where the initial value~$\mbox{$^3\!R_\nul(t_0)$}$ is given by
\begin{equation}\label{eq:init-R0}
   \mbox{$^3\!R_\nul(t_0)$} = -\dfrac{6k}{a^2(t_0)},
\end{equation}
in accordance with Eq.~(\ref{spRicci}). It should be emphasized
that Eq.~(\ref{momback}) is not an Einstein equation, since it is
equivalent to Eq.~(\ref{spRicci}). It will be used here as an
ancillary relation.

\subsubsection{Conservation laws}\label{nulde-conservation}

Upon substituting Eqs.~(\ref{metricFRW}) and~(\ref{emt}) into the
$0$-component of the conservation law, Eq.~(\ref{Tnulnu}), one
finds
\begin{equation}
   \dot{\varepsilon}_\nul +
    3H(\varepsilon_\nul+p_\nul)=0,
   \label{energyFRW}
\end{equation}
which is the relativistic background continuity equation. The
background momentum conservation laws (i.e., the background
relativistic Euler equations) are identically satisfied, as
follows by substituting (\ref{metricFRW}) and~(\ref{emt}) into the
spatial components of the conservation laws, Eq.~(\ref{Tinu}).

The background particle number density conservation law can be
found by substituting (\ref{u0}) and~(\ref{fes2}) into
Eq.~(\ref{deeltjes}). One gets
\begin{equation}
   \dot{n}_\nul + 3H n_\nul = 0.   \label{deel1}
\end{equation}
This concludes the derivation of the background equations.

What we have found is that for the Robertson-Walker
metric~(\ref{gFLRW}) the Einstein equations~(\ref{subeq:Ein-syn})
and conservation laws~(\ref{subeq:cons-laws}) and~(\ref{deeltjes})
reduce to the initial value condition~(\ref{endencon0}) and four
differential equations~(\ref{dyn0}), (\ref{momback}),
(\ref{energyFRW}) and~(\ref{deel1}) for the four unknown functions
$H$, $\mbox{$^3\!R_\nul$}$, $\varepsilon_\nul$ and $n_\nul$. If we
use Eqs.~(\ref{fes2}) and~(\ref{toestandback}), we can state that
we have found equations for all unknown background
quantities~(\ref{subeq:exp-scalar}) in zero order approximation,
namely $\theta_\nul$, $\vartheta_\nul$, $\mbox{$^3\!R_\nul$}$,
$\varepsilon_\nul$ and $n_\nul$. In Sec.~\ref{evo-scal} we derive
equations for the corresponding first order quantities
$\theta_\een$, $\vartheta_\een$, $\mbox{$^3\!R_\een$}$,
$\varepsilon_\een$ and $n_\een$.

\subsection{First order equations}   \label{eerste}

In this section we derive the first order perturbation equations
from the Einstein equations~(\ref{subeq:Ein-syn}) and conservation
laws~(\ref{subeq:cons-laws}) and~(\ref{deeltjes}). The procedure is,
by now, completely standard. We use the series expansion in~$\eta$
for the various quantities occurring in the Einstein equations and
conservation laws of energy-momentum, and we equate the coefficients
linear in~$\eta$ to obtain the `linearized' or first order
equations.

\subsubsection{Einstein equations}  \label{eerste-einstein}

Using the series expansions for $\mbox{$^3\!R$}$, (\ref{F9}),
$\varkappa^i{}_j$, (\ref{F6}), and $T^0{}_0$, (\ref{F4a}), in the
$(0,0)$-component of the constraint equation (\ref{Ein-syn1}), one
finds
\begin{equation}
  2\varkappa_{\nul k}^k \varkappa_{\een l}^l - \mbox{$^3\!R_\een$} -
  2\varkappa_{\nul l}^k \varkappa_{\een k}^l=
      2\kappa T_{\een 0}^0. \label{endenconpert}
\end{equation}
With the zero order equation (\ref{metricFRW}), the abbreviation
(\ref{dgam}) and the expression for $T^0_{\een 0}$, (\ref{pertemt}),
we may rewrite this equation in the form
\begin{equation}
   H\dot{h}^k{}_k + \tfrac{1}{2}\,\mbox{$^3\!R_\een$} =
         -\kappa\varepsilon_\een.         \label{endencon1}
\end{equation}
Employing the series expansions for $\varkappa^i{}_j$, (\ref{F6}),
and $T^0{}_i$, (\ref{F4a}), we find for the $(0,i)$-components of
the constraint equations (\ref{Ein-syn2})
\begin{equation}
   \varkappa^k_{\een k|i} - \varkappa^k_{\een i|k}=\kappa T_{\een i}^0,
        \label{momconspert}
\end{equation}
where we noted that
\begin{equation}
   (\varkappa^i{}_{j|k})_\een = \varkappa^i_{\een j|k},    \label{gamijk1}
\end{equation}
which is a consequence of
\begin{equation}
    \Gamma^i_{\een lk}\varkappa^l_{\nul j}-
           \Gamma^l_{\een jk}\varkappa^i_{\nul l}=0,
\end{equation}
which, in turn, is a direct consequence of Eq.~(\ref{metricFRW}).
From Eqs.~(\ref{dgam}) and~(\ref{pertemt}) we find
\begin{equation}
    \dot{h}^k{}_{k|i}-\dot{h}^k{}_{i|k} =
         2\kappa(\varepsilon_\nul + p_\nul) u_{\een i}.    \label{dR0i2}
\end{equation}

Finally, we consider the $(i,j)$-components of the Einstein
equations~(\ref{Ein-syn3}). Using the series expansions for
$\varkappa^i{}_j$, (\ref{F6}), $T^i{}_j$, (\ref{F4a}), and
$\mbox{$^3\!R^i{}_j$}$, (\ref{F8}), we find
\begin{equation}
     \dot{\varkappa}^i_{\een j}-\varkappa^i_{\een j}\varkappa^k_{\nul k}-
     \varkappa^i_{\nul j} \varkappa^k_{\een k} +
        \mbox{$^3\!R^i_{\een j}$} =
       \kappa(T^i_{\een j}-\tfrac{1}{2}\delta^i{}_j T^\mu_{\een\mu}).
\end{equation}
With Eqs.~(\ref{metricFRW}), (\ref{dgam}) and~(\ref{pertemt}), we
get
\begin{equation}
   \ddot{h}^i{}_j+3H\dot{h}^i{}_j +
    \delta^i{}_j H\dot{h}^k{}_k + 2\,\mbox{$^3\!R^i_{\een j}$}=
     -\kappa\delta^i{}_j(\varepsilon_\een-p_\een),  \label{ddhij}
\end{equation}
where $\mbox{$^3\!R^i_{\een j}$}$ is given by Eq.~(\ref{R1mixed}).

Note that the first order equations~(\ref{endencon1})
and~(\ref{ddhij}) are independent of the cosmological
constant~$\Lambda$: the effect of the non-zero cosmological
constant is accounted for by the zero order quantities
[cf.~Eqs.~(\ref{endencon0}) and~(\ref{dyn0})].

\subsubsection{Conservation laws}   \label{eerste-conservation}

We now consider the energy conservation law~(\ref{Tnulnu}). With
the help of the series expansions for $\varkappa^i{}_j$,
(\ref{F6}), and $T^\mu{}_\nu$, (\ref{F4a}), one finds for the
first order equation
\begin{equation}
   \dot{T}^{00}_\een+T^{0k}_\een{}_{|k}+\varkappa^k_{\nul l} T^l_{\een k} +
    \varkappa^k_{\een l} T^l_{\nul k}-\varkappa^k_{\nul k} T^{00}_\een -
     \varkappa^k_{\een k} T^{00}_\nul =0,
           \label{energypert}
\end{equation}
where we have used that for a three-vector $T^{0k}$ we have
\begin{equation}
    (T^{0k}{}_{|k})_\een = T^{0k}_\een{}_{|k},
\end{equation}
see Eq.~(\ref{div-1}). Employing Eqs.~(\ref{metricFRW}),
(\ref{emt}), (\ref{dgam}), (\ref{den1a}) and~(\ref{pertemt}) we
arrive at the first order energy conservation law
\begin{equation}
   \dot{\varepsilon}_\een+3H(\varepsilon_\een +p_\een)+
         (\varepsilon_\nul +p_\nul)(u^k_{\een|k}-\tfrac{1}{2}\dot{h}^k{}_k)=0,
               \label{enpert}
\end{equation}

Next, we consider the momentum conservation laws~(\ref{Tinu}).
With the series expansions for $\varkappa^i{}_j$, (\ref{F6}) and
$T^{\mu\nu}$, (\ref{F4a}), we find for the first order momentum
conservation law
\begin{equation}
  \dot{T}^{i0}_\een+(T^{ik}{}_{|k})_\een-2\varkappa^i_{\nul k}T^{k0}_\een-
    2\varkappa^i_{\een k} T^{k0}_\nul-\varkappa^k_{\nul k} T^{i0}_\een-
   \varkappa^k_{\een k} T^{i0}_\nul=0.         \label{Tinu1}
\end{equation}
Using that
\begin{equation}\label{eq:Tik-k}
   (T^{ik}{}_{|k})_\een=-g_\nul^{ik}p_{\een|k},
\end{equation}
and Eqs.~(\ref{metricFRW}), (\ref{emt}), (\ref{dgam})
and~(\ref{pertemt}) we arrive at
\begin{equation}
   \frac{1}{c}\frac{\dif}{\dif t}
   \Bigl[(\varepsilon_\nul+p_\nul) u^i_\een\Bigr]-
    g^{ik}_\nul p_{\een|k}+5H(\varepsilon_\nul+p_\nul) u^i_\een=0,
            \label{mom1}
\end{equation}
where we have also used that the covariant derivative of
$g^{ij}_\nul$ vanishes: $g^{ij}_{\nul |k}=0$.

Finally, we consider the particle number density conservation
law~(\ref{deeltjes}). With the expansions for $n$, (\ref{F2}),
$\theta$, (\ref{F3a}), and $u^\mu$, (\ref{F4}), it follows that the
first order equation reads
\begin{equation}
   \dot{n}_\nul u^0_\een + \dot{n}_\een u^0_\nul + n_{\nul,k}u^k_\een+
   n_{\een,k}u^k_\nul + n_\nul \theta_\een + n_\een \theta_\nul = 0.
       \label{deel2}
\end{equation}
With the help of Eqs.~(\ref{u0}), (\ref{fes2}) and~(\ref{du0}) we
find for the first order particle number conservation law
\begin{equation}
   \dot{n}_\een +3Hn_\een + n_\nul (u^k_{\een|k}-\tfrac{1}{2}\dot{h}^k{}_k) = 0,
         \label{deel3}
\end{equation}
where we employed Eq.~(\ref{fes5}) to eliminate $\theta_\een$.

\subsubsection{Summary}

In the preceding two Secs.~\ref{eerste-einstein}
and~\ref{eerste-conservation} we have found the equations which,
basically, describe the perturbations in a \textsc{flrw} universe,
in first approximation. They are Eqs.~(\ref{endencon1}),
(\ref{dR0i2}), (\ref{ddhij}), (\ref{enpert}), (\ref{mom1})
and~(\ref{deel3}). For convenience we repeat them here
\begin{subequations}
\label{subeq:basis}
\begin{eqnarray}
  && H\dot{h}^k{}_k + \tfrac{1}{2}\,\mbox{$^3\!R_\een$} =
         -\kappa\varepsilon_\een,     \label{basis-1} \\
  &&  \dot{h}^k{}_{k|i}-\dot{h}^k{}_{i|k} =
         2\kappa(\varepsilon_\nul + p_\nul) u_{\een i}, \label{basis-2} \\
  && \ddot{h}^i{}_j+3H\dot{h}^i{}_j+
       \delta^i{}_j H\dot{h}^k{}_k + 2\, \mbox{$^3\!R^i_{\een j}$}=
     -\kappa\delta^i{}_j(\varepsilon_\een-p_\een),
         \label{basis-3} \\
  && \dot{\varepsilon}_\een+3H(\varepsilon_\een +p_\een)+
       (\varepsilon_\nul +p_\nul)(u^k_{\een|k}-\tfrac{1}{2}\dot{h}^k{}_k)=0,
               \label{basis-4}  \\
  && \frac{1}{c}\frac{\dif}{\dif t}
   \Bigl[(\varepsilon_\nul+p_\nul) u^i_\een\Bigr]-
    g^{ik}_\nul p_{\een|k}+5H(\varepsilon_\nul+p_\nul) u^i_\een=0,
            \label{basis-5} \\
  &&  \dot{n}_\een+3Hn_\een+n_\nul
     (u^k_{\een|k}-\tfrac{1}{2}\dot{h}^k{}_k) = 0,  \label{basis-6}
\end{eqnarray}
\end{subequations}
where $\mbox{$^3\!R^i_{\een j}$}$ and $\mbox{$^3\!R_\een$}$ are
given by Eqs.~(\ref{R1mixed}) and~(\ref{driekrom}), respectively.
Hence, the equations~(\ref{subeq:basis}) essentially are fifteen
equations for the eleven ($6+3+1+1$) unknown functions $h^i{}_j$,
$u^i_\een$, $\varepsilon_\een$ and $n_\een$. The pressure~$p_\nul$
is given by an equation of state~(\ref{toestandback}), and the
perturbation to the pressure, $p_\een$, is given by
Eq.~(\ref{perttoes}). The system of equations is not overdetermined,
however, since the four equations (\ref{basis-1})
and~(\ref{basis-2}) are only conditions on the initial values. These
initial value conditions are fulfilled for all times $t$
automatically if they are satisfied at some (initial) time~$t=t_0$.

\section{Classification of the solutions of first order}
    \label{klasse}

The set of equations~(\ref{subeq:basis}), which are linear in their
(eleven) unknown functions, can be split up into three sets of
equations, which, together, are equivalent to the original set. We
will refer to these sets by their usual names of scalar, vector and
tensor perturbation equations. We will show that the vector and
tensor perturbations do not, in first order, contribute to the
physical perturbations~$\varepsilon_\een^\gi$ and~$n_\een^\gi$. As a
consequence, we only need, for our problem, the set of equations
which are related to the scalar perturbations. By considering only
the scalar part of the full set of perturbation equations we are
able to cast the perturbation equations into a set which is directly
related to the physical perturbations~$\varepsilon_\een^\gi$
and~$n_\een^\gi$. This is the subject of the next section,
Sec.~\ref{evo-scal}.

At the basis of the replacement of one set~(\ref{subeq:basis}) by
three sets of equations stands a theorem proved by
Stewart~\cite{York1974, Stewart}, which states that a symmetric
second rank tensor can be split up into three irreducible pieces,
and that a vector can be split up into two irreducible pieces. Here,
we will use this general theorem to obtain equations for the scalar
irreducible parts of the tensors $h^i{}_j$ and $\mbox{$^3\!R^i_{\een
j}$}$ and the vector $\vec{u}_\een$, namely $h^i_\parallel{}_j$,
$\mbox{$^3\!R^i_{\een\parallel j}$}$ and~$\vec{u}_{\een\parallel}$.

For the perturbation to the metric, a symmetric second rank
tensor, we have in particular
\begin{equation}\label{decomp-symh}
  h^i{}_j=h^i_\parallel{}_j + h^i_\perp{}_j + h^i_\ast{}_j,
\end{equation}
where, according to the theorem of Stewart~\cite{Stewart}, the
irreducible constituents $h^i_\parallel{}_j$, $h^i_\perp{}_j$ and
$h^i_\ast{}_j$ have the properties
\begin{subequations}
\label{subeq:dec-hij}
\begin{eqnarray}
    h^i_\parallel{}_j & = &
      \frac{2}{c^2}(\phi\delta^i{}_j+\zeta^{|i}{}_{|j}),
        \label{decomp-hij-par} \\
    h^k_{\perp}{}_{k} & = & 0,   \label{decomp-hij-perp} \\
    h^k_{\ast}{}_{k} & = & 0, \quad h^k_{\ast}{}_{i|k} = 0,
       \label{decomp-hij-ast}
\end{eqnarray}
\end{subequations}
with $\phi(t,\vec{x})$ and $\zeta(t,\vec{x})$ arbitrary functions.
The contravariant derivative $A^{|i}$ is defined as
$g_\nul^{ij}A_{|j}$. The functions~$h^i_\parallel{}_j$,
$h^i_{\perp}{}_{j}$ and~$h^i_{\ast}{}_{j}$ correspond to scalar,
vector and tensor perturbations, respectively.

In the same way, the perturbation to the Ricci tensor can be
decomposed into irreducible components, i.e.,
\begin{equation}\label{decomp-Rij}
   \mbox{$^3\!R^i_{\een j}$} = \mbox{$^3\!R^i_{\een\parallel j}$} +
     \mbox{$^3\!R^i_{\een\perp j}$}+\mbox{$^3\!R^i_{\een\ast j}$}.
\end{equation}
The tensors $\mbox{$^3\!R^i_{\een\parallel j}$}$,
$\mbox{$^3\!R^i_{\een\perp j}$}$ and $\mbox{$^3\!R^i_{\een\ast
j}$}$ have the properties comparable to~(\ref{subeq:dec-hij}),
i.e.,
\begin{subequations}
\label{subeq:prop-Rij}
\begin{eqnarray}
   \mbox{$^3\!R^i_{\een\parallel j}$} & = &
       \frac{2}{c^2}(\gamma\delta^i{}_j+\pi^{|i}{}_{|j}), \label{Rij-paral} \\
   \mbox{$^3\!R^k_{\een\perp k}$} & = & 0, \label{Rij-perp} \\
   \mbox{$^3\!R^k_{\een\ast k}$} & = & 0, \quad \mbox{$^3\!R^k_{\een\ast i|k}$}=0,
        \label{Rij-ast}
\end{eqnarray}
\end{subequations}
where~$\gamma(t,\vec{x})$ and~$\pi(t,\vec{x})$ are two arbitrary
functions. By now using Eq.~(\ref{R1mixed}) for each of the
irreducible parts we find
\begin{subequations}
\label{subeq:dec-Rij}
\begin{eqnarray}
 \mbox{$^3\!R^i_{\een\parallel j}$} & = & \dfrac{1}{c^2}
   \Bigl[\phi^{|i}{}_{|j}+\delta^i{}_j\phi^{|k}{}_{|k}-\zeta^{|k|i}{}_{|j|k}-
   \zeta^{|k}{}_{|j}{}^{|i}{}_{|k}+
    \zeta^{|k}{}_{|k}{}^{|i}{}_{|j}+
  \zeta^{|i}{}_{|j}{}^{|k}{}_{|k}+
  \tfrac{2}{3}\, \mbox{$^3\!R_\nul$}(\delta^i{}_j\phi+\zeta^{|i}{}_{|j})  \Bigr],
    \label{decomp-Rij-par} \\
  \mbox{$^3\!R^i_{\een\perp j}$} & = & -\tfrac{1}{2}g^{ip}_\nul(h_{\perp}^k{}_{p|j|k} +
   h_{\perp}^k{}_{j|p|k})+
   \tfrac{1}{2}g^{kl}_\nul h_{\perp}^i{}_{j|k|l} +
      \tfrac{1}{3}\,\mbox{$^3\!R_\nul$} h_{\perp}^i{}_j,  \label{decomp-Rij-perp} \\
  \mbox{$^3\!R^i_{\een\ast j}$} & = & -\tfrac{1}{2}g^{ip}_\nul(h_{\ast}^k{}_{p|j|k} +
   h_{\ast}^k{}_{j|p|k})+
    \tfrac{1}{2}g^{kl}_\nul h_{\ast}^i{}_{j|k|l} +
      \tfrac{1}{3}\,\mbox{$^3\!R_\nul$} h_{\ast}^i{}_j.
       \label{decomp-Rij-ast}
\end{eqnarray}
\end{subequations}
Combining the expressions~(\ref{subeq:prop-Rij})
and~(\ref{subeq:dec-Rij}), we can derive relations, to be obeyed
by~$\gamma$, $\pi$, $h_{\perp}^i{}_j$, and~$h_{\ast}^i{}_j$.
Firstly, from property~(\ref{Rij-paral}) and~(\ref{decomp-Rij-par})
it follows that
\begin{subequations}
\label{subeq:Rij-mu-nu}
\begin{eqnarray}
\gamma & = & \tfrac{1}{2}(\phi^{|k}{}_{|k}+\tfrac{2}{3}\,\mbox{$^3\!R_\nul$}\phi),  \\
    \pi^{|i}{}_{|j} & = & \tfrac{1}{2}(\phi^{|i}{}_{|j}-\zeta^{|k|i}{}_{|j|k}-
     \zeta^{|k}{}_{|j}{}^{|i}{}_{|k}+
    \zeta^{|k}{}_{|k}{}^{|i}{}_{|j}+
    \zeta^{|i}{}_{|j}{}^{|k}{}_{|k}+\tfrac{2}{3}\,
    \mbox{$^3\!R_\nul$}\zeta^{|i}{}_{|j}).
\end{eqnarray}
\end{subequations}
In a flat \textsc{flrw} universe, Eqs.~(\ref{subeq:Rij-mu-nu})
reduce to
\begin{subequations}
\label{subeq:Rij-mu-nu-flat}
\begin{eqnarray}
\gamma & = & \tfrac{1}{2}\phi^{|k}{}_{|k},  \\
    \pi^{|i}{}_{|j} & = & \tfrac{1}{2}\phi^{|i}{}_{|j},
\end{eqnarray}
\end{subequations}
implying that, for a flat \textsc{flrw} universe,
\begin{equation}\label{eq:flat-mu-is-nu}
    \gamma = \pi^{|k}{}_{|k},
\end{equation}
whereas there is no such restriction on the functions~$\phi$
and~$\zeta$ in a flat \textsc{flrw} universe.

Secondly, combining expressions~(\ref{Rij-perp})
and~(\ref{decomp-Rij-perp}) it follows that $h^i_\perp{}_j$ has
the property
\begin{equation}\label{eq:hklkl}
  h_{\perp}^{kl}{}_{|k|l} = 0,
\end{equation}
in addition to the property~(\ref{decomp-hij-perp}). In
Sec.~\ref{vector} we show that this additional condition is needed
to allow for the decomposition (\ref{decomp-u}).

Finally, combining Eqs.~(\ref{Rij-ast})
and~(\ref{decomp-Rij-ast}), we find that $h^i_\ast{}_j$ must obey
\begin{equation}\label{ast-add}
  \tilde{g}^{kl}(h^m_\ast{}_{k|i|m|l}+h^m_\ast{}_{i|k|m|l}-
      h^m_\ast{}_{i|k|l|m})=0,
\end{equation}
in addition to~(\ref{decomp-hij-ast}). The relations~(\ref{ast-add})
are, however, fulfilled identically for \textsc{flrw} universes.
This can easily be shown. First, we recall the well-known relation
that the difference of the covariant derivatives~$A^{i\cdots
j}{}_{k\cdots l|p|q}$ and~$A^{i\cdots j}{}_{k\cdots l|q|p}$ of an
arbitrary tensor can be expressed in terms of the curvature and the
tensor itself (Weinberg~\cite{c8}, Chap.~6, Sec.~5)
\begin{eqnarray}\label{eq:commu-Riemann}
   \lefteqn{ A^{i\cdots j}{}_{k\cdots l|p|q}-A^{i\cdots j}{}_{k\cdots l|q|p} = } \nonumber\\
  && +\, A^{i\cdots j}{}_{s\cdots l}\,\mbox{$^3\!R^s_{\nul kpq}$}+\cdots
    +A^{i\cdots j}{}_{k\cdots s}\,\mbox{$^3\!R^s_{\nul lpq}$} \nonumber \\
  && -\, A^{s\cdots j}{}_{k\cdots l}\,\mbox{$^3\!R^i_{\nul spq}$}-\cdots
    -A^{i\cdots s}{}_{k\cdots l}\,\mbox{$^3\!R^j_{\nul spq}$},
\end{eqnarray}
where~$\mbox{$^3\!R^i_{\nul jkl}$}$ is the Riemann tensor for the
spaces of constant time. At the right-hand side, there is a term
with a plus sign for each lower index and a term with a minus sign
for each upper index.

We apply this identity taking for~$A$ the second rank
tensor~$h^i_\ast{}_j$ to obtain
\begin{equation}\label{eq:pieter1}
   h^m_\ast{}_{k|i|m}-h^m_\ast{}_{k|m|i}=
      h^m_\ast{}_s\,\mbox{$^3\!R^s_{\nul kim}$}-
       h^s_\ast{}_k\,\mbox{$^3\!R^m_{\nul sim}$}.
\end{equation}
Now note that $h^m_\ast{}_{k|m|i}$ vanishes in view
of~(\ref{decomp-hij-ast}). Next, we take the covariant derivative
of~(\ref{eq:pieter1}) with respect to~$x^l$, and contract
with~$\tilde{g}^{kl}$
\begin{equation}\label{eq:commu-h-ast}
   \tilde{g}^{kl}h^m_\ast{}_{k|i|m|l}=
      \tilde{g}^{kl}(h^m_\ast{}_s\,\mbox{$^3\!R^s_{\nul kim}$}-
       h^s_\ast{}_k\,\mbox{$^3\!R^m_{\nul sim}$})_{|l}.
\end{equation}
Next, using the expression which one has for the Riemann tensor of a
maximally symmetric three-space,
\begin{equation}\label{eq:Riemann}
    \mbox{$^3\!R^a_{\nul bcd}$}=k\left(\delta^a{}_c \tilde{g}_{bd}-
         \delta^a{}_d\tilde{g}_{bc}\right),
\end{equation}
(where~$k=0,\pm1$ is the curvature constant) we find
\begin{equation}\label{eq:first-term}
    \tilde{g}^{kl}h^m_\ast{}_{k|i|m|l}=0,
\end{equation}
i.e., the first term of~(\ref{ast-add}) vanishes. The second and
third term can similarly be expressed in the curvature
\begin{equation}\label{eq:commu-h-ast-2}
    h^m_\ast{}_{i|k|m|l}-h^m_\ast{}_{i|k|l|m}=
    h^m_\ast{}_{s|k}\,\mbox{$^3\!R^s_{\nul iml}$}+
    h^m_\ast{}_{i|s}\,\mbox{$^3\!R^s_{\nul kml}$}-
    h^s_\ast{}_{i|k}\,\mbox{$^3\!R^m_{\nul sml}$},
\end{equation}
where the general property~(\ref{eq:commu-Riemann}) has been used.
Upon substituting the Riemann tensor~(\ref{eq:Riemann}) and
contracting with~$\tilde{g}^{kl}$, we then arrive at
\begin{equation}\label{eq:sec-third-term}
    \tilde{g}^{kl}(h^m_\ast{}_{i|k|m|l}-h^m_\ast{}_{i|k|l|m})=0,
\end{equation}
i.e., the second and third term of~(\ref{ast-add}) together vanish.
Hence, for \textsc{flrw} universes, Eq.~(\ref{ast-add}) is
identically fulfilled. Consequently, the decomposition
(\ref{decomp-hij-ast}) imposes no additional condition on the
irreducible part $h^i_{\ast j}$ of the perturbation $h^i{}_j$.

The three-vector $\vec{u}_\een$ can be uniquely split up according
to~\cite{Stewart}
\begin{equation}
  \vec{u}_{\een} = \vec{u}_{\een\parallel} +
               \vec{u}_{\een\perp},      \label{decomp-u}
\end{equation}
where $\vec{u}_{\een\parallel}$ is the \emph{longitudinal} part of
$\vec{u}_{\een}$, with the properties
\begin{equation}
   \vec{\tilde{\nabla}}\wedge(\vec{u}_{\een\parallel})=0, \quad
   \vec{\tilde{\nabla}}\cdot\vec{u}_\een=
      \vec{\tilde{\nabla}}\cdot\vec{u}_{\een\parallel},
       \label{longitudinal}
\end{equation}
and $\vec{u}_{\een\perp}$ is the \emph{transverse} part of
$\vec{u}_{\een}$, with the properties
\begin{equation}
   \vec{\tilde{\nabla}}\cdot\vec{u}_{\een\perp}=0, \quad
   \vec{\tilde{\nabla}}\wedge\vec{u}_\een=
       \vec{\tilde{\nabla}}\wedge\vec{u}_{\een\perp},
       \label{transversal}
\end{equation}
where the divergence of the vector $\vec{u}_\een$ is defined by,
Eq.~(\ref{driediv}),
\begin{equation}
     \vec{\tilde{\nabla}}\cdot\vec{u}_\een:=u^k_{\een}{}_{|k}=\vartheta_\een,
             \label{divergence}
\end{equation}
and the rotation of the vector $\vec{u}_\een$ is defined by
\begin{equation}
   (\vec{\tilde{\nabla}}\wedge\vec{u}_\een)_i:=
          \epsilon_i{}^{jk}u_{\een j|k}=\epsilon_i{}^{jk}u_{\een j,k},
     \label{rotation}
\end{equation}
where $\epsilon_i{}^{jk}$ is the Levi-Civita tensor with
$\epsilon_1{}^{23}=+1$. In Eq.~(\ref{rotation}) we could replace
the covariant derivative by the ordinary partial derivative
because of the symmetry of $\Gamma^i{}_{jk}$.

Having decomposed the tensors $h^i{}_j$, $\mbox{$^3\!R^i_{\een j}$}$
and $u^i_\een$ in a scalar~$\parallel$, a vector~$\perp$ and a
tensor part~$\ast$, we can now decompose the set of
equations~(\ref{subeq:basis}) into three independent sets. The
recipe is simple: all we have to do is to append a
subindex~$\parallel\,$, $\perp$ or~$\ast$ to the relevant tensorial
quantities in equations~(\ref{subeq:basis}). This will be the
subject of the Secs.~\ref{tensor}, \ref{vector} and~\ref{scalar}
below.

\subsection{Tensor perturbations}   \label{tensor}

We will show that tensor perturbations are not coupled to, i.e.,
do not give rise to, density perturbations.

Upon substituting $h^i{}_j=h^i_\ast{}_j$ and $\mbox{$^3\!R^i_{\een
j}$}= \mbox{$^3\!R^i_{\een\ast j}$}$ into the perturbation
equations~(\ref{subeq:basis}) and using the
properties~(\ref{decomp-hij-ast}) and~(\ref{Rij-ast}), we find
from equations~(\ref{basis-1}), (\ref{basis-2})
and~(\ref{basis-4})
\begin{equation}\label{nul-ast}
  \varepsilon_\een=0, \quad p_\een=0, \quad n_\een=0, \quad
  \vec{u}_\een=0,
\end{equation}
where we have also used Eq.~(\ref{perttoes}). With
(\ref{nul-ast}), Eqs.~(\ref{basis-5}) and~(\ref{basis-6}) are
identically satisfied. The only surviving equation
is~(\ref{basis-3}), which now reads
\begin{equation}
   \ddot{h}_{\ast}^i{}_j + 3H\dot{h}_{\ast}^i{}_j +
     2\, \mbox{$^3\!R^i_{\een\ast j}$} = 0,  \label{ddhij-tensor}
\end{equation}
where~$\mbox{$^3\!R^i_{\een\ast j}$}$ is given by
Eq.~(\ref{decomp-Rij-ast}). Using Eqs.~(\ref{fes5}),
(\ref{den1a}), (\ref{decomp-hij-ast}) and~(\ref{nul-ast}) it
follows from Eqs.~(\ref{subeq:gi-en}) that
\begin{equation}\label{eq:nul-tensor}
  \varepsilon^\gi_\een=0, \quad n^\gi_\een=0,
\end{equation}
so that tensor perturbations do not, in first order, contribute to
physical energy density and particle number density perturbations.
Hence, the Eqs.~(\ref{ddhij-tensor}) do not play a role in this
context, where we are interested in energy density and particle
number density perturbations only.

The equations~(\ref{ddhij-tensor}) have a wave equation like form
with an extra term. The extra term~$3H\dot{h}_{\ast}^i{}_j$ in
these equations is due to the expansion of the universe.
Therefore, these tensor perturbations are sometimes called
\emph{gravitational waves}. The six components~$h^i_{\ast j}$
satisfy the four equations~(\ref{decomp-hij-ast}), leaving us with
two independent functions~$h^i_{\ast j}$. They are related to
linearly and circularly polarized waves.

\subsection{Vector perturbations} \label{vector}

We will show that, just like tensor perturbations, vector
perturbations are not coupled to density perturbations.

Upon replacing $h^i{}_j$ by $h^i_{\perp j}$ and
$\mbox{$^3\!R^i_{\een j}$}$ by $\mbox{$^3\!R^i_{\een\perp}{}_j$}$
in the perturbation equations~(\ref{subeq:basis}), and using the
expressions~(\ref{decomp-hij-perp}) and~(\ref{Rij-perp}), we find
from Eq.~(\ref{basis-1}) and the trace of Eq.~(\ref{basis-3})
\begin{equation}\label{nul-perp}
  \varepsilon_\een=0, \quad p_\een=0, \quad n_\een=0,
\end{equation}
where we have also used Eq.~(\ref{perttoes}).

Since $h^i_{\perp j}$ is traceless and raising the index with
$g^{ij}_\nul$ in Eq.~(\ref{basis-2}) we get
\begin{equation}\label{basis-2-raise}
  \dot{h}^{kj}_\perp{}_{|k}+2Hh^{kj}_\perp{}_{|k}=
  2\kappa(\varepsilon_\nul+p_\nul)u^j_{\een},
\end{equation}
where we have used Eqs.~(\ref{def-gammas}) and~(\ref{metricFRW}).
We now calculate the covariant derivative of~(\ref{basis-2-raise})
with respect to~$x^j$, and use~(\ref{eq:hklkl}) to obtain
\begin{equation}\label{div-0}
      \vec{\tilde{\nabla}}\cdot\vec{u}_{\een}=0,
\end{equation}
where we made use of the fact that the time derivative and the
covariant derivative commute. With
Eqs.~(\ref{decomp-u})--(\ref{transversal}) we see that only the
transverse part of~$\vec{u}_\een$, namely $\vec{u}_{\een\perp}$,
plays a role in vector perturbations. From (\ref{decomp-hij-perp})
and~(\ref{nul-perp}) it follows that the equations~(\ref{basis-4})
and~(\ref{basis-6}) are identically satisfied. The only surviving
equations are~(\ref{basis-2}), (\ref{basis-3}) and~(\ref{basis-5}),
which now read
\begin{subequations}
\label{subeq:vector}
\begin{eqnarray}
   && \dot{h}_{\perp}^k{}_{i|k} =
         -2\kappa(\varepsilon_\nul + p_\nul) u_{\een\perp i},
           \label{const-vector} \\
   && \ddot{h}_{\perp}^i{}_j + 3H\dot{h}_{\perp}^i{}_j +
     2\, \mbox{$^3\!R^i_{\een\perp j}$} = 0, \label{ddhij-vector} \\
   &&  \frac{1}{c}\frac{\dif}{\dif t}
     \Bigl[(\varepsilon_\nul+p_\nul)u^i_{\een\perp}\Bigr] +
     5H(\varepsilon_\nul+p_\nul) u^i_{\een\perp}=0,
            \label{basis-5-perp}
\end{eqnarray}
\end{subequations}
where $\mbox{$^3\!R^i_{\een\perp j}$}$ is given by
Eq.~(\ref{decomp-Rij-perp}).

Using Eqs.~(\ref{fes5}), (\ref{den1a}), (\ref{nul-perp})
and~(\ref{div-0}) we get from Eqs.~(\ref{subeq:gi-en})
\begin{equation}\label{eq:nul-vector}
  \varepsilon^\gi_\een=0, \quad n^\gi_\een=0,
\end{equation}
implying that also vector perturbations do not, in first order,
contribute to physical energy density and particle number density
perturbations. Hence, the equations~(\ref{subeq:vector}) do not
play a role when we are interested in energy density and particle
number density perturbations, as we are here. Vector perturbations
are also called \emph{vortices}.

Since vector perturbations obey
$\vec{\tilde{\nabla}}\cdot\vec{u}_{\een\perp}=0$, they have two
degrees of freedom. As a consequence, the tensor $h^i_{\perp j}$
has also two degrees of freedom. These degrees of freedom are
related to clockwise and counter-clockwise rotation of matter.

\subsection{Scalar perturbations}\label{scalar}

Differentiation of Eqs.~(\ref{basis-2}) covariantly with respect
to~$x^j$ we obtain
\begin{equation}
    \dot{h}_{\parallel}^k{}_{k|i|j}-\dot{h}_{\parallel}^k{}_{i|k|j} =
         2\kappa(\varepsilon_\nul + p_\nul)
         u_{\een\parallel i|j}.
    \label{eq:feiko1}
\end{equation}
Interchanging~$i$ and~$j$ in this equation, and subtracting the
resulting equation from~(\ref{eq:feiko1}) we get
\begin{equation}
    \dot{h}_{\parallel}^k{}_{i|k|j} - \dot{h}_{\parallel}^k{}_{j|k|i} =
         -2\kappa(\varepsilon_\nul + p_\nul)
         (u_{\een\parallel i|j}-u_{\een\parallel j|i}),
    \label{dR0i2-rot}
\end{equation}
where we have used that
$\dot{h}_{\parallel}^k{}_{k|i|j}=\dot{h}_{\parallel}^k{}_{k|j|i}$.
Using that $\vec{\tilde{\nabla}}\wedge\vec{u}_{\een\parallel}=0$,
we find from Eq.~(\ref{decomp-hij-par}) that the function~$\zeta$
must obey the equations
\begin{equation}
  \dot{\zeta}^{|k}{}_{|i|k|j}-\dot{\zeta}^{|k}{}_{|j|k|i}=0.
      \label{eq:rest-zeta}
\end{equation}
These equations are fulfilled identically in \textsc{flrw}
universes. This can be seen as follows. We first rewrite these
equations by interchanging the covariant derivatives in the form
\begin{equation}
   (\dot{\zeta}^{|k}{}_{|i|k|j}-\dot{\zeta}^{|k}{}_{|i|j|k})-
    (\dot{\zeta}^{|k}{}_{|j|k|i}-\dot{\zeta}^{|k}{}_{|j|i|k})+
    (\dot{\zeta}^{|k}{}_{|i|j}-\dot{\zeta}^{|k}{}_{|j|i})_{|k}=0.
     \label{eq:verwissel}
\end{equation}
Next, we use Eq.~(\ref{eq:commu-Riemann}) and substitute the Riemann
tensor~(\ref{eq:Riemann}) into the resulting expression. Using that
$\dot{\zeta}_{|i|j}=\dot{\zeta}_{|j|i}$, we find that the left-hand
sides of the Eqs.~(\ref{eq:verwissel}) vanish. As a consequence, the
Eqs.~(\ref{eq:rest-zeta}) are identities. Therefore, the
decomposition (\ref{decomp-hij-par}) imposes no additional condition
on the irreducible part $h^i_{\parallel j}$ of the perturbation
$h^i{}_j$.

The evolution equations~(\ref{subeq:basis}) for scalar
perturbations read
\begin{subequations}
\label{subeq:scalar}
\begin{eqnarray}
 &&  H\dot{h}_{\parallel}^k{}_k +
         \tfrac{1}{2}\,\mbox{$^3\!R_{\een\parallel}$}=
         -\kappa\varepsilon_\een,  \label{basis-1-scal} \\
 &&  \dot{h}_{\parallel}^k{}_{k|i}-\dot{h}_{\parallel}^k{}_{i|k}=
          2\kappa(\varepsilon_\nul + p_\nul) u_{\een\parallel i},
         \label{basis-2-scal} \\
 &&   \ddot{h}_{\parallel}^i{}_j+3H\dot{h}_{\parallel}^i{}_j+
      \delta^i{}_j H\dot{h}_{\parallel}^k{}_k +
        2\,\mbox{$^3\!R^i_{\een\parallel j}$}=
     -\kappa\delta^i{}_j(\varepsilon_\een-p_\een),  \label{basis-3-scal} \\
 &&  \dot{\varepsilon}_\een+3H(\varepsilon_\een +p_\een)+
    (\varepsilon_\nul +p_\nul)[(u^k_{\een\parallel}){}_{|k}-
     \tfrac{1}{2}\dot{h}^k_\parallel{}_k]=0,  \label{basis-4-scal}  \\
 &&  \frac{1}{c}\frac{\dif}{\dif t}
   \Bigl[(\varepsilon_\nul+p_\nul) u^i_{\een\parallel}\Bigr]-
    g^{ik}_\nul p_{\een|k}+5H(\varepsilon_\nul+p_\nul)u^i_{\een\parallel}=0,
            \label{basis-5-scal} \\
 &&    \dot{n}_\een+3Hn_\een+n_\nul
     [(u^k_{\een\parallel}){}_{|k}-\tfrac{1}{2}\dot{h}^k_\parallel{}_k] = 0,
         \label{basis-6-scal}
\end{eqnarray}
\end{subequations}
where the perturbations to the metric and the Ricci tensor are given
by Eqs.~(\ref{decomp-hij-par}) and (\ref{decomp-Rij-par}),
respectively. In the tensorial and vectorial case we found
$\varepsilon_\een=0$ and $n_\een=0$, implying that
$\varepsilon^\gi_\een=0$ and $n^\gi_\een=0$, which made the
tensorial and vectorial equations irrelevant for our purpose. Such a
conclusion cannot be drawn from the equations~(\ref{subeq:scalar}).
Perturbations with $\varepsilon_\een\neq0$ and $n_\een\neq0$ are
usually referred to as scalar perturbations.

In Sec.~\ref{evo-scal} we rewrite the set of
equations~(\ref{subeq:scalar}) in such a way that they determine the
evolution of the quantities~(\ref{subeq:gi-en}). Since the
perturbation equations contain only the
components~$h^i_{\parallel}{}_j$, it follows that
\emph{relativistic} energy density and particle number density
perturbations are characterized by \emph{two} potentials, $\phi$
and~$\zeta$.

\subsection{Summary} \label{subsec:summary}

In the foregoing three Secs.~\ref{tensor}--\ref{scalar} we have
written down, using the decompositions~(\ref{decomp-symh}),
(\ref{decomp-Rij}) and~(\ref{decomp-u}), the perturbation
equations for gravitational waves, Eqs.~(\ref{ddhij-tensor}),
vortex perturbations, Eqs.~(\ref{subeq:vector}) and scalar
perturbations, Eqs.~(\ref{subeq:scalar}). For gravitational waves
and vortices it was shown that $\varepsilon^\gi_\een=0$ and
$n^\gi_\een=0$. Therefore, we consider from now on only the
Eqs.~(\ref{subeq:scalar}) for the scalar perturbations.

An important consequence of the theorem of Stewart,
Eqs.~(\ref{decomp-symh})--(\ref{subeq:dec-hij})
and~(\ref{decomp-Rij})--(\ref{subeq:prop-Rij}), is that we need only
two potentials~$\phi$ and~$\zeta$ (instead of six
potentials~$h^i{}_j$) to describe the evolution of the scalar
perturbations. In Sec.~\ref{nrl} we will show that, in a flat
\textsc{flrw} universe, the potential~$\zeta$ automatically
disappears from the perturbed Einstein equations and that
$\phi(t,\vec{x})$ can be related to the well-known, time-independent
potential $\varphi(\vec{x})$ encountered in the Poisson equation
$\nabla^2\varphi(\vec{x})=4\pi G\varrho_\een(\vec{x})$, where
$\varrho_\een(\vec{x})=
\varepsilon_\een^\gi(t_\mathrm{p},\vec{x})/c^2$ is the mass density
at~$\vec{x}$. In fact, we will show that in the non-relativistic
limit we have $\varphi(\vec{x})=\phi(\vec{x})/a^2(t_\mathrm{p})$,
where $a(t)$ is the scale factor occurring in (\ref{m2}). In these
equations~$t_\mathrm{p}$ stands for the present time.

In the definitions~(\ref{subeq:gi-en}) of the physical
perturbations~$\varepsilon^\gi_\een$ and~$n^\gi_\een$ the background
quantities~$\varepsilon_\nul$, $n_\nul$ and~$\theta_\nul=3H$ occur.
The evolution of these quantities is governed by the background
Einstein equations given in Sec.~\ref{nulde}. Of the first order
quantities~$\varepsilon_\een$, $n_\een$ and~$\theta_\een$
in~(\ref{subeq:gi-en}), only the first two do explicitly show up in
the first order equations~(\ref{subeq:scalar}). In
Sec.~\ref{evo-scal} we rewrite this set of equations in such a way
that also the first order quantity~$\theta_\een$ does explicitly
occur in the set of perturbation equations. The result is that the
evolution of the physical quantities~(\ref{subeq:gi-en}) is
completely determined by the background- and first order equations.

\section{Scalar first order equations}
\label{evo-scal}

In this quite technical section we rewrite the scalar perturbation
equations~(\ref{subeq:scalar}) in terms of quantities $\theta_\een$,
$\mbox{$^3\!R_{\een\parallel}$}$, $\vartheta_\een$,
$\varepsilon_\een$ and $n_\een$, which are suitable to describe
exclusively the scalar perturbations. The result is that the first
order quantities occurring in the definitions~(\ref{subeq:gi-en}) do
explicitly occur in the set of equations.

Eliminating the quantity $\dot{h}^k_\parallel{}_k$ from
Eq.~(\ref{basis-1-scal}) with the help of Eq.~(\ref{fes5}) yields
\begin{equation}
  2H(\theta_\een-\vartheta_\een)-\tfrac{1}{2}\,\mbox{$^3\!R_{\een\parallel}$}=
       \kappa\varepsilon_\een. \label{theta1}
\end{equation}
Thus the $(0,0)$-component of the constraint equations becomes an
algebraic equation which relates the first order quantities
$\theta_\een$, $\vartheta_\een$, $\mbox{$^3\!R_{\een\parallel}$}$
and $\varepsilon_\een$.

It now takes some steps to rewrite the three constraint
equations~(\ref{basis-2-scal}) in a suitable form. Firstly,
multiplying both sides by~$g^{ij}_\nul$ and taking the covariant
divergence with respect to the index~$j$ we find
\begin{equation}
  g_\nul^{ij} (\dot{h}^k_\parallel{}_{k|i|j}-\dot{h}^k_\parallel{}_{i|k|j}) =
    2\kappa(\varepsilon_\nul+p_\nul) \vartheta_\een,  \label{dR0i4}
\end{equation}
where we have also used Eq.~(\ref{den1a}). The left-hand side will
turn up as a part of the first order derivative of the curvature
$\mbox{$^3\!R_{\een\parallel}$}$. In fact, differentiating
Eq.~(\ref{driekrom}) with respect to~$ct$ and recalling the fact
that the connection coefficients $\Gamma^k_{\nul ij}$,
Eqs.~(\ref{con3FRW}), are independent of time, one gets
\begin{equation}
  \mbox{$^3\!\dot{R}_{\een\parallel}$} =
-2H\,\mbox{$^3\!R_{\een\parallel}$} +
   g_\nul^{ij} (\dot{h}^k_\parallel{}_{k|i|j}-\dot{h}^k_\parallel{}_{i|k|j})+
     \tfrac{1}{3}\,\mbox{$^3\!R_\nul$} \dot{h}^k_\parallel{}_k,
   \label{dR0i6}
\end{equation}
where we have used Eqs.~(\ref{def-gammas}), (\ref{metricFRW}),
(\ref{momback}) and
\begin{equation}
    g_\nul^{ij}h^k_\parallel{}_{i|j|k} = g_\nul^{ij}h^k_\parallel{}_{i|k|j},
\end{equation}
which is a consequence of $g^{ij}_{\nul|k}=0$ and the symmetry of
$h_\parallel^{ij}$. Next, combining Eqs.~(\ref{dR0i4})
and~(\ref{dR0i6}), and, finally, eliminating
$\dot{h}^k_\parallel{}_k$ with the help of Eq.~(\ref{fes5}), one
arrives at
\begin{equation}
 \mbox{$^3\!\dot{R}_{\een\parallel}$}+
        2H\,\mbox{$^3\!R_{\een\parallel}$}-
   2\kappa(\varepsilon_\nul + p_\nul)\vartheta_\een+
     \tfrac{2}{3}\,\mbox{$^3\!R_\nul$}(\theta_\een-\vartheta_\een)=0.
     \label{mompert}
\end{equation}
In this way we managed to recast the three $(0,i)$-components of
the constraint equations in the form of one ordinary differential
equation for the local perturbation,
$\mbox{$^3\!R_{\een\parallel}$}$, to the spatial curvature.

We now consider the dynamical equations~(\ref{basis-3-scal}).
Taking the trace of these equations and using Eq.~(\ref{fes5}) to
eliminate the quantity~$\dot{h}^k_{\parallel}{}_k$, we arrive at
\begin{equation}
   \dot{\theta}_\een-\dot{\vartheta}_\een +
      6H(\theta_\een-\vartheta_\een)-\mbox{$^3\!R_{\een\parallel}$}=
       \tfrac{3}{2}\kappa(\varepsilon_\een-p_\een).
        \label{eq:168a}
\end{equation}
Thus, for scalar perturbations, the three dynamical Einstein
equations~(\ref{basis-3-scal}) with $i=j$ reduce to one ordinary
differential equation for the difference
$\theta_\een-\vartheta_\een$. For $i\neq j$ the dynamical Einstein
equations are not coupled to scalar perturbations. As a
consequence, they need not be considered.

Taking the covariant derivative of Eq.~(\ref{basis-5-scal}) with
respect to the metric $g_{\nul ij}$ and using Eq.~(\ref{den1a}),
we get
\begin{equation}
  \frac{1}{c}\frac{\dif}{\dif t}
     \Bigl[(\varepsilon_\nul+p_\nul)\vartheta_\een\Bigr]-
    g^{ik}_\nul p_{\een|k|i}+5H(\varepsilon_\nul+p_\nul)\vartheta_\een = 0,   \label{mom2}
\end{equation}
where we have used that the operations of taking the time derivative
and the covariant derivative commute, since the connection
coefficients $\Gamma^k_{\nul ij}$, (\ref{con3FRW}), are independent
of time. With Eq.~(\ref{energyFRW}), we can rewrite Eq.~(\ref{mom2})
in the form
\begin{equation}
  \dot{\vartheta}_\een +
 H\left(2-3\frac{\dot{p}_\nul}{\dot{\varepsilon}_\nul}\right)\vartheta_\een+
     \frac{1}{\varepsilon_\nul+p_\nul}\dfrac{\tilde{\nabla}^2 p_\een}{a^2}=0,
                \label{mom3}
\end{equation}
where $\tilde{\nabla}^2$ is the generalized Laplace operator
which, for an arbitrary function $f(t,\vec{x})$ and with respect
to an arbitrary three-dimensional metric
$\tilde{g}^{ij}(\vec{x})$, is defined by
\begin{equation}\label{Laplace}
 \tilde{\nabla}^2 f:=\tilde{g}^{ij} f_{|i|j}.
\end{equation}
Thus, the three first order momentum conservation laws
(\ref{basis-5-scal}) reduce to one ordinary differential equation
for the divergence~$\vartheta_\een$.

Finally, we consider the conservation laws~(\ref{basis-4-scal})
and~(\ref{basis-6-scal}). Eliminating the quantity
$\dot{h}^k_\parallel{}_k$ from these equations with the help of
Eq.~(\ref{fes5}), we get
\begin{equation}\label{basis-4-theta}
  \dot{\varepsilon}_\een+3H(\varepsilon_\een +p_\een)+
     (\varepsilon_\nul +p_\nul)\theta_\een = 0,
\end{equation}
and
\begin{equation}\label{basis-6-theta}
   \dot{n}_\een+3Hn_\een+n_\nul\theta_\een = 0.
\end{equation}
The algebraic equation (\ref{theta1}) and the five \emph{ordinary}
differential equations (\ref{mompert}), (\ref{eq:168a}),
(\ref{mom3}), (\ref{basis-4-theta}) and~(\ref{basis-6-theta}) is a
system of six equations for the five quantities $\theta_\een$,
$\mbox{$^3\!R_\een$}$, $\vartheta_\een$, $\varepsilon_\een$ and
$n_\een$, respectively. This system is, however, not overdetermined,
since the constraint equation~(\ref{theta1}) is only an initial
value condition. It can easily be shown by differentiation of
Eq.~(\ref{theta1}) with respect to time and eliminating the time
derivatives with the help of Eqs.~(\ref{endencon0}), (\ref{dyn0}),
(\ref{mompert}), (\ref{eq:168a}) and~(\ref{basis-4-theta}) that the
general solution of the system (\ref{theta1}), (\ref{mompert}),
(\ref{mom3}), (\ref{basis-4-theta}) and~(\ref{basis-6-theta}) is
also a solution of Eq.~(\ref{eq:168a}). Therefore, we do not
consider Eq.~(\ref{eq:168a}) anymore and we take the algebraic
equation (\ref{theta1}) and the four ordinary differential equations
(\ref{mompert}), (\ref{mom3}), (\ref{basis-4-theta})
and~(\ref{basis-6-theta}) for the five quantities $\theta_\een$,
$\mbox{$^3\!R_\een$}$, $\vartheta_\een$, $\varepsilon_\een$ and
$n_\een$, respectively, as the basis of our perturbation theory. For
scalar perturbations this system of equations is equivalent to the
full set (\ref{subeq:basis}) of linear Einstein equations and
conservation laws.

\section{Summary quantities and equations}  \label{resume}

\subsection{Zero order equations}

The Einstein equations and conservation laws for the background of
an \textsc{flrw} universe are given by~(\ref{dyn0}),
(\ref{energyFRW}), (\ref{deel1}) and~(\ref{momback}):
\begin{subequations}
\label{subeq:einstein-flrw}
\begin{eqnarray}
    \dot{H} & = & -\tfrac{1}{6}\,\mbox{$^3\!R_\nul$}
        -\tfrac{1}{2}\kappa\varepsilon_\nul(1+w),   \label{FRW1}   \\
    \dot{\varepsilon}_\nul & = & -3H\varepsilon_\nul(1+w), \label{FRW2} \\
       \dot{n}_\nul & = & -3Hn_\nul,   \label{FRW2a}  \\
  \mbox{$^3\!\dot{R}_\nul$} & = & -2H\,\mbox{$^3\!R_\nul$}, \label{FRW3a}
\end{eqnarray}
\end{subequations}
and the constraint equation~(\ref{endencon0})
\begin{equation}\label{FRW3}
    3H^2 = \tfrac{1}{2}\,\mbox{$^3\!R_\nul$} +
        \kappa\varepsilon_\nul + \Lambda.
\end{equation}
In order to arrive at Eq.~(\ref{FRW1}) we eliminated the term
$3H^2$ from Eq.~(\ref{dyn0}) with the help of
Eq.~(\ref{endencon0}). Furthermore, we introduced the abbreviation
\begin{equation}
    w(t) := \frac{p_\nul(t)}{\varepsilon_\nul(t)}. \label{begam1}
\end{equation}
The set (\ref{subeq:einstein-flrw}) consists of four differential
equations with respect to time for the four unknown quantities
$\varepsilon_\nul(t)$, $n_\nul(t)$, $\theta_\nul(t)=3H(t)$, and
$\mbox{$^3\!R_\nul$}(t)$. Recall $\vartheta_\nul=0$,
Eq.~(\ref{fes2}). The pressure $p_\nul(t)$ is related to the energy
density $\varepsilon_\nul(t)$ and the particle number density
$n_\nul(t)$ via the equation of state~(\ref{toestandback}). The
algebraic equation~(\ref{FRW3}) is a constraint on the initial
values.

\subsection{First order equations}\label{foe}

The first order equations describing density perturbations are
given by the set of four differential
equations~(\ref{basis-4-theta}), (\ref{basis-6-theta}),
(\ref{mom3}) and~(\ref{mompert})
\begin{subequations}
\label{subeq:pertub-flrw}
\begin{eqnarray}
  &&  \dot{\varepsilon}_\een + 3H(\varepsilon_\een + p_\een)+
         \varepsilon_\nul(1 + w)\theta_\een=0,  \label{FRW4} \\
  &&  \dot{n}_\een + 3H n_\een +
         n_\nul\theta_\een=0, \label{FRW4a} \\
  && \dot{\vartheta}_\een+H(2-3\beta^2)\vartheta_\een+
   \frac{1}{\varepsilon_\nul(1+w)}\dfrac{\tilde{\nabla}^2p_\een}{a^2}=0,  \label{FRW5}\\
  &&  \mbox{$^3\!\dot{R}_{\een\parallel}$}+
     2H\,\mbox{$^3\!R_{\een\parallel}$}-
    2\kappa \varepsilon_\nul(1 + w)\vartheta_\een
       +\tfrac{2}{3}\,\mbox{$^3\!R_\nul$}(\theta_\een-\vartheta_\een)=0,
                \label{FRW6}
\end{eqnarray}
\end{subequations}
together with one constraint equation, Eq.~(\ref{theta1})
\begin{equation}\label{con-sp-1}
    2H(\theta_\een-\vartheta_\een)-
       \tfrac{1}{2}\,\mbox{$^3\!R_{\een\parallel}$} = \kappa\varepsilon_\een,
\end{equation}
for the five unknown functions~$\varepsilon_\een$, $n_\een$,
$\vartheta_\een$, $\mbox{$^3\!R_{\een\parallel}$}$,
and~$\theta_\een$, respectively. These are the first order
perturbations to the background quantities~$\varepsilon_\nul$,
$n_\nul$, $\vartheta_\nul=0$, $\mbox{$^3\!R_\nul$}$,
and~$\theta_\nul=3H$. The first order perturbation to the pressure
is given by the perturbed equation of state~(\ref{perttoes}).

As has been shown in Sec.~\ref{evo-scal}, the
equations~(\ref{subeq:pertub-flrw})--(\ref{con-sp-1}) comprise the
conservation laws and the constraint equations. As a consequence,
the general solution of these equations is also a solution of the
dynamical Einstein equation~(\ref{eq:168a}). Therefore, we may, in
the study of scalar perturbations, replace the full set of
perturbation equations~(\ref{subeq:scalar}) by the
set~(\ref{subeq:pertub-flrw})--(\ref{con-sp-1}). Note that only
three scalars $\varepsilon$, $n$ and~$\theta:=u^\mu{}_{;\mu}$ play a
role in a density perturbation theory. The only non-trivial
gauge-invariant combinations which can be constructed from these
scalars and their first order perturbations are the
combinations~(\ref{subeq:gi-en}). These combinations do always
exist, since, in a non-static universe, we have
$\dot{\theta}_\nul(t)\neq0$ for all times~$t$, as follows from
Eq.~(\ref{FRW1}).

The operator $\tilde{\nabla}^2$, occurring in Eq.~(\ref{FRW5}), is
the generalized Laplace operator defined by Eq.~(\ref{Laplace}).
The quantity~$\beta(t)$ occurring in Eq.~(\ref{FRW5}) is defined
by
\begin{equation}
  \beta(t):=\sqrt{\dfrac{\dot{p}_\nul(t)}{\dot{\varepsilon}_\nul(t)}}.
          \label{begam2}
\end{equation}

Eliminating the time derivatives of~$\varepsilon_\nul$
and~$n_\nul$ from Eqs.~(\ref{subeq:gi-en}) with the help of the
background equations~(\ref{subeq:einstein-flrw}), we arrive~at
\begin{subequations}
\label{subeq:gi-en-flrw}
\begin{eqnarray}
  \varepsilon^\gi_\een &=& \varepsilon_\een-\dfrac{6H\varepsilon_\nul(1+w)}
       {\mbox{$^3\!R_\nul$}+3\kappa\varepsilon_\nul(1+w)}\theta_\een,  \\
  n^\gi_\een &=& n_\een-\dfrac{6Hn_\nul}
     {\mbox{$^3\!R_\nul$}+3\kappa\varepsilon_\nul(1+w)}\theta_\een,
\end{eqnarray}
\end{subequations}
where we have used that $\theta_\nul=3H$, Eq.~(\ref{fes2}). We
have achieved now that the set of perturbation
equations~(\ref{subeq:pertub-flrw}) and~(\ref{con-sp-1}), together
with the background equations~(\ref{subeq:einstein-flrw})
and~(\ref{FRW3}) determine the evolution of the physical
quantities~(\ref{subeq:gi-en}), or, equivalently,
(\ref{subeq:gi-en-flrw}). In principle, we are ready. However, the
solution of the set of equations~(\ref{subeq:pertub-flrw})
and~(\ref{con-sp-1}) is gauge dependent (see
Appendix~\ref{giofoe}). Indeed, one may easily check, by a direct
calculation, that these equations are invariant under the
transformation
\begin{subequations}
\label{subeq:gauge-transform}
\begin{eqnarray}
\varepsilon_\een & \rightarrow &
    \hat{\varepsilon}_\een=\varepsilon_\een+\psi\dot{\varepsilon}_\nul,
    \label{eq:gte} \\
n_\een & \rightarrow & \hat{n}_\een=n_\een+\psi\dot{n}_\nul, \\
\vartheta_\een & \rightarrow & \hat{\vartheta}_\een =
   \vartheta_\een-\frac{\tilde{\nabla}^2\psi}{a^2}, \\
   \mbox{$^3\!R_{\een\parallel}$} & \rightarrow & \mbox{$^3\!\hat{R}_{\een\parallel}$}=
    \mbox{$^3\!R_{\een\parallel}$}+
     4H\left(\frac{\tilde{\nabla}^2\psi}{a^2} -
       \tfrac{1}{2}\,\mbox{$^3\!R_\nul$}\psi\right), \\
\theta_\een & \rightarrow &
    \hat{\theta}_\een=\theta_\een+\psi\dot{\theta}_\nul,
\end{eqnarray}
\end{subequations}
where $\psi$ is time independent in synchronous coordinates, see
Eq.~(\ref{xi-syn}). By switching from the variables
$\varepsilon_\een$, $n_\een$, $\vartheta_\een$,
$\mbox{$^3\!R_{\een\parallel}$}$ and~$\theta_\een$ to the
variables~$\varepsilon_\een^\gi$ and~$n_\een^\gi$, we will arrive at
a set of equations for~$\varepsilon_\een^\gi$ and~$n_\een^\gi$ with
a unique, i.e., gauge-invariant solution. This will be the subject
of Sec.~\ref{thirdstep}. First, however, we derive some auxiliary
equations related to the entropy, pressure and temperature.

\section{Entropy, pressure and temperature}
\label{sec:pertub-therm}

\subsection{Gauge-invariant entropy perturbations}  \label{sec:eqs-sgi}

The second law of thermodynamics reads
\begin{equation}\label{eq:sec-law-thermo}
    \dif E = T\,\dif S - p\,\dif V + \mu\, \dif N,
\end{equation}
where $E$, $S$ and $N$ are the energy, the entropy and the number
of particles of a system with volume~$V$, and where~$\mu$, the
thermodynamic ---or chemical--- potential, is the energy needed to
add one particle to the system. In terms of the energy per
particle $E/N=\varepsilon/n$ and the entropy per particle $s=S/N$
the law~(\ref{eq:sec-law-thermo}) can be rewritten
\begin{equation}\label{eq:sec-law-2}
    \dif\biggl(\dfrac{\varepsilon}{n}N
    \biggr)=T\,\dif(sN)-p\,\dif\biggl(\dfrac{N}{n}
    \biggr)+\mu\,\dif N,
\end{equation}
where $\varepsilon$ and $n$ are the energy and particle number
densities. With the Euler relation $\mu=(\varepsilon+p)n^{-1}-Ts$,
the second law can be cast in a form without~$\mu$ and~$N$. In
fact, using the rule $\dif(fg)=f\dif g+g\dif f$ we immediately
find from Eq.~(\ref{eq:sec-law-2})
\begin{equation}\label{eq:TdS}
  T\,\dif s = \dif \biggl(\dfrac{\varepsilon}{n}\biggr) +
         p\; \dif \biggl(\dfrac{1}{n}\biggr),
\end{equation}
where~$\mu$ and~$N$ have canceled, indeed. The thermodynamic
relation~(\ref{eq:TdS}) is true for a system in thermodynamic
equilibrium. For a non-equilibrium system that is `not too far' from
equilibrium, the equation~(\ref{eq:TdS}) may be replaced by
\begin{equation}\label{eq:sec-law-noneq}
    T \dfrac{\dif s}{\dif t}=\dfrac{\dif}{\dif t}\biggl(\dfrac{\varepsilon}{n}\biggr)+
p\,\dfrac{\dif}{\dif t}\biggl(\dfrac{1}{n}\biggr),
\end{equation}
where $\dif/\dif t$ is the time derivative in a local comoving
Lorentz system. Now, using
$\varepsilon=\varepsilon_\nul+\varepsilon_\een$,
$s=s_\nul+s_\een$, $p=p_\nul+p_\een$ and $n=n_\nul+n_\een$, we
find from Eq.~(\ref{eq:sec-law-noneq})
\begin{equation}\label{eq:TdS-back}
 T_\nul\dfrac{\dif s_\nul}{\dif t} =
    \dfrac{\dif}{\dif t} \biggl(\dfrac{\varepsilon_\nul}{n_\nul}\biggr) +
         p_\nul\,\dfrac{\dif}{\dif t} \biggl(\dfrac{1}{n_\nul}\biggr),
\end{equation}
where we neglected time derivatives of first order quantities.
With the help of Eqs.~(\ref{FRW2}), (\ref{FRW2a})
and~(\ref{begam1}) we find that the right-hand side of
Eq.~(\ref{eq:TdS-back}) vanishes. Hence,
\begin{equation}\label{eq:entropy-back}
  \dfrac{\dif s_\nul}{\dif t} = 0,
\end{equation}
implying that, in zero order, the expansion takes place without
generating entropy: $s_\nul$ is constant in time. Hence, in view of
Eq.~(\ref{sigmahat3}), which is valid for any scalar, and
Eq.~(\ref{eq:entropy-back}), the first order perturbation~$s_\een$
is automatically a gauge-invariant quantity, i.e.,
$\hat{s}_\een=s_\een$, in contrast to~$\varepsilon_\een$
and~$n_\een$, which had to be redefined according to
Eqs.~(\ref{subeq:gi-en}). Apparently, the entropy per
particle~$s_\een$ is such a combination of~$\varepsilon_\een$
and~$n_\een$ that it need not be redefined. This can be made
explicit by noting that in the linear approximation we are
considering in this article, the second law of
thermodynamics~(\ref{eq:TdS}) should hold for zero order and first
order quantities separately. In particular, Eq.~(\ref{eq:TdS})
implies
\begin{equation}\label{eq:TdS-1}
    T_\nul s_\een=\dfrac{1}{n_\nul}\left(\varepsilon_\een-
      \dfrac{\varepsilon_\nul+p_\nul}{n_\nul}n_\een\right),
\end{equation}
where we neglected products of differentials and first order
quantities, and where we replaced~$\dif\varepsilon$ and~$\dif n$
by~$\varepsilon_\een$ and~$n_\een$, respectively. We now note that
the linear combination in the right-hand side of
Eq.~(\ref{eq:TdS-1}) has the property
\begin{equation}\label{eq:lin-gi}
  \varepsilon_\een-\dfrac{\varepsilon_\nul+p_\nul}{n_\nul}n_\een=
  \varepsilon^\gi_\een-\dfrac{\varepsilon_\nul+p_\nul}{n_\nul}n^\gi_\een,
\end{equation}
as may immediately be verified with the help of
Eqs.~(\ref{subeq:gi-en}), (\ref{FRW2}) and~(\ref{FRW2a}). The
right-hand side of Eq.~(\ref{eq:lin-gi}) being gauge-invariant, the
left-hand side must be gauge-invariant. This observation makes
explicit the gauge invariance of the first order approximation to
the entropy per particle, $s_\een$. In order to stress the gauge
invariance of the correction~$s_\een$ to the (constant) entropy per
particle, $s_\nul$, we will write~$s^\gi_\een$, rather
than~$s_\een$:
\begin{equation}\label{eq:Sgi-is-S}
  s^\gi_\een := s_\een.
\end{equation}
From Eqs.~(\ref{eq:TdS-1})--(\ref{eq:Sgi-is-S}) we then get
\begin{equation}\label{eq:TdS-1-gi}
  T_\nul s^\gi_\een = \dfrac{1}{n_\nul}\left(
     \varepsilon^\gi_\een-\frac{\varepsilon_\nul(1+w)}{n_\nul}n^\gi_\een\right),
\end{equation}
where $w$ is the quotient of zero order pressure and zero order
energy density defined by Eq.~(\ref{begam1}). Notice that for the
internal logic of our reasoning it is not essential at all to use
the second law of thermodynamics. One may simply
consider~(\ref{eq:TdS-1-gi}) as the defining expression for a
certain linear combination of~$\varepsilon^\gi_\een$
and~$n^\gi_\een$, and replace everywhere in the equations below the
product $T_\nul s^\gi_\een$ by the right-hand side of
Eq.~(\ref{eq:TdS-1-gi}), without changing anything. However, the
second law of thermodynamics yields a physical interpretation of the
particular linear combination of~$\varepsilon^\gi_\een$
and~$n^\gi_\een$ that we will encounter in the equations of the next
section. In fact, we rewrite Eq.~(\ref{eq:TdS-1-gi}) in the form
\begin{equation}\label{eq:hulp-entropy}
   T_\nul s^\gi_\een =
      -\dfrac{\varepsilon_\nul(1+w)}{n_\nul^2}\sigma^\gi_\een,
\end{equation}
where the gauge independent, entropy related
quantity~$\sigma^\gi_\een$ is given by
\begin{equation}\label{eq:entropy-gi}
  \sigma_\een^\gi := n_\een^\gi -
  \frac{n_\nul}{\varepsilon_\nul(1+w)}\varepsilon_\een^\gi.
\end{equation}
The quantity~$\sigma^\gi_\een$ occurs as the source term in the
gauge-invariant evolution equations~(\ref{subeq:eerste}) below.
Equation~(\ref{eq:entropy-gi}) implies~(\ref{eq:ngi}).

\subsection{Gauge-invariant pressure perturbations}

We will now derive a gauge-invariant expression for the physical
pressure perturbations. To that end, we first calculate the time
derivative of the background pressure. From Eq.~(\ref{toestandback})
we have
\begin{equation}\label{eq:pnul-t}
  \dot{p}_\nul=p_n\dot{n}_\nul +p_\varepsilon
    \dot{\varepsilon}_\nul.
\end{equation}
where $p_\varepsilon$ and $p_n$ are the partial derivatives given
by Eqs.~(\ref{perttoes1}) and~(\ref{eq:pn-pe-back}). Multiplying
both sides of this expressions by $\theta_\een/\dot{\theta}_\nul$
and subtracting the result from Eq.~(\ref{perttoes}) we get
\begin{equation}\label{eq:press-gi}
    p_\een-\dfrac{\dot{p}_\nul}{\dot{\theta}_\nul}\theta_\een=
    p_n n^\gi_\een + p_\varepsilon \varepsilon^\gi_\een,
\end{equation}
where we have used Eqs.~(\ref{subeq:gi-en}) to rewrite the
right-hand side. Since~$p_n$ and~$p_\varepsilon$ depend on the
background quantities~$\varepsilon_\nul$ and~$n_\nul$ only, the
right-hand side is gauge-invariant. Hence, the quantity~$p^\gi_\een$
defined by
\begin{equation}\label{eq:pgi}
  p^\gi_\een := p_\een -
  \dfrac{\dot{p}_\nul}{\dot{\theta}\nul}\theta_\een,
\end{equation}
is gauge-invariant. We thus obtain the gauge-invariant counterpart
of Eq.~(\ref{perttoes})
\begin{equation}\label{eq:pgi-2}
  p_\een^\gi=p_\varepsilon\varepsilon_\een^\gi+p_n n_\een^\gi.
\end{equation}
We will now rewrite this equation in a slightly different form.
From Eqs.~(\ref{begam2}) and~(\ref{eq:pnul-t}) we obtain
$\beta^2=p_\varepsilon+p_n(\dot{n}_\nul/\dot{\varepsilon}_\nul)$.
Using Eqs.~(\ref{FRW2}) and~(\ref{FRW2a}) we find
\begin{equation}\label{eq:begam3}
  \beta = \sqrt{p_\varepsilon+\frac{n_\nul p_n}{\varepsilon_\nul(1+w)}}.
\end{equation}
With this expression and Eqs.~(\ref{eq:entropy-gi})
and~(\ref{eq:begam3}) we can express the gauge-invariant
pressure~(\ref{eq:pgi-2}) in terms of the energy density
perturbation~$\varepsilon^\gi_\een$ and the entropy related
quantity~$\sigma^\gi_\een$ rather than~$\varepsilon^\gi_\een$ and
the particle number density perturbation~$n^\gi_\een$
\begin{equation}\label{eq:pgi-3}
   p_\een^\gi=\beta^2 \varepsilon^\gi_\een + p_n \sigma^\gi_\een,
\end{equation}
a nice result which we will not use, however. Compare this
relation with the one derived by
Mukhanov~\emph{et~al.}~\cite{mfb1992}, their equation~(5.3).

\subsection{Gauge-invariant temperature perturbations}

Finally, we will derive an expression for the gauge-invariant
temperature perturbation~$T^\gi_\een$ with the help of
Eq.~(\ref{subeq:de-dp-a}). For the time derivative of the energy
density~$\varepsilon_\nul(n_\nul,T_\nul)$ we have
\begin{equation}\label{eq:de-dp-time}
  \dot{\varepsilon}_\nul =
   \left(\dfrac{\partial \varepsilon}{\partial n} \right)_{\!T}\dot{n}_\nul +
   \left(\dfrac{\partial \varepsilon}{\partial T} \right)_{\!n}\dot{T}_\nul.
\end{equation}
Replacing the infinitesimal quantities in
Eq.~(\ref{subeq:de-dp-a}) by perturbations, we find
\begin{equation}\label{eq:de-dp-gd}
  \varepsilon_\een =
   \left(\dfrac{\partial \varepsilon}{\partial n} \right)_{\!T}n_\een +
   \left(\dfrac{\partial \varepsilon}{\partial T} \right)_{\!n}T_\een.
\end{equation}
Multiplying both sides of Eq.~(\ref{eq:de-dp-time}) by
$\theta_\een/\dot{\theta}_\nul$ and subtracting the result from
Eq.~(\ref{eq:de-dp-gd}) we get
\begin{equation}\label{eq:de-dp-gi}
  \varepsilon^\gi_\een =
   \left(\dfrac{\partial \varepsilon}{\partial n} \right)_{\!T}n^\gi_\een +
   \left(\dfrac{\partial \varepsilon}{\partial T}
   \right)_{\!n}\left(T_\een-\dfrac{\dot{T}_\nul}{\dot{\theta}_\nul}\theta_\een\right),
\end{equation}
where we have used Eqs.~(\ref{subeq:gi-en}). Hence, the quantity
\begin{equation}\label{eq:Tgi}
  T^\gi_\een := T_\een -
     \dfrac{\dot{T}_\nul}{\dot{\theta}_\nul}\theta_\een,
\end{equation}
is gauge-invariant. Thus, Eq.~(\ref{eq:de-dp-gi}) can be written as
\begin{equation}\label{eq:de-dp-dT-gi}
  \varepsilon^\gi_\een =
   \left(\dfrac{\partial \varepsilon}{\partial n} \right)_{\!T}n^\gi_\een +
   \left(\dfrac{\partial \varepsilon}{\partial T}
   \right)_{\!n}T^\gi_\een,
\end{equation}
implying that $T^\gi_\een$ can be interpreted as the gauge-invariant
temperature perturbation. An equivalent form
of~(\ref{eq:de-dp-dT-gi}) is given by Eq.~(\ref{eq:T-pert}). We thus
have expressed the perturbation in the absolute temperature as a
function of the perturbations in the energy density and particle
number density for a given equation of state of the form
$\varepsilon=\varepsilon(n,T)$ and $p=p(n,T)$. This equation will be
used in Sec.~\ref{thirdstep} to derive an expression for the
fluctuations in the background temperature, $\delta_\mathrm{T}$, a
measurable quantity.

Finally, we give the evolution equation for the background
temperature~$T_\nul(t)$. From Eq.~(\ref{eq:de-dp-time}) it follows
\begin{equation}\label{eq:evo-T0}
    \dot{T}_\nul=\dfrac{-3H\left[\varepsilon_\nul(1+w)-
    \left(\dfrac{\partial \varepsilon}{\partial n}\right)_{\!T}n_\nul\right]}
    {\left(\dfrac{\partial \varepsilon}{\partial T}\right)_{\!n}},
\end{equation}
where we have used Eqs.~(\ref{FRW2}) and~(\ref{FRW2a}). This
equation will be used to follow the time development of the
background temperature once~$\varepsilon_\nul(t)$ and~$n_\nul(t)$
are found from the zero order Einstein equations.

\section{Manifestly gauge-invariant first order equations} \label{thirdstep}

The five perturbation equations~(\ref{subeq:pertub-flrw})
and~(\ref{con-sp-1}) form a set of five equations for the five
unknown quantities~$\varepsilon_\een$, $n_\een$, $\vartheta_\een$,
$\mbox{$^3\!R_{\een\parallel}$}$ and~$\theta_\een$. This system of
equations can be reduced in the following way. In order to arrive
at non-zero expressions for the physical
quantities~$\varepsilon^\gi_\een$ and~$n^\gi_\een$ we have chosen
in Eqs.~(\ref{subeq:ent-gi}) $\omega=\theta$, Eq.~(\ref{omega1}),
implying that~$\theta^\gi_\een=0$, Eq.~(\ref{thetagi}). As a
consequence, we do not need the gauge dependent
quantity~$\theta_\een$. Eliminating the quantity $\theta_\een$
from equations~(\ref{subeq:pertub-flrw}) with the help of
Eq.~(\ref{con-sp-1}), we arrive at the set of four first order
differential equations
\begin{subequations}
\label{subeq:pertub-gi}
\begin{eqnarray}
 && \dot{\varepsilon}_\een + 3H(\varepsilon_\een + p_\een)+
     \varepsilon_\nul(1 + w)\Bigl[\vartheta_\een +\frac{1}{2H}\left(
 \kappa\varepsilon_\een+\tfrac{1}{2}\,\mbox{$^3\!R_{\een\parallel}$}\right)\Bigr]=0,
         \label{FRW4gi} \\
 &&   \dot{n}_\een + 3H n_\een+
    n_\nul \Bigl[\vartheta_\een +\frac{1}{2H}\left(\kappa\varepsilon_\een +
      \tfrac{1}{2}\,\mbox{$^3\!R_{\een\parallel}$}\right)\Bigr]=0,  \label{FRW4agi} \\
 &&  \dot{\vartheta}_\een+H(2-3\beta^2)\vartheta_\een +
        \frac{1}{\varepsilon_\nul(1+w)}\dfrac{\tilde{\nabla}^2p_\een}{a^2}=0,    \label{FRW5gi}\\
 &&  \mbox{$^3\!\dot{R}_{\een\parallel}$} +
2H\,\mbox{$^3\!R_{\een\parallel}$}
      -2\kappa\varepsilon_\nul(1 + w)\vartheta_\een+
      \frac{\mbox{$^3\!R_\nul$}}{3H} \left(\kappa\varepsilon_\een +
 \tfrac{1}{2}\,\mbox{$^3\!R_{\een\parallel}$} \right)=0,     \label{FRW6gi}
\end{eqnarray}
\end{subequations}
for the four quantities~$\varepsilon_\een$, $n_\een$,
$\vartheta_\een$ and~$\mbox{$^3\!R_{\een\parallel}$}$.

The system of equations~(\ref{subeq:pertub-gi}) is now cast in a
suitable form to arrive at a system of manifestly gauge-invariant
equations for the physical quantities~$\varepsilon^\gi_\een$
and~$n^\gi_\een$, since we then can immediately calculate these
quantities. Indeed, eliminating the quantity~$\theta_\een$ from
Eqs.~(\ref{subeq:gi-en-flrw}) with the help of Eq.~(\ref{con-sp-1}),
we get
\begin{subequations}
\label{subeq:pertub-gi-e-n}
\begin{eqnarray}
 &&   \varepsilon_\een^\gi  =
     \frac{ \varepsilon_\een \,\mbox{$^3\!R_\nul$} -
   3\varepsilon_\nul(1 + w) (2H\vartheta_\een +
  \tfrac{1}{2}\,\mbox{$^3\!R_{\een\parallel}$}) }
   {\mbox{$^3\!R_\nul$}+3\kappa\varepsilon_\nul(1 + w)},
        \label{Egi} \\
 &&  n_\een^\gi = n_\een-\dfrac{3n_\nul(\kappa\varepsilon_\een+2H\vartheta_\een+
            \tfrac{1}{2}\,\mbox{$^3\!R_{\een\parallel}$})}
         {\mbox{$^3\!R_\nul$}+3\kappa\varepsilon_\nul(1+w)}.  \label{nu2}
\end{eqnarray}
\end{subequations}
The quantities ~$\varepsilon^\gi_\een$ and~$n^\gi_\een$ are now
completely determined by the system of background
equations~(\ref{subeq:einstein-flrw})--(\ref{FRW3}) and the first
order equations~(\ref{subeq:pertub-gi}).

\subsection{Evolution equations for density perturbations}

Instead of calculating~$\varepsilon^\gi_\een$ and~$n^\gi_\een$ in
the way described above, we proceed by first making explicit the
gauge invariance of the theory. To that end, we rewrite the system
of the four differential equations~(\ref{subeq:pertub-gi}) for the
gauge dependent variables~$\varepsilon_\een$, $n_\een$,
$\vartheta_\een$ and~$\mbox{$^3\!R_{\een\parallel}$}$ into a system
of equations for the gauge-invariant
variables~$\varepsilon_\een^\gi$ and~$n_\een^\gi$. It is, however,
of convenience to use the entropy related
perturbation~$\sigma_\een^\gi$, defined by
Eq.~(\ref{eq:entropy-gi}), rather than the particle number density
perturbation~$n_\een^\gi$. The result is
\begin{subequations}
\label{subeq:eerste}
\begin{eqnarray}
 && \ddot{\varepsilon}^\gi_\een+a_1\dot{\varepsilon}^\gi_\een+
  a_2\varepsilon^\gi_\een = a_3 \sigma^\gi_\een,
      \label{eq:vondst2}  \\
 && \dot{\sigma}^\gi_\een = a_4 \sigma^\gi_\een,   \label{eq:vondst1}
\end{eqnarray}
\end{subequations}
where~$\sigma^\gi_\een$ is a short hand notation for the
combination of~$\varepsilon^\gi_\een$ and~$n^\gi_\een$ given in
Eq.~(\ref{eq:entropy-gi}). The derivation of these equations is
given in detail in Appendix~\ref{sec:deriv-egi}. The coefficients
$a_1,\ldots,a_4$ occurring in Eqs.~(\ref{subeq:eerste}) are
given~by
\begin{widetext}
\begin{subequations}
\label{subeq:coeff}
\begin{eqnarray}
  a_1 & = & \dfrac{\kappa\varepsilon_\nul(1+w)}{H}
  -2\dfrac{\dot{\beta}}{\beta}+H(4-3\beta^2)+
   \mbox{$^3\!R_\nul$}\left(\dfrac{1}{3H} + \dfrac{2H(1+3\beta^2)}
  {\mbox{$^3\!R_\nul$}+3\kappa\varepsilon_\nul(1+w)}\right), \label{eq:alpha-1} \\
  a_2 & = & \kappa\varepsilon_\nul(1+w)-
  4H\dfrac{\dot{\beta}}{\beta}+2H^2(2-3\beta^2)+
 \mbox{$^3\!R_\nul$}\left(\dfrac{1}{2}+
\dfrac{5H^2(1+3\beta^2)-2H\dfrac{\dot{\beta}}{\beta}}
  {\mbox{$^3\!R_\nul$}+3\kappa\varepsilon_\nul(1+w)}\right)
    -\beta^2\left(\frac{\tilde{\nabla}^2}{a^2}-\tfrac{1}{2}\,\mbox{$^3\!R_\nul$}\right),
         \label{eq:alpha-2} \\
  a_3 & = & \Biggl\{\dfrac{-18H^2}{\mbox{$^3\!R_\nul$}+3\kappa\varepsilon_\nul(1+w)}
  \Biggl[\varepsilon_\nul p_{\varepsilon n}(1+w)+
   \dfrac{2p_n}{3H}\dfrac{\dot{\beta}}{\beta}-
   \beta^2p_n+p_\varepsilon p_n+n_\nul
   p_{nn}\Biggr]+p_n\Biggr\}
     \left(\frac{\tilde{\nabla}^2}{a^2}-\tfrac{1}{2}\,\mbox{$^3\!R_\nul$}\right),
       \label{eq:alpha-3}\\
  a_4 & = & -3H\left(1-\frac{n_\nul
  p_n}{\varepsilon_\nul(1+w)}\right), \label{eq:alpha-4}
\end{eqnarray}
\end{subequations}
\end{widetext}
where the functions~$w(t)$ and $\beta(t)$ are given by
Eqs.~(\ref{begam1}) and~(\ref{begam2}), respectively. In the
derivation of the above results, we used
Eqs.~(\ref{subeq:einstein-flrw}) and~(\ref{FRW3}). The
abbreviations~$p_n$ and~$p_\varepsilon$ are given by
Eq.~(\ref{perttoes1}). Furthermore, we used the abbreviations
\begin{equation}
    p_{nn} := \frac{\partial^2 p}{\partial n^2}, \quad
    p_{\varepsilon n} :=
           \frac{\partial^2 p}{\partial \varepsilon\,\partial n}.
        \label{pne}
\end{equation}

The equations~(\ref{subeq:eerste}) contain only gauge-invariant
quantities and the coefficients are scalar functions. Thus, these
equations are \emph{manifestly gauge-invariant}.

The equations~(\ref{subeq:eerste}) are equivalent to one equation
of the third order, whereas one would expect that the four first
order equations~(\ref{subeq:pertub-gi}) would be equivalent to one
equation of the fourth order. This observation reflects the fact
that the solutions of the first order equations are gauge
dependent, while the solutions~$\varepsilon^\gi_\een$
and~$\sigma^\gi_\een$ of Eq.~(\ref{subeq:eerste}) are gauge
independent. One `degree of freedom', say, the gauge
function~$\psi$ has disappeared from the scene altogether.

The Eq.~(\ref{eq:vondst1}) can be solved
\begin{equation}\label{eq:sigmagi}
    \sigma^\gi_\een(t,\vec{x})=\sigma^\gi_\een(t_0,\vec{x})
  \exp \left\{-3\int^t_{t_0}H(\tau)\left(1-\dfrac{n_\nul(\tau) p_n(\tau)}
    {\varepsilon_\nul(\tau)\left[1+w(\tau)\right]}  \right) \dif\tau\right\}.
\end{equation}
Inserting this expression into Eq.~(\ref{eq:vondst2}) we obtain the
final equation for the perturbation~(\ref{eq:epsilongi}).

The equations~(\ref{subeq:eerste}) constitute the main result of
this article. In view of Eq.~(\ref{eq:entropy-gi}), they
essentially are two differential equations for the
perturbations~$\varepsilon_\een^\gi$ and~$n_\een^\gi$ to the
energy density~$\varepsilon_\nul(t)$ and the particle number
density~$n_\nul(t)$, respectively, for \textsc{flrw} universes
with $k=-1,0,+1$. They describe the evolution of the energy
density perturbation~$\varepsilon_\een^\gi$ and the particle
number density perturbation~$n_\een^\gi$ for \textsc{flrw}
universes filled with a fluid which is described by an equation of
state of the form ${p=p(n,\varepsilon)}$, the precise form of
which is left unspecified.

\subsection{Evolution equations for contrast functions}

In the study of the evolution of density perturbations it is of
convenience to use a quantity which measures the perturbation to
the density relative to the background densities. To that end we
define the gauge-invariant contrast functions $\delta_\varepsilon$
and $\delta_n$ by
\begin{equation}\label{eq:contrast}
  \delta_\varepsilon(t,\vec{x}):=
      \dfrac{\varepsilon^\gi_\een(t,\vec{x})}{\varepsilon_\nul(t)}, \quad
  \delta_n(t,\vec{x}):=
      \dfrac{n^\gi_\een(t,\vec{x})}{n_\nul(t)}.
\end{equation}
Using these quantities, Eqs.~(\ref{subeq:eerste}) can be
rewritten~as (see Appendix~\ref{app:contrast})
\begin{subequations}
\label{subeq:final}
\begin{eqnarray}
  && \ddot{\delta}_\varepsilon + b_1 \dot{\delta}_\varepsilon +
      b_2 \delta_\varepsilon =
      b_3 \left(\delta_n - \frac{\delta_\varepsilon}{1+w}\right),
              \label{sec-ord}  \\
  && \frac{1}{c}\frac{\dif}{\dif t}
      \left(\delta_n - \frac{\delta_\varepsilon}{1 + w}\right) =
     \frac{3Hn_\nul p_n}{\varepsilon_\nul(1 + w)}
     \left(\delta_n - \frac{\delta_\varepsilon}{1 + w}\right),
                 \label{fir-ord}
\end{eqnarray}
\end{subequations}
where the coefficients $b_1$, $b_2$ and~$b_3$ are given by
\begin{widetext}
\begin{subequations}
\label{subeq:coeff-contrast}
\begin{eqnarray}
  b_1 & = & \dfrac{\kappa\varepsilon_\nul(1+w)}{H}
  -2\dfrac{\dot{\beta}}{\beta}-H(2+6w+3\beta^2)+
 \mbox{$^3\!R_\nul$}\left(\dfrac{1}{3H}+
  \dfrac{2H(1+3\beta^2)}
  {\mbox{$^3\!R_\nul$}+3\kappa\varepsilon_\nul(1+w)}\right), \\
  b_2 & = & -\tfrac{1}{2}\kappa\varepsilon_\nul(1+w)(1+3w)+
  H^2\left(1-3w+6\beta^2(2+3w)\right) \nonumber \\
&&
+\,6H\dfrac{\dot{\beta}}{\beta}\left(w+\dfrac{\kappa\varepsilon_\nul(1+w)}
{\mbox{$^3\!R_\nul$}+3\kappa\varepsilon_\nul(1+w)}\right)-
 \mbox{$^3\!R_\nul$}\left(\tfrac{1}{2}w+
\dfrac{H^2(1+6w)(1+3\beta^2)}{\mbox{$^3\!R_\nul$}+3\kappa\varepsilon_\nul(1+w)}\right)
  -\beta^2\left(\frac{\tilde{\nabla}^2}{a^2}-\tfrac{1}{2}\,\mbox{$^3\!R_\nul$}\right), \\
  b_3 & = & \Biggl\{\dfrac{-18H^2}{\mbox{$^3\!R_\nul$}+3\kappa\varepsilon_\nul(1+w)}
  \Biggl[\varepsilon_\nul p_{\varepsilon n}(1+w)+
   \dfrac{2p_n}{3H}\dfrac{\dot{\beta}}{\beta}-
   \beta^2p_n+p_\varepsilon p_n+n_\nul p_{nn}\Biggr]+
   p_n\Biggr\}\dfrac{n_\nul}{\varepsilon_\nul}
     \left(\frac{\tilde{\nabla}^2}{a^2}-\tfrac{1}{2}\,\mbox{$^3\!R_\nul$}\right).
\end{eqnarray}
\end{subequations}
\end{widetext}
In Sec.~\ref{sec:an-exam} we use the Eqs.~(\ref{subeq:final}) to
study the evolution of small energy density perturbations and
particle number perturbations in \textsc{flrw} universes.

The entropy perturbation~(\ref{eq:TdS-1-gi}) reads, in terms of
the contrast functions~(\ref{eq:contrast}),
\begin{equation}\label{eq:S-contrast}
     s^\gi_\een = -\dfrac{\varepsilon_\nul(1+w)}{n_\nul T_\nul}
  \left( \delta_n - \frac{\delta_\varepsilon}{1 + w} \right).
\end{equation}

Finally, if we define the relative temperature
perturbation~$\delta_\mathrm{T}$ by
\begin{equation}\label{eq:delta-T}
  \delta_\mathrm{T}(t,\vec{x}):=\dfrac{T^\gi_\een(t,\vec{x})}{T_\nul(t)},
\end{equation}
then the relative temperature perturbation~(\ref{eq:T-pert}) is
given by
\begin{equation}\label{eq:rel-T-pert}
  \delta_\mathrm{T} = \dfrac{\varepsilon_\nul\delta_\varepsilon-
       \left(\dfrac{\partial \varepsilon}{\partial n}\right)_{\!T}n_\nul\delta_n}
        {T_\nul\left(\dfrac{\partial \varepsilon}{\partial T}\right)_{\!n}}.
\end{equation}
We thus have found the relative temperature perturbation as a
function of the relative perturbations in the energy density and
particle number density for an equation of state of the form
$\varepsilon=\varepsilon(n,T)$ and $p=p(n,T)$ (see
Appendix~\ref{sec:eq-state}). The quantity
$\delta_\mathrm{T}(t,\vec{x})$ is a measurable quantity in the
cosmic background radiation.

\section{Non-relativistic limit in an expanding universe} \label{nrl}

It is well known that if the gravitational field is weak and
velocities are small with respect to the velocity of light
($v/c\rightarrow0$), the system of Einstein equations and
conservation laws may reduce to the single field equation of the
Newtonian theory of gravity, namely the Poisson
equation~(\ref{poisson}). Since in the Newtonian theory the
gravitational field is described by only one, time-independent,
potential $\varphi(\vec{x})$, one cannot obtain the Newtonian limit
by simply taking the limit $v/c\rightarrow0$, since in a
relativistic theory the gravitational field is described, in
general, by six potentials, namely the components $h_{ij}(x)$ of the
metric. In this article we have used the
decomposition~(\ref{decomp-symh}). Moreover, we have shown in
Sec.~\ref{klasse} that $h^i_{\parallel j}$ given by
(\ref{decomp-hij-par}) describes the evolution of density
perturbations. By using this decomposition we have reduced the
number of potentials to only two, namely $\phi(x)$ and $\zeta(x)$.
At first sight, it would follow from the evolution
equations~(\ref{subeq:scalar}) for scalar perturbations that a
further reduction of the number of potentials by one is impossible.
However, we have rewritten the system of equations
(\ref{subeq:scalar}) for scalar perturbations into an equivalent
system (\ref{subeq:pertub-gi}). As a result, the perturbation to the
metric, $h^i_{\parallel j}$, enters the system
(\ref{subeq:pertub-gi}) only via the trace
\begin{equation}\label{RnabEE}
   \mbox{$^3\!R_{\een\parallel}$} =
   \dfrac{2}{c^2}\Bigl[2\phi^{|k}{}_{|k}-
  \zeta^{|k|l}{}_{|l|k}+\zeta^{|k}{}_{|k}{}^{|l}{}_{|l}+
  \tfrac{1}{3}\,\mbox{$^3\!R_\nul$}(3\phi+\zeta^{|k}{}_{|k})\Bigr],
\end{equation}
of the perturbation to the Ricci tensor (\ref{decomp-Rij-par}). This
shows explicitly that density perturbations are, in a non-flat
\textsc{flrw} universe, described by two potentials $\phi(x)$ and
$\zeta(x)$. Hence, in non-flat \textsc{flrw} universes the
perturbation equations~(\ref{subeq:pertub-gi}) do not reduce to the
Newtonian theory of gravity. However, for a \emph{flat}
\textsc{flrw} universe, i.e., a universe characterized by $k=0$,
implying, in view of~(\ref{spRicci}), that
\begin{equation}\label{eq:k-0-lam-0}
    \mbox{$^3\!R_\nul$}=0,
\end{equation}
the perturbation to the Ricci scalar, Eq.~(\ref{RnabEE}), reduces to
\begin{equation}\label{RnabEE-0}
  \mbox{$^3\!R_{\een\parallel}$} =
       -\dfrac{4}{c^2}\dfrac{\nabla^2\phi}{a^2},
\end{equation}
where~$\nabla^2$ is the usual Laplace operator. Hence, in a
\emph{flat} \textsc{flrw} universe we need only \emph{one}
potential~$\phi$ to describe density perturbations. Therefore, we
consider the only candidate \textsc{flrw} universe in which the
Newtonian limit could be obtained, namely a \emph{flat}
\textsc{flrw} universe.

Usually, one considers the non-relativistic limit in the case of a
\emph{static} gravitational field. In this section we will show that
our treatment of perturbations leads to the Poisson equation also
for an \emph{expanding} universe, i.e., for a non-static
gravitational field. Thus, in the approach presented in this article
there is no necessity to take the static limit at all. In fact, we
will show that the set of perturbation equations
(\ref{subeq:pertub-gi})--(\ref{subeq:pertub-gi-e-n}) reduce to
Eq.~(\ref{poisson}) without the assumption that the expansion is
negligibly slow or absent.

Let us first consider the background equations. For a flat
\textsc{flrw} universe, the zero order
equations~(\ref{subeq:einstein-flrw}) and~(\ref{FRW3}) reduce~to
\begin{subequations}
\label{subeq:einstein-flrw-flat}
\begin{eqnarray}
    \dot{H} & = & -\tfrac{1}{2}\kappa\varepsilon_\nul(1+w),  \label{FRW1-flat}   \\
    \dot{\varepsilon}_\nul & = & -3H\varepsilon_\nul(1+w), \label{FRW2-flat} \\
       \dot{n}_\nul & = & -3Hn_\nul,   \label{FRW2a-flat}
\end{eqnarray}
\end{subequations}
and the constraint equation
\begin{equation}\label{FRW3-flat}
    3H^2 = \kappa\varepsilon_\nul + \Lambda.
\end{equation}

We now consider the perturbation equations. Upon substituting
Eq.~(\ref{RnabEE-0}) into Eqs.~(\ref{subeq:pertub-gi}) and putting
$\mbox{$^3\!R_\nul$}=0$, we arrive at the set of perturbation
equations
\begin{subequations}
\label{subeq:pertub-gi-flat}
\begin{eqnarray}
 && \dot{\varepsilon}_\een + 3H(\varepsilon_\een + p_\een)+
    \varepsilon_\nul(1 + w)\left[\vartheta_\een +\frac{1}{2H}\left(
    \kappa\varepsilon_\een-\dfrac{2}{c^2}
     \dfrac{\nabla^2\phi}{a^2}\right)\right]=0, \label{FRW4gi-flat} \\
 && \dot{n}_\een + 3H n_\een+
    n_\nul \left[\vartheta_\een +\frac{1}{2H}\left(\kappa\varepsilon_\een -
      \dfrac{2}{c^2}\dfrac{\nabla^2\phi}{a^2}\right)\right]=0,  \label{FRW4agi-flat} \\
 && \dot{\vartheta}_\een+H(2-3\beta^2)\vartheta_\een+
        \frac{1}{\varepsilon_\nul(1+w)}\dfrac{\nabla^2p_\een}{a^2}=0,   \label{FRW5gi-flat}\\
 && \nabla^2\dot{\phi}+
       \dfrac{4\pi G}{c^2}a^2\varepsilon_\nul(1+w)\vartheta_\een=0,  \label{FRW6gi-flat}
\end{eqnarray}
\end{subequations}
for the four quantities~$\varepsilon_\een$, $n_\een$
$\vartheta_\een$ and~$\phi$. Similarly, we obtain from
Eqs.~(\ref{subeq:pertub-gi-e-n})
\begin{subequations}
\label{subeq:pertub-gi-e-n-flat}
\begin{eqnarray}
 &&  \nabla^2\left(\phi+
     \dfrac{2H\dot{\phi}}{(3H^2-\Lambda)(1+w)}\right)=
    \dfrac{4\pi G}{c^2}a^2\varepsilon_\een^\gi,  \label{Egi-flat} \\
 && n_\een^\gi = n_\een-
       \dfrac{n_\nul}{\varepsilon_\nul(1+w)}
       (\varepsilon_\een-\varepsilon^\gi_\een),   \label{nu2-flat}
\end{eqnarray}
\end{subequations}
where we have eliminated~$\vartheta_\een$ with the help of
Eq.~(\ref{FRW6gi-flat}) and~$\varepsilon_\nul$ with the help of
Eq.~(\ref{FRW3-flat}). The scale factor of the universe $a(t)$
follows from the Einstein equations via $H:=\dot{a}/a$.

We now consider the limits (see Appendix~\ref{sec:gauge-newton})
\begin{equation}\label{eq:nrl-limit}
  u_{\een\parallel}^i:=\dfrac{U_{\een\parallel}^i}{c} = 0, \quad p=0.
\end{equation}
In this limit, the kinetic energy is small compared to the rest
energy of a particle [see, for example, Eq.~(\ref{state-mat})], so
that the energy density of the universe in the limit
(\ref{eq:nrl-limit}) is
\begin{equation}\label{eq:rest-energy}
    \varepsilon = n m c^2,
\end{equation}
where $mc^2$ is the rest energy of a particle with mass~$m$ and~$n$
the particle number density. Thus, in the  limit
(\ref{eq:nrl-limit}), the background
equations~(\ref{subeq:einstein-flrw-flat})--(\ref{FRW3-flat}) take
the simple form
\begin{subequations}
\label{subeq:einstein-flrw-newt}
\begin{eqnarray}
    \dot{H} & = & -\tfrac{1}{2}\kappa\varepsilon_\nul,   \label{FRW1-newt}   \\
    \dot{\varepsilon}_\nul & = & -3H\varepsilon_\nul,  \label{FRW2-newt} \\
    \dot{n}_\nul & = & -3H n_\nul,  \label{FRW2a-newt}
\end{eqnarray}
\end{subequations}
while the constraint equation~(\ref{FRW3-flat}) reads
\begin{equation}\label{FRW3-newt}
    3H^2 = \kappa\varepsilon_\nul + \Lambda.
\end{equation}
Note that in the limit (\ref{eq:nrl-limit}), Eqs.~(\ref{FRW2-newt})
and~(\ref{FRW2a-newt}) are identical, since then
$\varepsilon_\nul=n_\nul mc^2$, in view of
Eq.~(\ref{eq:rest-energy}).

We now consider the perturbation equations
(\ref{subeq:pertub-gi-flat}) and (\ref{subeq:pertub-gi-e-n-flat}) in
the limit (\ref{eq:nrl-limit}). Substituting
$\vartheta_\een:=(u^k_{\een\parallel})_{|k}=0$,
$w:=p_\nul/\varepsilon_\nul=0$ and $p_\een=0$ into these equations
we find that Eq.~(\ref{FRW5gi-flat}) is identically satisfied,
whereas the remaining equations~(\ref{subeq:pertub-gi-flat}) reduce
to
\begin{subequations}
\label{subeq:pertub-gi-flat-newt}
\begin{eqnarray}
&&   \dot{\varepsilon}_\een + 3H\varepsilon_\een+
     \dfrac{\varepsilon_\nul}{2H}\left(
 \kappa\varepsilon_\een-\dfrac{2}{c^2}\dfrac{\nabla^2\phi}{a^2}\right)=0,
                   \label{FRW4gi-flat-newt} \\
&&   \dot{n}_\een + 3H n_\een+
     \dfrac{n_\nul}{2H}\left(
 \kappa\varepsilon_\een-\dfrac{2}{c^2}\dfrac{\nabla^2\phi}{a^2}\right)=0,
                   \label{FRW4agi-flat-newt} \\
&&    \nabla^2\dot{\phi}=0,   \label{FRW6gi-flat-newt}
\end{eqnarray}
\end{subequations}
and Eqs.~(\ref{subeq:pertub-gi-e-n-flat}) become
\begin{subequations}
\label{subeq:pertub-gi-flat-e-n-newt}
\begin{eqnarray}
&&  \nabla^2\left(\phi+
     \dfrac{2H\dot{\phi}}{(3H^2-\Lambda)}\right)=
    \dfrac{4\pi G}{c^2}a^2\varepsilon_\een^\gi,        \label{Egi-flat-newt} \\
&&  n_\een^\gi = \dfrac{\varepsilon^\gi_\een}{mc^2}.
\label{nu2-flat-newt}
\end{eqnarray}
\end{subequations}
Note that in the limit (\ref{eq:nrl-limit}),
Eqs.~(\ref{FRW4gi-flat-newt}) and~(\ref{FRW4agi-flat-newt}) are
identical, since then $\varepsilon_\een=n_\een mc^2$, in view of
Eq.~(\ref{eq:rest-energy}). In
Eqs.~(\ref{FRW6gi-flat-newt})--(\ref{subeq:pertub-gi-flat-e-n-newt})
the quantities~$\varepsilon_\een$ and~$n_\een$ do not occur. As a
consequence, these equations can be solved for the
quantities~$\phi$, $\varepsilon^\gi_\een$ and~$n^\gi_\een$. In
Appendix~\ref{sec:gauge-newton} we show that $\varepsilon_\een$
and~$n_\een$ are gauge dependent also in the limit
(\ref{eq:nrl-limit}), so that Eqs.~(\ref{FRW4gi-flat-newt})
and~(\ref{FRW4agi-flat-newt}) need not be considered.

We now consider Eqs.~(\ref{FRW6gi-flat-newt})
and~(\ref{Egi-flat-newt}) in some detail. By substituting
Eq.~(\ref{FRW6gi-flat-newt}) into Eq.~(\ref{Egi-flat-newt}) we
obtain
\begin{equation}
  \nabla^2\phi(\vec{x})=
    \dfrac{4\pi G}{c^2}a^2(t)\varepsilon_\een^\gi(t,\vec{x}).
    \label{Egi-poisson}
\end{equation}
Since $\nabla^2\phi$ is independent of time,
Eq.~(\ref{FRW6gi-flat-newt}), this equation is equivalent to
\begin{equation}
  \nabla^2\phi(\vec{x})=
    \dfrac{4\pi G}{c^2}a^2(t_\mathrm{p})
    \varepsilon_\een^\gi(t_\mathrm{p},\vec{x}),
    \label{Egi-poisson-present}
\end{equation}
where $t_\mathrm{p}$ indicates the present time. This Einstein
equation can be rewritten in a form that closely resembles the
Poisson equation, by introducing the potential~$\varphi$
\begin{equation}\label{eq:ident}
   \varphi(\vec{x}) :=\frac{\phi(\vec{x})}{a^2(t_\mathrm{p})}.
\end{equation}
Inserting (\ref{eq:ident}) into~(\ref{Egi-poisson-present}) we
obtain the result~(\ref{poisson}): the Einstein
equation~(\ref{Egi-poisson}) for the \emph{time dependent}
perturbation $\varrho^\gi_\een(t,\vec{x})$ is, \emph{at fixed}
$t=t_\mathrm{p}$, identical to the usual, time independent, Poisson
equation. Hence, as argued in Sec.~\ref{frommat-non}, the
perturbations~$\varepsilon^\gi_\een$ and~$n^\gi_\een$ may indeed be
interpreted as physical perturbations.

\section{Analytical examples of the perturbation theory}
\label{sec:an-exam}

In this section we consider two particular simple cases as an
example: a flat \textsc{flrw} universe with a vanishing cosmological
constant in its radiation-dominated and matter-dominated stage. In
order to keep this article self-contained, we start with the zero
order equations and their solutions, although nothing is new here.
We need these solutions to obtain explicit forms for the first order
equations.

\subsection{Radiation-dominated phase of the flat universe}
\label{sec:rad}

In the radiation-dominated era we have $p=\frac{1}{3}\varepsilon$,
so that, according to (\ref{begam1}), $w=\frac{1}{3}$. Next to this
the cosmological constant is put equal to zero: $\Lambda=0$.
Furthermore, a flat universe ($k=0$) is considered, implying, with
(\ref{spRicci}), that $\mbox{$^3\!R_\nul$}(t)=0$.

\subsubsection{Zero-order equations and their solutions}

The zero-order equations (\ref{subeq:einstein-flrw}) then reduce to
\begin{subequations}
\label{subeq:rad-R0}
\begin{eqnarray}
   \dot{H} & = & -\tfrac{2}{3} \kappa\varepsilon_\nul,
            \label{dyn} \\
    \dot{\varepsilon}_\nul & = & -4H
         \varepsilon_\nul,  \label{eq:behoud} \\
    \dot{n}_\nul & = & -3Hn_\nul,  \label{eq:n-behoud}
\end{eqnarray}
\end{subequations}
while the constraint equation (\ref{FRW3}) becomes
\begin{equation}\label{con}
    H^2 = \tfrac{1}{3}\kappa\varepsilon_\nul.
\end{equation}
The general solutions of these equations are
\begin{subequations}
\label{subeq:rad-R0-sol}
\begin{eqnarray}
   H(t) & = & \tfrac{1}{2} \left(ct\right)^{-1}=
      H(t_0)\left(\frac{t}{t_0}\right)^{-1},  \label{sol1a} \\
   \varepsilon_{\scriptscriptstyle(0)}(t) & = &
       \frac{3}{4\kappa} \left(ct\right)^{-2}=
       \varepsilon_\nul(t_0)\left(\frac{t}{t_0}\right)^{-2},  \label{sol1c}  \\
   n_\nul(t) & = & n_\nul(t_0) \left(\frac{a(t)}{a(t_0)}\right)^{-3}. \label{sol1d}
\end{eqnarray}
\end{subequations}
The initial values $H(t_0)$ and $\varepsilon_\nul(t_0)$ are related
by the initial value condition (\ref{con}).

Using the definition of the Hubble parameter $H:=\dot{a}/a$ we find
from (\ref{sol1a}) that
\begin{equation}\label{sol1b}
   a(t) = a(t_0) \left(\frac{t}{t_0}\right)^{\tfrac{1}{2}},
\end{equation}
where $t_0$ is the time at which the radiation-dominated era sets
in.

\subsubsection{First-order equations and their solutions}

The zero-order solutions (\ref{subeq:rad-R0-sol}) can now be
substituted into the coefficients (\ref{subeq:coeff-contrast}) of
the equations (\ref{subeq:final}). Recalling that
$p=\frac{1}{3}\varepsilon$ in the radiation-dominated regime,
$p_n=0$ and $p_\varepsilon=\frac{1}{3}$, so that $\beta=1/\sqrt{3}$,
see (\ref{eq:begam3}). Since $\tilde{\nabla}^2=\nabla^2$ for a flat
universe, the coefficients $b_1$, $b_2$ and $b_3$ reduce to
\begin{equation}
\label{subeq:coeff-contrast-flat}
  b_1 = -H, \quad
  b_2 = -\frac{1}{3}\frac{\nabla^2}{a^2}+
     \tfrac{2}{3}\kappa\varepsilon_\nul, \quad
  b_3  = 0,
\end{equation}
where we have used (\ref{con}). For the first-order
equations~(\ref{subeq:final}) this yields the simple forms
\begin{subequations}
\label{subeq:final-rad}
\begin{eqnarray}
   \ddot{\delta}_\varepsilon-H\dot{\delta}_\varepsilon+
  \left(-\frac{1}{3}\frac{\nabla^2}{a^2}+
   \tfrac{2}{3}\kappa\varepsilon_\nul\right)
   \delta_\varepsilon & = & 0,   \label{eq:delta-rad} \\
    \frac{1}{c}\frac{\dif}{\dif t}
        \left(\delta_n-\tfrac{3}{4}\delta_\varepsilon\right) & = & 0,
            \label{eq:entropy-rad}
\end{eqnarray}
\end{subequations}
where $H:=\dot{a}/a$, (\ref{Hubble}). The solution of equation
(\ref{eq:entropy-rad}) expresses the fact that particle number
density perturbations are coupled to perturbations in the radiation,
i.e.,
\begin{equation}\label{eq:coupled-rad}
    \delta_n(t,\vec{x})=\tfrac{3}{4}\delta_\varepsilon(t,\vec{x}).
\end{equation}
This relation is, in the extreme case of radiation-domination,
\emph{independent} of both the nature of the particles and the scale
of the perturbation.

Equation (\ref{eq:delta-rad}) may be solved by Fourier analysis of
the function $\delta_\varepsilon$. Writing
\begin{equation}\label{pw12}
    \delta_\varepsilon(t,\vec{x}) =
    \delta_\varepsilon(t,\vec{q})\me^{\mi {\vecs{q}} \cdot {\vecs{x}}},\quad
    \delta_n(t,\vec{x}) =
    \delta_n(t,\vec{q})\me^{\mi {\vecs{q}} \cdot {\vecs{x}}},
\end{equation}
with $q=|\vec{q}|=2\pi/\lambda$, where $\lambda$ is the wavelength
of the perturbation and ${\mi^2=-1}$, yielding
\begin{equation}\label{Fourier12}
   \nabla^2\delta_\varepsilon(t,\vec{x}) =
       -q^2\delta_\varepsilon(t,\vec{q}), \quad
   \nabla^2\delta_n(t,\vec{x}) =
       -q^2\delta_n(t,\vec{q}),
\end{equation}
so that the evolution equation (\ref{eq:delta-rad}) for the
amplitude $\delta_\varepsilon(t,\vec{q})$ reads
\begin{equation}
  \ddot{\delta}_\varepsilon -
    H(t_0)\left(\frac{t}{t_0}\right)^{-1} \dot{\delta}_\varepsilon
    +\left[\frac{1}{3}\frac{q^2}{a^2(t_0)}\left(\frac{t}{t_0}\right)^{-1} +
    2H^2(t_0)\left(\frac{t}{t_0}\right)^{-2}\right]\delta_\varepsilon =0,  \label{delta-pnue}
\end{equation}
where we have used (\ref{con})--(\ref{subeq:rad-R0-sol}). This
equation will be rewritten in such a way that the coefficients
become dimensionless. To that end a dimensionless time variable is
introduced, defined by
\begin{equation}
   \tau := \frac{t}{t_0}, \quad t \ge t_0.  \label{tau}
\end{equation}
This definition implies
\begin{equation}
   \frac{\dif^n}{c^n\dif t^n}=\left(\frac{1}{ct_0}\right)^n\frac{\dif^n}{\dif\tau^n}=
   \left[2H(t_0)\right]^n
   \frac{\dif^n}{\dif\tau^n}, \quad n=1,2,\ldots\,,  \label{dtau-n}
\end{equation}
where we have used (\ref{sol1a}). Using (\ref{sol1a}), (\ref{tau})
and (\ref{dtau-n}), equation (\ref{delta-pnue}) for the density
contrast $\delta_\varepsilon(\tau,\vec{q})$ can be written as
\begin{equation}
    \delta_\varepsilon^{\prime\prime} - \frac{1}{2\tau}\delta_\varepsilon^\prime +
  \left(\frac{\mu_\mathrm{r}^2}{4\tau} + \frac{1}{2\tau^2}\right)\delta_\varepsilon=0,
         \label{delta-pnue-tau}
\end{equation}
where a prime denotes differentiation with respect to $\tau$. The
constant $\mu_\mathrm{r}$ is given by
\begin{equation}
     \mu_\mathrm{r} := \frac{q}{a(t_0)}\frac{1}{H(t_0)}\frac{1}{\sqrt{3}}\,.   \label{xi}
\end{equation}
The general solution of equation (\ref{delta-pnue-tau}) is a linear
combination of the functions
$J_{\pm\frac{1}{2}}(\mu_\mathrm{r}\sqrt{\tau})\tau^{3/4}$, where
$J_{+\frac{1}{2}}(x)=\sqrt{2/(\pi x)}\sin x$ and
$J_{-\frac{1}{2}}(x)=\sqrt{2/(\pi x)}\cos x$ are Bessel Functions of
the first kind:
\begin{equation}
 \delta_\varepsilon(\tau,\vec{q}) =
      \Bigl[A_1(\vec{q})\sin\left(\mu_\mathrm{r}\sqrt{\tau}\right) +
         A_2(\vec{q})\cos\left(\mu_\mathrm{r}\sqrt{\tau}\right)\Bigr]\sqrt{\tau}, \label{nu13}
\end{equation}
where the functions $A_1(\vec{q})$ and $A_2(\vec{q})$ are given by
\begin{subequations}
\label{subeq:C1-C2}
\begin{eqnarray}
   A_1(\vec{q}) & = & \delta_\varepsilon(t_0,\vec{q})\sin\mu_\mathrm{r} - \frac{\cos\mu_\mathrm{r}}{\mu_\mathrm{r}}
     \left[\delta_\varepsilon(t_0,\vec{q})-\frac{\dot{\delta}_\varepsilon(t_0,\vec{q})}{H(t_0)}\right],
        \label{C1} \\
   A_2(\vec{q}) & = & \delta_\varepsilon(t_0,\vec{q})\cos\mu_\mathrm{r} + \frac{\sin\mu_\mathrm{r}}{\mu_\mathrm{r}}
     \left[\delta_\varepsilon(t_0,\vec{q})-\frac{\dot{\delta}_\varepsilon(t_0,\vec{q})}{H(t_0)}\right],
      \label{C2}
\end{eqnarray}
\end{subequations}
where we have used that
\begin{equation}
     \delta_\varepsilon(t_0,\vec{q}) =
     \delta_\varepsilon(\tau=1,\vec{q}), \quad
     \dot{\delta}_\varepsilon(t_0,\vec{q}) =
         2 \delta_\varepsilon^\prime(\tau=1,\vec{q})H(t_0),
\end{equation}
as follows from (\ref{dtau-n}).

For large-scale perturbations, ${\lambda\rightarrow\infty}$, the
magnitude of the wave vector $|\vec{q}|=2\pi/\lambda$ vanishes.
Writing ${\delta_\varepsilon(t)\equiv\delta_\varepsilon(t,q=0)}$ and
${\dot{\delta}_\varepsilon(t)\equiv\dot{\delta}_\varepsilon(t,q=0)}$,
we find from (\ref{xi})--(\ref{subeq:C1-C2}) that, for ${t\ge t_0}$,
\begin{equation}
    \delta_\varepsilon(t) = -\left[\delta_\varepsilon(t_0)-
       \frac{\dot{\delta}_\varepsilon(t_0)}{H(t_0)}\right]
       \frac{t}{t_0}
    +\left[2\delta_\varepsilon(t_0)
    - \frac{\dot{\delta}_\varepsilon(t_0)}{H(t_0)}\right]
     \left(\frac{t}{t_0}\right)^{\tfrac{1}{2}}. \label{delta-H-rad}
\end{equation}
The growth rates proportional to $t$ and $t^{1/2}$ have been derived
from the full set of linearized Einstein equations and conservation
laws by a large number of researchers. See Lifshitz and Khalatnikov
\cite{c15}, (8.11), Adams and Canuto \cite{adams-canuto1975},
(4.5b), Olson \cite{olson1976}, page 329, Peebles \cite{c11},
(86.20), and Kolb and Turner \cite{kolb}, (9.121). The complete
solution (\ref{delta-H-rad}), however, has never been presented in
literature. From this solution it follows that large-scale
perturbations grow only if the initial growth rate is large enough,
i.e.,
\begin{equation}\label{eq:ls-grow}
    \dot{\delta}_\varepsilon(t_0) \ge
    \delta_\varepsilon(t_0)H(t_0),
\end{equation}
otherwise the perturbations are decaying.

In the small-scale limit ${\lambda\rightarrow0}$ (or, equivalently,
${|\vec{q}|\rightarrow\infty}$) we find, using
(\ref{xi})--(\ref{subeq:C1-C2}), that
\begin{equation}\label{dc-small}
   \delta_\varepsilon(t,\vec{q}) \approx \delta_\varepsilon(t_0,\vec{q})
      \left(\frac{t}{t_0}\right)^{\tfrac{1}{2}}\cos\left[\mu_\mathrm{r}-
               \mu_\mathrm{r}\left(\frac{t}{t_0}\right)^{\tfrac{1}{2}}\right].
\end{equation}
Due to the pressure gradients, which play a role only on small
scales, a smaller growth rate than in the case of large-scale
perturbations is found.

In contrast to the solution which results from the `standard
equation', Eq.~(\ref{eq:delta-rad-peacock}),  which yields
oscillating density perturbations (\ref{eq:peacock-sol}) with a
\emph{decreasing} amplitude, our equation leads to a solution which
shows that  small-scale perturbations oscillate with an
\emph{increasing} amplitude.

Although the growth rates $t$ and $t^{1/2}$ occurring in the
large-scale solution (\ref{delta-H-rad}) are well-known, the general
solution (\ref{nu13}), and, in particular, the large-scale solution
(\ref{delta-H-rad}) and the small-scale solution (\ref{dc-small}),
have not been found earlier.

The approach presented in this article yields an evolution
equation~(\ref{eq:delta-rad}) which differs from the standard
equation~(\ref{eq:delta-rad-peacock}). This difference will be
explained in detail in Sec.~\ref{sec:stan-th}\@.

\subsection{Matter-dominated phase of the flat universe}
\label{matter}

Once protons and electrons recombine to yield hydrogen at a
temperature around $4000\,\mathrm{K}$, the radiation pressure
becomes negligible, and the equations of state reduce to those of a
non-relativistic monatomic perfect gas [Weinberg \cite{c8},
equations (15.8.20) and (15.8.21)]
\begin{equation}
  \varepsilon(n,T) = nm_{\rm H}c^2+\tfrac{3}{2}nk_{\rm B}T, \quad
  p(n,T) = nk_{\rm B}T,    \label{state-mat}
\end{equation}
where $k_{\rm B}$ is Boltzmann's constant, $m_{\rm H}$ the proton
mass, and $T$ the temperature of the matter.

\subsubsection{Zero-order equations and their solutions}

The maximum temperature in the matter-dominated era occurs around
time $t_0$ of the decoupling of matter and radiation:
${T(t_0)\approx4000\,\mathrm{K}}$. Hence, from equation
(\ref{state-mat}) it follows that the pressure is negligible with
respect to the energy density, i.e.,
\begin{equation}\label{eq:p-neg-e}
    \frac{p}{\varepsilon}\approx \frac{k_{\rm B}T}{m_{\rm H}c^2} \le
3.7\times 10^{-10}.
\end{equation}
This implies that, to a good approximation,
$\varepsilon_{\scriptscriptstyle(0)}\pm p_{\scriptscriptstyle(0)}
\approx \varepsilon_{\scriptscriptstyle(0)}$ and
$\varepsilon_{\scriptscriptstyle(0)}\approx
n_{\scriptscriptstyle(0)}m_{\rm H} c^2$. Hence, in an
\emph{un}perturbed flat \textsc{flrw} universe, the pressure can be
neglected with respect to the energy density. The above facts, yield
that the Einstein equations and conservation laws
(\ref{subeq:einstein-flrw})--(\ref{FRW3}) for a flat \textsc{flrw}
universe reduce~to
\begin{subequations}
\label{subeq:matp0}
\begin{eqnarray}
   \dot{H} & = & -\tfrac{1}{2}\kappa\varepsilon_\nul, \label{mat1p0} \\
   \dot{\varepsilon}_\nul & = & -3H\varepsilon_\nul, \label{mat2p0} \\
   \dot{n}_\nul & = & -3Hn_\nul,   \label{mat4p0}
\end{eqnarray}
\end{subequations}
and the constraint equation
\begin{equation}
     H^2 = \tfrac{1}{3}\kappa\varepsilon_\nul, \label{mat3p0}
\end{equation}
where we have put the cosmological constant $\Lambda$ equal to zero.
The general solutions of the zero-order Einstein equations and
conservation laws are
\begin{subequations}
\label{subeq:matp0-sol}
\begin{eqnarray}
 H(t) & = & \tfrac{2}{3} \left(ct\right)^{-1}=
   H(t_0)\left(\frac{t}{t_0}\right)^{-1}, \label{matsol1a} \\
   \varepsilon_\nul(t) & = &
     \frac{4}{3\kappa} \left(ct\right)^{-2}=
     \varepsilon_\nul(t_0)\left(\frac{t}{t_0}\right)^{-2},
     \label{matsol1c} \\
     n_\nul(t) & = &
     n_\nul(t_0)\left(\frac{t}{t_0}\right)^{-2}. \label{matsol1d}
\end{eqnarray}
\end{subequations}
The initial values $H(t_0)$ and $\varepsilon_\nul(t_0)$ are related
by the initial value condition (\ref{mat3p0}).

Using the definition of the Hubble parameter $H:=\dot{a}/a$, it is
found from (\ref{matsol1a}) for the scale factor that
\begin{equation}\label{matsol1b}
    a(t) = a(t_0) \left(\frac{t}{t_0}\right)^{\tfrac{2}{3}},
\end{equation}
where $t_0$ is the time at which the matter-dominated era sets in.

\subsubsection{First-order equations and their solutions}

Using Eqs.~(\ref{subeq:final-pn-pe}) we find from the equations of
state~(\ref{state-mat})
\begin{equation}
   p_\varepsilon = \tfrac{2}{3}, \quad p_n=-\tfrac{2}{3} m_{\rm H}c^2,
     \label{pepn-mat}
\end{equation}
so that, to a good approximation, using (\ref{eq:begam3})
\begin{equation}
   \beta(t) \approx \frac{v_\mathrm{s}(t)}{c}=\sqrt{\frac{5}{3}
        \frac{k_{\rm B}T_\nul(t)}{m_{\rm H}c^2}},
\label{coef-nu1}
\end{equation}
where $v_\mathrm{s}$ is the speed of sound. Differentiating
(\ref{coef-nu1}) with respect to time yields
\begin{equation}
   \frac{\dot{\beta}}{\beta} = \frac{\dot{T}_\nul}{2T_\nul}.
     \label{coef-a}
\end{equation}
For the time development of the background temperature it is found
from Eqs.~(\ref{eq:evo-T0}) and~(\ref{state-mat}), this time
\emph{not} neglecting the pressure with respect to the energy
density, that
\begin{equation}\label{eq:bd-b}
    \dot{T}_\nul=-2HT_\nul.
\end{equation}
(Note that putting $w=0$ in Eq.~(\ref{eq:evo-T0}) yields the
incorrect result $\dot{T}_\nul=0$.) Combining (\ref{coef-a}) and
(\ref{eq:bd-b}) results in
\begin{equation}
   \frac{\dot{\beta}}{\beta} = -H.
     \label{eq:db-b-H}
\end{equation}
For the evolution of the background temperature it is found that
\begin{equation}
    T_\nul(t)= T_\nul(t_0) \left(\frac{a(t)}{a(t_0)}\right)^{-2},
       \label{e-T3}
\end{equation}
where we have used that $H:=\dot{a}/a$. Using (\ref{pepn-mat}) it is
found for equation (\ref{fir-ord})
\begin{equation}\label{eq:entropy-dust}
      \frac{1}{c}\frac{\dif}{\dif t}
        \left(\delta_n-\delta_\varepsilon\right)=
        -2H\left(\delta_n-\delta_\varepsilon\right).
\end{equation}
A solution to this equation is
\begin{equation}
  \delta_n(t,\vec{x})=\delta_\varepsilon(t,\vec{x}),
      \label{eq:entropy-dust-sol}
\end{equation}
where we have used that in the matter-dominated phase nearly all
energy resides in the rest mass of the particles. After substituting
(\ref{pepn-mat}), (\ref{coef-nu1}) and (\ref{eq:db-b-H}) into the
coefficients (\ref{subeq:coeff-contrast}) it is found that
\begin{equation}\label{eq:coeff-dust}
    b_1=3H, \quad b_2=-\tfrac{5}{6}\kappa\varepsilon_\nul-
      \frac{v_\mathrm{s}^2}{c^2}\frac{\nabla^2}{a^2}, \quad
      b_3=-\frac{2}{3}\frac{\nabla^2}{a^2},
\end{equation}
considering that for a flat universe $\tilde{\nabla}^2=\nabla^2$.
For the evolution equation for density perturbations,
(\ref{sec-ord}), this results in the simple form
\begin{equation}\label{eq:delta-dust}
  \ddot{\delta}_\varepsilon + 3H\dot{\delta}_\varepsilon-
  \left(\frac{v_\mathrm{s}^2}{c^2}\frac{\nabla^2}{a^2}+
   \tfrac{5}{6}\kappa\varepsilon_\nul\right)
   \delta_\varepsilon =0,
\end{equation}
where we have used the result (\ref{eq:entropy-dust-sol}). Using
(\ref{pw12}) and (\ref{Fourier12}) the evolution equation for the
amplitude $\delta_\varepsilon(t,\vec{q})$ can be rewritten as
\begin{equation}\label{eq:delta-dust-amp}
   \ddot{\delta}_\varepsilon+
  3H(t_0)\left(\frac{t}{t_0}\right)^{-1}\dot{\delta}_\varepsilon
  + H^2(t_0)\left[\mu_\mathrm{m}^2\left(\frac{t}{t_0}\right)^{-\tfrac{8}{3}}-
   \frac{5}{2}\left(\frac{t}{t_0}\right)^{-2}
  \right]\delta_\varepsilon=0,
\end{equation}
having incorporated (\ref{mat3p0})--(\ref{matsol1b}),
(\ref{coef-nu1}) and (\ref{e-T3}). The constant $\mu_\mathrm{m}$ is
given by
\begin{equation}\label{eq:const-mu}
    \mu_\mathrm{m}=\frac{q}{a(t_0)}\frac{1}{H(t_0)}\frac{v_\mathrm{s}(t_0)}{c}, \quad
    v_\mathrm{s}(t_0)=
        \sqrt{\frac{5}{3}\frac{k_{\rm B}T_\nul(t_0)}{m_{\rm H}}}.
\end{equation}
Using the dimensionless time variable (\ref{tau}), it is found from
(\ref{matsol1a}) that
\begin{equation}
   \frac{\dif^n}{c^n\dif t^n}=\left(\frac{1}{ct_0}\right)^n\frac{\dif^n}{\dif\tau^n}=
   \left[\frac{3}{2}H(t_0)\right]^n
   \frac{\dif^n}{\dif\tau^n}, \quad n=1,2,\ldots\,.    \label{dtau-n-dust}
\end{equation}
Using this expression, equation (\ref{eq:delta-dust-amp}) can be
rewritten in the form
\begin{equation}\label{eq:dust-dimless}
    \delta_\varepsilon^{\prime\prime}+\frac{2}{\tau}\delta_\varepsilon^\prime+
    \left(\frac{4\mu_\mathrm{m}^2}{9\tau^{8/3}}-\frac{10}{9\tau^2}\right)\delta_\varepsilon=0,
\end{equation}
where a prime denotes differentiation with respect to $\tau$. Hence,
the general solution of equation (\ref{eq:dust-dimless}) is
\begin{equation}\label{eq:matter-physical}
 \delta_\varepsilon(\tau,\vec{q}) =
    \left[B_1(\vec{q}) J_{+\frac{7}{2}}\bigl(2\mu_\mathrm{m}\tau^{-1/3}\bigr)+
        B_2(\vec{q}) J_{-\frac{7}{2}}\bigl(2\mu_\mathrm{m}\tau^{-1/3}\bigr)\right]\tau^{-1/2},
\end{equation}
where $B_1(\vec{q})$ and $B_2(\vec{q})$ are arbitrary functions and
$J_{\pm\nu}(x)$ is the Bessel function of the first kind. In other
words, in the matter-dominated era density perturbations oscillate
with a slowly decaying amplitude.

In the large-scale limit ($|\vec{q}|\rightarrow0$), it is found
that, transforming back from $\tau$ to $t$,
\begin{equation}
    \delta_\varepsilon(t) =
    \frac{1}{7}\left[5\delta_\varepsilon(t_0)+
     \frac{2\dot{\delta}_\varepsilon(t_0)}{H(t_0)}\right]
 \left(\frac{t}{t_0}\right)^{\tfrac{2}{3}}
  +\frac{2}{7}\left[\delta_\varepsilon(t_0)-
      \frac{\dot{\delta}_\varepsilon(t_0)}{H(t_0)}\right]
      \left(\frac{t}{t_0}\right)^{-\tfrac{5}{3}},
  \label{eq:new-dust-53}
\end{equation}
having used that
\begin{equation}
     \delta_\varepsilon(t_0,\vec{q}) =
     \delta_\varepsilon(\tau=1,\vec{q}), \quad
     \dot{\delta}_\varepsilon(t_0,\vec{q}) =
         \tfrac{3}{2} \delta_\varepsilon^\prime(\tau=1,\vec{q})H(t_0),
\end{equation}
as follows from (\ref{dtau-n-dust}). The solution proportional to
$t^{2/3}$ is a standard result. Since $\delta_\varepsilon$ is
gauge-invariant, the standard non-physical gauge mode proportional
to $t^{-1}$ is absent from our theory. Instead, a physical mode
proportional to $t^{-5/3}$ is found.

The approach presented in this article yields an evolution
equation~(\ref{eq:delta-dust}) which differs from the standard
equation~(\ref{eq:delta-dust-standard}). This difference will be
explained in detail in the next section.

\section{Relation to earlier work} \label{sec:stan-th}

In this section it is shown that, in contrast to the approach
developed in this article, the standard theory of cosmological
density perturbations results in \emph{decaying} small-scale density
perturbations in a radiation-dominated flat \textsc{flrw} universe.
Furthermore, we consider the standard Newtonian treatment for
density perturbations in a flat \textsc{flrw} matter-dominated
universe in the light of our equations, which are refined compared
to the usual equations of earlier treatments.

The difference between our treatment and the standard approach is
that the solutions of the equations of the standard approach contain
the gauge function $\psi(\vec{x})$, whereas our solutions are
gauge-invariant.

\subsection{Radiation-dominated universe} \label{sec:rad-stand}

The standard equation for the density contrast function $\delta$
which can be found e.g., in the textbook of Peacock
\cite{peacock1999}, equation (15.25), is given by
\begin{equation}\label{eq:delta-rad-peacock}
  \ddot{\delta}+2H\dot{\delta}-
  \left(\frac{1}{3}\frac{\nabla^2}{a^2}+
   \tfrac{4}{3}\kappa\varepsilon_\nul\right) \delta=0.
\end{equation}
This equation is derived by using special relativistic fluid
mechanics and the Newtonian theory of gravity with a relativistic
source term. In agreement with the text under equation (15.25) of
this textbook, the term $-\frac{1}{3}\nabla^2\delta/a^2$ has been
added. The same result, equation (\ref{eq:delta-rad-peacock}), can
be found in Weinberg's classic \cite{c8}, equation (15.10.57) with
$p=\frac{1}{3}\rho$ and $v_\mathrm{s}=1/\sqrt{3}$. Using
(\ref{pw12}), it is found for the general solution of equation
(\ref{eq:delta-rad-peacock}) that
\begin{equation}\label{eq:peacock-sol}
    \delta(\tau,\vec{q})=\frac{8C_1(\vec{q})}{\mu_\mathrm{r}^2}
            J_2\bigl(\mu_\mathrm{r}\sqrt{\tau}\bigr)+
           \psi(\vec{q})\pi \mu_\mathrm{r}^2 H(t_0)
               Y_2\bigl(\mu_\mathrm{r}\sqrt{\tau}\bigr),
\end{equation}
where $C_1(\vec{q})$ and $\psi(\vec{q})$ are arbitrary functions
(the integration `constants'), $\tau$ is given by (\ref{tau}), the
constant $\mu_\mathrm{r}$ is given by (\ref{xi}) and $J_\nu(x)$ and
$Y_\nu(x)$ are Bessel functions of the first and second kind,
respectively. The factors $8/\mu_\mathrm{r}^2$ and
$\pi\mu_\mathrm{r}^2H(t_0)$ have been inserted for convenience. The
derivation of the solution (\ref{eq:peacock-sol}) runs along the
same lines as the derivation of (\ref{nu13}). Thus, the standard
equation (\ref{eq:delta-rad-peacock}) yields oscillating density
perturbations with a \emph{decaying} amplitude.

For large-scale
perturbations ($|\vec{q}|\rightarrow0$, or, equivalently,
$\mu_\mathrm{r}\rightarrow0$), the asymptotic expressions for the
Bessel functions $J_2$ and $Y_2$ are given by
\begin{equation}\label{eq:J2Y2-0}
    J_2\bigl(\mu_\mathrm{r}\sqrt{\tau}\bigr)\approx
             \frac{\mu_\mathrm{r}^2}{8}\tau, \quad
    Y_2\bigl(\mu_\mathrm{r}\sqrt{\tau}\bigr)
             \approx-\frac{4}{\pi\mu_\mathrm{r}^2}\tau^{-1}.
\end{equation}
Substituting these expressions into (\ref{eq:peacock-sol}), it is
found for large-scale perturbations that
\begin{equation}\label{eq:delta-rad-peacock-sol}
    \delta(\tau,\vec{q})=C_1(\vec{q}) \tau - 4H(t_0)\psi(\vec{q})\tau^{-1}.
\end{equation}
Large-scale perturbations can also be obtained from the standard
equation (\ref{eq:delta-rad-peacock}) by substituting
$\nabla^2\delta=0$, i.e.,
\begin{equation}\label{eq:delta-rad-peacock-ls}
  \ddot{\delta}+2H\dot{\delta}-
   \tfrac{4}{3}\kappa\varepsilon_\nul \delta=0.
\end{equation}
The general solution of this equation is, using
Eqs.~(\ref{subeq:rad-R0-sol}), given by
(\ref{eq:delta-rad-peacock-sol}). Thus far, the functions
$C_1(\vec{q})$ and $\psi(\vec{q})$ are the integration `constants'
which can be determined by the initial values $\delta(t_0,\vec{q})$
and $\dot{\delta}(t_0,\vec{q})$. However, equation
(\ref{eq:delta-rad-peacock-ls}) can also be derived from the
linearized Einstein equations and conservation laws for scalar
perturbations (\ref{subeq:pertub-gi}): see the derivation in
Appendix~\ref{app:standard-equation}\@. As a consequence, equation
(\ref{eq:delta-rad-peacock-ls}) is found to be also  a
\emph{relativistic} equation and the quantity
$\delta=\varepsilon_\een/\varepsilon_\nul$ is gauge-dependent.
Therefore, the second term in the solution
(\ref{eq:delta-rad-peacock-sol}) is not a physical mode, but equal
to the gauge mode
\begin{equation}\label{eq:gauge-rad}
    \delta_\mathrm{gauge}(\tau,\vec{q})=
    \psi(\vec{q})\frac{\dot{\varepsilon}_\nul}{\varepsilon_\nul}=
    -4H(t_0)\psi(\vec{q})\tau^{-1},
\end{equation}
as follows from (\ref{eq:gte}), (\ref{eq:behoud}) and (\ref{sol1a}).
Consequently, $\psi(\vec{q})$ should \emph{not} be interpreted as an
integration constant, but as a gauge function, which cannot be
determined by imposing initial value conditions. Thus, the general
solution (\ref{eq:peacock-sol}) of the standard equation
(\ref{eq:delta-rad-peacock}) depends on the gauge function
$\psi(\vec{q})$ and has, as a consequence, no physical significance.
This, in turn, implies that the standard equation
(\ref{eq:delta-rad-peacock}) does \emph{not} describe the evolution
of density perturbations. Here the negative effect of the gauge
function is clearly seen: up till now it was commonly accepted that
small-scale perturbations in the radiation-dominated era of a flat
\textsc{flrw} universe oscillate with a \emph{decaying} amplitude,
according to (\ref{eq:peacock-sol}). The approach presented in this
article reveals, however, that small-scale density perturbations
oscillate with an \emph{increasing} amplitude, according to
(\ref{dc-small}).

\subsection{Matter-dominated universe}
\label{sec:matter}

The standard perturbation equation of the Newtonian theory of
gravity is derived from \emph{approximate}, \emph{non-relativistic}
equations. It reads
\begin{equation}\label{eq:delta-dust-standard}
  \ddot{\delta} + 2H\dot{\delta}-
  \left(\frac{v_\mathrm{s}^2}{c^2}\frac{\nabla^2}{a^2}+
   \tfrac{1}{2}\kappa\varepsilon_\nul\right)
   \delta =0,
\end{equation}
where $v_\mathrm{s}$ is the speed of sound. (See, e.g., Weinberg
\cite{c8}, Sec.~15.9, or Peacock \cite{peacock1999}, Sec.~15.2.)
Using (\ref{pw12}), the general solution of equation
(\ref{eq:delta-dust-standard}) is found to be
\begin{equation}\label{eq:matter-non-physical}
  \delta_\varepsilon(\tau,\vec{q}) =
    \left[\frac{15}{8}\sqrt{\frac{\pi}{\mu_\mathrm{m}^5}}D_1(\vec{q})
    J_{+\frac{5}{2}}\bigl(2\mu_\mathrm{m}\tau^{1/3}\bigr)-
        4\psi(\vec{q})\sqrt{\pi\mu_\mathrm{m}^5}H(t_0)
        J_{-\frac{5}{2}}\bigl(2\mu_\mathrm{m}\tau^{1/3}\bigr)\right]\tau^{-1/6},
\end{equation}
where $D_1(\vec{q})$ and $\psi(\vec{q})$ are arbitrary functions
(the `constants' of integration) and $J_{\pm\nu}(x)$ is the Bessel
function of the first kind. The factors
$\frac{15}{8}\sqrt{\pi/\mu_\mathrm{m}^5}$ and
$4H(t_0)\sqrt{\pi\mu_\mathrm{m}^5}$ have been inserted for
convenience. The constant $\mu_\mathrm{m}$ is given by
(\ref{eq:const-mu}) and $\tau$ is given by (\ref{tau}). The
derivation of the solution (\ref{eq:matter-non-physical}) runs along
the same lines as the derivation of (\ref{eq:matter-physical}).

We now consider large-scale perturbations characterized by
$\nabla^2\delta=0$ (i.e., $|\vec{q}|\rightarrow0$) or perturbations
of all scales in the non-relativistic limit (i.e.,
$v_\mathrm{s}/c\rightarrow0$). Both limits imply
$\mu_\mathrm{m}\rightarrow0$, as follows from (\ref{eq:const-mu}).
The asymptotic expressions for the Bessel functions in the limit
$\mu_\mathrm{m}\rightarrow0$ are given by
\begin{equation}\label{eq:limit-J5/2}
  J_{+\frac{5}{2}}\bigl(2\mu_\mathrm{m}\tau^{1/3}\bigr)\approx
       \frac{8}{15}\sqrt{\frac{\mu_\mathrm{m}^5}{\pi}}\tau^{5/6}, \quad
    J_{-\frac{5}{2}}\bigl(2\mu_\mathrm{m}\tau^{1/3}\bigr)\approx
       \frac{3}{4\sqrt{\pi\mu_\mathrm{m}^5}}\tau^{-5/6}.
\end{equation}
Substituting these expressions into the general solution
(\ref{eq:matter-non-physical}), results in
\begin{equation}\label{eq:delta-standard-sol}
    \delta(\tau,\vec{q})=D_1(\vec{q}) \tau^{2/3} -3 H(t_0)\psi(\vec{q})\tau^{-1}.
\end{equation}
In the limit $\mu_\mathrm{m}\rightarrow0$, equation
(\ref{eq:delta-dust-standard}) reduces to
\begin{equation}\label{eq:delta-standard}
    \ddot{\delta}+2H\dot{\delta}-\tfrac{1}{2}\kappa\varepsilon_\nul\delta=0.
\end{equation}
Using (\ref{matsol1a})--(\ref{matsol1d}) we find the general
solution (\ref{eq:delta-standard-sol}) of this equation. Thus far,
the functions $D_1(\vec{q})$ and $\psi(\vec{q})$ are the integration
`constants' which can be determined by the initial values
$\delta(t_0,\vec{q})$ and $\dot{\delta}(t_0,\vec{q})$. However,
equation (\ref{eq:delta-standard}) can also be derived from the
general theory of relativity, and is, as a consequence, a
\emph{relativistic} equation. In fact, this equation follows for
large-scale perturbations from equations (\ref{subeq:pertub-gi})
(see Appendix~\ref{app:standard-equation} for a derivation), whereas
in the non-relativistic limit it follows from equations
(\ref{FRW4gi-flat-newt}) and (\ref{FRW6gi-flat-newt}). In both
cases, however, it is based on the gauge-dependent quantity
$\delta=\varepsilon_\een/\varepsilon_\nul$. (In
Appendix~\ref{sec:gauge-newton} it is shown that if a variable,
e.g., $\varepsilon_\een$, is gauge-dependent, it is also
gauge-dependent in the non-relativistic limit.) As a consequence,
the second term of (\ref{eq:delta-standard-sol}) is equal to the
gauge mode
\begin{equation}\label{eq:gauge-matter}
    \delta_\mathrm{gauge}(\tau,\vec{q})=
    \psi(\vec{q})\frac{\dot{\varepsilon}_\nul}{\varepsilon_\nul}=
    -3H(t_0)\psi(\vec{q})\tau^{-1},
\end{equation}
as follows from (\ref{eq:gte}), (\ref{mat2p0}) and (\ref{matsol1a}).
Therefore, $\psi(\vec{q})$ should \emph{not} be interpreted as an
integration constant, but as a gauge function, which cannot be
determined by imposing initial value conditions. Since the solution
(\ref{eq:matter-non-physical}) of equation
(\ref{eq:delta-dust-standard}) depends on the gauge function
$\psi(\vec{q})$ it has no physical significance. Consequently, the
standard equation (\ref{eq:delta-dust-standard}) does \emph{not}
describe the evolution of density perturbations. Again we encounter
the negative effect of the gauge function: up till now it was
commonly accepted that for small-scale density perturbations (i.e.,
density perturbations with wave lengths much smaller than the
particle horizon) a Newtonian treatment suffices and gauge
ambiguities do not occur and that the evolution of density
perturbations in the Newtonian regime is described by the standard
equation (\ref{eq:delta-dust-standard}). The approach presented in
this article reveals, however, that in a fluid with an equation of
state (\ref{state-mat}), density perturbations are described by the
\emph{relativistic} equation (\ref{eq:delta-dust}),
\emph{independent} of the scale of the perturbations.


\appendix

\section{Equations of state for the energy density and pressure}
\label{sec:eq-state}

We have used an equation of state for the pressure of the form
$p=p(n,\varepsilon)$. In general, however, this equation of state is
given in the form of two equations for the energy
density~$\varepsilon$ and the pressure~$p$ which contain also the
absolute temperature~$T$:
\begin{equation}\label{eq:state-e-p}
  \varepsilon=\varepsilon(n,T), \quad  p=p(n,T).
\end{equation}
In principle it is possible to eliminate $T$ from the two
equations~(\ref{eq:state-e-p}) to get $p=p(n,\varepsilon)$, so
that our choice of the form $p=p(n,\varepsilon)$ is justified. In
practice, however, it may in general be difficult to eliminate the
temperature~$T$ from the equations~(\ref{eq:state-e-p}). However,
this is not necessary, since the partial
derivatives~$p_\varepsilon$ and~$p_n$ (\ref{perttoes1}), the only
quantities that are actually needed, can be found in an
alternative way. From Eq.~(\ref{eq:state-e-p}) it follows
\begin{subequations}
\label{subeq:de-dp}
\begin{eqnarray}
&& \dif \varepsilon =
   \left(\dfrac{\partial \varepsilon}{\partial n} \right)_{\!T}\dif n +
   \left(\dfrac{\partial \varepsilon}{\partial T} \right)_{\!n}\dif T,
      \label{subeq:de-dp-a}  \\
&& \dif p =
   \left(\dfrac{\partial p}{\partial n} \right)_{\!T}\dif n +
   \left(\dfrac{\partial p}{\partial T} \right)_{\!n}\dif T.
      \label{subeq:de-dp-b}
\end{eqnarray}
\end{subequations}
From Eq.~(\ref{subeq:de-dp-b}) it follows that the partial
derivatives~(\ref{perttoes1}) are
\begin{subequations}
\label{subeq:pn-pe}
\begin{eqnarray}
&& p_n=
   \left(\dfrac{\partial p}{\partial n} \right)_{\!T} +
   \left(\dfrac{\partial p}{\partial T} \right)_{\!n}
   \left(\dfrac{\partial T}{\partial n} \right)_{\!\varepsilon},
      \label{subeq:pn-pe-a}  \\
&& p_\varepsilon=
   \left(\dfrac{\partial p}{\partial T} \right)_{\!n}
   \left(\dfrac{\partial T}{\partial \varepsilon} \right)_{\!n}.
      \label{subeq:pn-pe-b}
\end{eqnarray}
\end{subequations}
From Eq.~(\ref{subeq:de-dp-a}) it follows
\begin{subequations}
\label{eq:dT-dn-de}
\begin{eqnarray}
  \left(\dfrac{\partial T}{\partial n} \right)_{\!\varepsilon} & = &
  -\left(\dfrac{\partial \varepsilon}{\partial n} \right)_{\!T}
  \left(\dfrac{\partial \varepsilon}{\partial T} \right)_{\!n}^{-1}, \\
  \left(\dfrac{\partial T}{\partial \varepsilon}\right)_{\!n} & = &
  \left(\dfrac{\partial \varepsilon}{\partial T} \right)_{\!n}^{-1}.
\end{eqnarray}
\end{subequations}
Upon substituting the expressions~(\ref{eq:dT-dn-de}) into
Eqs.~(\ref{subeq:pn-pe}), we find for the partial derivatives
defined by Eq.~(\ref{perttoes1})
\begin{subequations}
\label{subeq:final-pn-pe}
\begin{eqnarray}
&& p_n=
   \left(\dfrac{\partial p}{\partial n} \right)_{\!T} -
   \left(\dfrac{\partial p}{\partial T} \right)_{\!n}
   \left(\dfrac{\partial \varepsilon}{\partial n} \right)_{\!T}
   \left(\dfrac{\partial \varepsilon}{\partial T} \right)_{\!n}^{-1},
      \label{subeq:final-pn-pe-a}  \\
&& p_\varepsilon=
   \left(\dfrac{\partial p}{\partial T} \right)_{\!n}
   \left(\dfrac{\partial \varepsilon}{\partial T}
   \right)_{\!n}^{-1},
      \label{subeq:final-pn-pe-b}
\end{eqnarray}
\end{subequations}
where $\varepsilon$ and~$p$ are given by~(\ref{eq:state-e-p}). In
order to calculate the second order derivative $p_{nn}$ replace
$p$ in Eq.~(\ref{subeq:final-pn-pe-a}) by $p_n$. For
$p_{\varepsilon\varepsilon}$ replace $p$ in
Eq.~(\ref{subeq:final-pn-pe-b}) by $p_\varepsilon$. Finally, for
$p_{\varepsilon n}\equiv p_{n\varepsilon}$, replace $p$ in
Eq.~(\ref{subeq:final-pn-pe-a}) by $p_\varepsilon$ or,
equivalently, replace $p$ in Eq.~(\ref{subeq:final-pn-pe-b}) by
$p_n$.

\section{Gauge invariance of the first order equations}
\label{giofoe}

If we go over from one synchronous system of reference with
coordinates $x$ to another synchronous system of reference with
coordinates $\hat{x}$ given by Eq.~(\ref{func}), we have
\begin{equation}
   \xi_{\mu;0} + \xi_{0;\mu} = 0,    \label{kil-syn}
\end{equation}
as follows from the transformation rule (\ref{killing}) and the
synchronicity condition (\ref{sync-cond}). From this equation we
find, using Eqs.~(\ref{def-gammas}), (\ref{con1}), (\ref{con2})
and~(\ref{metricFRW})  that $\xi^\mu(t,\vec{x})$ must be of the
form
\begin{equation}
   \xi^0=\psi(\vec{x}), \quad
     \xi^i=\tilde{g}^{ik}\partial_k \psi(\vec{x})
       \int^{ct} \!\! \frac{\dif\tau}{a^2(\tau)} + \chi^i(\vec{x}),
               \label{xi-syn}
\end{equation}
where $\psi(\vec{x})$ and $\chi^i(\vec{x})$ are \emph{arbitrary}
functions ---of the first order---  of the spatial coordinates
$\vec{x}$. The fact that the gauge function $\psi$ does not depend
on the coordinate $t$ anymore, as it did in general coordinates, see
Eq.~(\ref{defpsi}), is a consequence of the choice of synchronous
coordinates for the original coordinates as well as for the
transformed system of reference. It does not imply that the
gauge-invariant quantities $\varepsilon^\gi_\een$ and $n^\gi_\een$
are independent of transformations within a plane of synchronicity
only. In fact, they have been shown to be gauge-invariant, i.e.,
invariant under \emph{arbitrary} infinitesimal coordinate
transformations~(\ref{func}), and the fact that we are considering,
from now on, only transformations that transform a synchronous
system of reference to another synchronous system of reference does
\emph{not} take away, of course, this more general property of being
invariant under arbitrary transformations~(\ref{func}).

The energy density perturbation transforms according to
(\ref{e-ijk}), where $\varepsilon_\nul$ is a solution of
Eq.~(\ref{FRW2}). Similarly, the particle number density
transforms according to (\ref{n-ijk}) where $n_\nul$ is a solution
of Eq.~(\ref{FRW2a}). Finally, as follows from (\ref{sigmahat3}),
the fluid expansion scalar $\theta$, Eq.~(\ref{exp1}), transforms
as
\begin{equation}
  \hat{\theta}_\een(x) = \theta_\een(x) +
        \psi(\vec{x})\dot{\theta}_\nul(t) , \label{tr-theta}
\end{equation}
where $\theta_\nul=3H$ is a solution of Eq.~(\ref{FRW1}).

From~(\ref{sigmahat3}) with $\sigma=p$, $\varepsilon$, or~$n$
and~(\ref{perttoes}) we find for the transformation rule for the
first order perturbations to the pressure
\begin{equation}\label{perttoes-hat}
  \hat{p}_\een = p_\varepsilon \hat{\varepsilon}_\een +
                        p_n \hat{n}_\een.
\end{equation}

The transformation rule~(\ref{transvec}) with $V^\mu$ the
four-velocity $u^\mu$ implies
\begin{equation}\label{eq:trans-alg-umu}
    \hat{u}^\mu_\een=u^\mu_\een-\xi^\mu{}_{,0},
\end{equation}
where we have used that $u^\mu_\nul=\delta^\mu{}_0$, Eq. (\ref{u0}).
From the transformation rule (\ref{eq:trans-alg-umu}) it follows
that $u^\mu_\een$ transforms under synchronicity preserving
transformations (\ref{xi-syn}) as
\begin{subequations}
\label{eq:trans-u0-ui}
\begin{eqnarray}
  && \hat{u}^0_\een(x)=u^0_\een(x) =0, \\
  &&      \hat{u}^i_{\een\parallel}(x)=u^i_{\een\parallel}(x)-
        \frac{1}{a^2(t)}\tilde{g}^{ik}(\vec{x})
        \partial_k\psi(\vec{x}). \label{eq:trans-u0-ui-b}
\end{eqnarray}
\end{subequations}

We want to determine the transformation rules for~$\vartheta_\een$
and~$\mbox{$^3\!R_{\een\parallel}$}$. Since the
quantities~$\mbox{$^3\!R$}$, (\ref{drieR}), and $\vartheta$,
(\ref{driediv}), are both non-scalars under general space-time
transformations, the transformation rule~(\ref{sigmahat3}) is not
applicable to determine the transformation of their first order
perturbations under infinitesimal space-time transformations
$x^\mu\rightarrow\hat{x}^\mu$, (\ref{func}).
Since~$u^i_{\een\parallel}$ satisfies Eq.~(\ref{basis-5-scal}),
and since~$u^i_{\een\parallel}$ transforms according
to~(\ref{eq:trans-u0-ui}), and since we know
that~$\hat{u}^i_{\een\parallel}$ satisfies
Eq.~(\ref{basis-5-scal}) with hats, one may verify,
using~(\ref{perttoes-hat}), that
\begin{equation}
   \hat{\vartheta}_\een(x) :=
   \vartheta_\een(x)-\frac{\tilde{\nabla}^2\psi(\vec{x})}{a^2(t)},
          \label{theta-ijk}
\end{equation}
satisfies Eq.~(\ref{FRW5}) with hatted quantities. The quantity
$\hat{\vartheta}_\een$ is defined in analogy to $\vartheta_\een$
in~(\ref{den1a})
\begin{equation}
    \hat{\vartheta}_\een=(\hat{u}_{\een\parallel}^k){}_{|k}.
\end{equation}
Apparently, $\vartheta_\een$ transforms according
to~(\ref{theta-ijk}) under arbitrary infinitesimal synchronicity
preserving space-time transformations. Similarly, one may verify
that
\begin{equation}
       \mbox{$^3\!\hat{R}_{\een\parallel}(x)$} :=
       \mbox{$^3\!R_{\een\parallel}(x)$} +
      4H(t)\left(\frac{\tilde{\nabla}^2\psi(\vec{x})}{a^2(t)} -
       \tfrac{1}{2}\,\mbox{$^3\!R_\nul(t)$}\psi(\vec{x})\right),
            \label{drie-ijk}
\end{equation}
satisfies Eq.~(\ref{con-sp-1}). Apparently, (\ref{drie-ijk}) is
the transformation rule under arbitrary infinitesimal
synchronicity preserving space-time transformations. An
alternative way to find the results~(\ref{theta-ijk})
and~(\ref{drie-ijk}) is to write
$\hat{\vartheta}_\een=\vartheta_\een-f$ and
$\mbox{$^3\!\hat{R}_{\een\parallel}$}=\mbox{$^3\!R_{\een\parallel}$}-g$,
where $f$ and $g$ are unknown functions, to substitute, thereupon,
$\hat{\vartheta}_\een$ and $\mbox{$^3\!\hat{R}_{\een\parallel}$}$
into Eqs.~(\ref{FRW5}) and~(\ref{con-sp-1}), and to determine $f$
and $g$ such that the old equations (\ref{FRW5})
and~(\ref{con-sp-1}) reappear. In fact, our approach to define
$\hat{\vartheta}_\een$, (\ref{theta-ijk}), and
$\mbox{$^3\!\hat{R}_{\een\parallel}$}$, (\ref{drie-ijk}), is
nothing but a shortcut to this procedure.

It may now easily be verified by substitution that if
$\varepsilon_\een$, $n_\een$, $\theta_\een$, $\vartheta_\een$, and
\mbox{$^3\!R_{\een\parallel}$} are solutions of the system
(\ref{subeq:pertub-flrw})--(\ref{con-sp-1}), then the quantities
$\hat{\varepsilon}_\een$, (\ref{e-ijk}), $\hat{n}_\een$,
(\ref{n-ijk}), $\hat{\theta}_\een$, (\ref{tr-theta}),
$\hat{\vartheta}_\een$, (\ref{theta-ijk}),
and~\mbox{$^3\!\hat{R}_{\een\parallel}$}, (\ref{drie-ijk}), are,
for an arbitrary function~$\psi(\vec{x})$, also solutions of this
system.

\section{Gauge dependence in the non-relativistic limit}
\label{sec:gauge-newton}

If we require that the limit~(\ref{eq:nrl-limit}) holds true before
and after a gauge transformation, we find from the transformation
rule~(\ref{eq:trans-u0-ui-b}) for~$u^i_{\parallel\een}$ that
$\partial_k\psi(\vec{x})=0$, or, equivalently,
\begin{equation}\label{eq:psi-K}
  \psi(\vec{x})=C,
\end{equation}
where~$C$ is an arbitrary, small, constant. In view of
Eq.~(\ref{eq:psi-K}), the gauge dependent
functions~$\varepsilon_\een$ and~$n_\een$ transform under a gauge
transformation in the non-relativistic limit according to
[cf.~Eqs.~(\ref{subeq:split-e-n})]
\begin{subequations}
\label{subeq:split-e-n-nrl}
\begin{eqnarray}
 \varepsilon_\een & \rightarrow & \hat{\varepsilon}_\een =
      \varepsilon_\een + C\dot{\varepsilon}_\nul, \label{e-ijk-nrl} \\
 n_\een & \rightarrow & \hat{n}_{\een} =  n_{\een} + C\dot{n}_{\nul}.
         \label{n-ijk-nrl}
\end{eqnarray}
\end{subequations}
Since in an expanding universe the time derivatives
$\dot{\varepsilon}_\nul(t)\neq0$, (\ref{FRW2-newt}), and
$\dot{n}_{\nul}(t)\neq0$, (\ref{FRW2a-newt}), the
functions~$\varepsilon_\een(x)$ and~$n_\een(x)$ do still depend
upon the gauge, so that these quantities have, also in the
non-relativistic limit, no physical significance. Therefore,
Eqs.~(\ref{FRW4gi-flat-newt}) and~(\ref{FRW4agi-flat-newt}) need
not be considered in the non-relativistic limit: these equations
are not part of the Newtonian theory of gravity. The only
remaining \emph{physical} equations are
Eqs.~(\ref{FRW6gi-flat-newt})--(\ref{subeq:pertub-gi-flat-e-n-newt}).

Finally, it follows from equations (\ref{xi-syn}) and
(\ref{eq:psi-K}) that the transformation (\ref{func}) reduces in the
non-relativistic limit to
\begin{equation}\label{eq:gauge-trans-newt}
    x^0 \rightarrow x^0 - C, \quad x^i \rightarrow
    x^i-\chi^i(\vec{x}).
\end{equation}
In other words, and not surprisingly, time and space transformations
are decoupled: time coordinates may be shifted, whereas spatial
coordinates may be chosen arbitrarily.

\section{Derivation of the manifestly gauge-invariant perturbation equations}
\label{sec:deriv-egi}

In this appendix we derive the perturbation
equations~(\ref{subeq:eerste}) and the evolution
equations~(\ref{subeq:final}).

\subsection{Derivation of the evolution equation for the entropy}

With the help of Eqs.~(\ref{FRW4gi}) and~(\ref{FRW4agi}) and
Eqs.~(\ref{subeq:einstein-flrw})--(\ref{begam1}) one may verify
that
\begin{equation}\label{hulp2}
  \frac{1}{c}\frac{\dif}{\dif t}
      \left(n_\een-\frac{n_\nul}{\varepsilon_\nul(1+w)}\varepsilon_\een\right)=-
    3H\left(1-\frac{n_\nul p_n}{\varepsilon_\nul(1+w)}\right)
\left(n_\een-\dfrac{n_\nul}{\varepsilon_\nul(1+w)}\varepsilon_\een\right).
\end{equation}
In view of Eq.~(\ref{eq:lin-gi}) one may
replace~$\varepsilon_\een$ and~$n_\een$ by~$\varepsilon^\gi_\een$
and~$n^\gi_\een$. Using Eq.~(\ref{eq:entropy-gi}) yields
Eq.~(\ref{eq:vondst1}) with coefficient~(\ref{eq:alpha-4}) of the
main text.

\subsection{Derivation of the evolution equation for the energy density perturbation}

\begin{table*}
\renewcommand{\arraystretch}{2.25}
\caption{\label{eq:aij}The coefficients $\alpha_{ij}$ figuring in
the equations~\ref{subeq:nieuw}.} \footnotesize
\[  \begin{array}{llll} \hline\hline
  \alpha_{11}=3H(1+p_\varepsilon)+\dfrac{\kappa\varepsilon_\nul(1+w)}{2H} &
      \alpha_{12}=3Hp_n & \alpha_{13}=\varepsilon_\nul(1+w) &
      \alpha_{14}=\dfrac{\varepsilon_\nul(1+w)}{4H} \\
  \alpha_{21}=\dfrac{\kappa n_\nul}{2H} & \alpha_{22}=3H & \alpha_{23}=n_\nul &
  \alpha_{24}=\dfrac{n_\nul}{4H} \\
  \alpha_{31}=\dfrac{p_\varepsilon}{\varepsilon_\nul(1+w)}\dfrac{\tilde{\nabla}^2}{a^2} &
       \alpha_{32}=\dfrac{p_n}{\varepsilon_\nul(1+w)}\dfrac{\tilde{\nabla}^2}{a^2} &
       \alpha_{33}=H(2-3\beta^2) & \alpha_{34}=0 \\
  \alpha_{41}=\dfrac{\kappa\,\mbox{$^3\!R_\nul$}}{3H} & \alpha_{42}=0 &
  \alpha_{43}=-2\kappa\varepsilon_\nul(1+w) &
        \alpha_{44}=2H+\dfrac{\mbox{$^3\!R_\nul$}}{6H} \\
  \alpha_{51}=\dfrac{-\,\mbox{$^3\!R_\nul$}}{\mbox{$^3\!R_\nul$}+
         3\kappa\varepsilon_\nul(1+w)} & \alpha_{52}=0 &
     \alpha_{53}=\dfrac{6\varepsilon_\nul H(1+w)}{\mbox{$^3\!R_\nul$}+3\kappa\varepsilon_\nul(1+w)} &
     \alpha_{54}=\dfrac{\tfrac{3}{2}\varepsilon_\nul(1+w)}
     {\mbox{$^3\!R_\nul$}+3\kappa\varepsilon_\nul(1+w)}  \\  \hline\hline
\end{array}  \]
\end{table*}
We will now derive Eq.~(\ref{eq:vondst2}). To that end, we rewrite
the system (\ref{subeq:pertub-gi}) and Eq.~(\ref{Egi}) in the form
\begin{subequations}
\label{subeq:nieuw}
\begin{eqnarray}
 && \dot{\varepsilon}_\een+\alpha_{11}\varepsilon_\een+
    \alpha_{12}n_\een+
   \alpha_{13}\vartheta_\een+\alpha_{14}\,\mbox{$^3\!R_{\een\parallel}$} = 0, \label{nieuw1} \\
 &&  \dot{n}_\een+\alpha_{21}\varepsilon_\een +
    \alpha_{22}n_\een+
   \alpha_{23}\vartheta_\een+\alpha_{24}\,\mbox{$^3\!R_{\een\parallel}$} = 0, \label{nieuw2} \\
 && \dot{\vartheta}_\een +\alpha_{31}\varepsilon_\een +
     \alpha_{32}n_\een+
   \alpha_{33}\vartheta_\een+\alpha_{34}\,\mbox{$^3\!R_{\een\parallel}$} = 0, \label{nieuw3} \\
 && \mbox{$^3\!\dot{R}_{\een\parallel}$}+\alpha_{41}\varepsilon_\een+
     \alpha_{42}n_\een+
   \alpha_{43}\vartheta_\een+\alpha_{44}\,\mbox{$^3\!R_{\een\parallel}$} = 0, \label{nieuw4} \\
 && \varepsilon^\gi_\een+\alpha_{51}\varepsilon_\een+\alpha_{52}n_\een+
    \alpha_{53}\vartheta_\een+\alpha_{54}\,\mbox{$^3\!R_{\een\parallel}$} = 0, \label{nieuw5}
\end{eqnarray}
\end{subequations}
where the coefficients~$\alpha_{ij}(t)$ are given in
Table~\ref{eq:aij}.

In calculating the coefficients~$a_1$, $a_2$ and~$a_3$,
(\ref{subeq:coeff}) in the main text, we use that the time
derivative of the quotient~$w$, defined by Eq.~(\ref{begam1}) is
given by
\begin{equation}\label{eq:td-w}
  \dot{w} = 3H(1+w)(w-\beta^2),
\end{equation}
as follows from Eqs.~(\ref{FRW2}) and~(\ref{begam2}). Moreover, it
is of convenience \emph{not} to expand the function~$\beta(t)$
defined by Eq.~(\ref{begam2}) since this will complicate
substantially the expressions for the coefficients~$a_1$, $a_2$
and~$a_3$.

\subparagraph{Step 1.} We first eliminate the
quantity~$\mbox{$^3\!R_{\een\parallel}$}$ from
Eqs.~(\ref{subeq:nieuw}). Differentiating Eq.~(\ref{nieuw5}) with
respect to time and eliminating the time derivatives
~$\dot{\varepsilon}_\een$, $\dot{n}_\een$, $\dot{\vartheta}_\een$
and~$\mbox{$^3\!\dot{R}_{\een\parallel}$}$ with the help of
Eqs.~(\ref{nieuw1})--(\ref{nieuw4}), we arrive at the equation
\begin{equation}\label{eq:equiv}
   \dot{\varepsilon}^\gi_\een + p_1\varepsilon_\een+p_2 n_\een+
   p_3\vartheta_\een+p_4\,\mbox{$^3\!R_{\een\parallel}$}=0,
\end{equation}
where the coefficients $p_1(t),\ldots,p_4(t)$ are given by
\begin{equation}\label{eq:coef-pi}
  p_i=\dot{\alpha}_{5i}-\alpha_{51}\alpha_{1i}-
         \alpha_{52}\alpha_{2i}-\alpha_{53}\alpha_{3i}-\alpha_{54}\alpha_{4i}.
\end{equation}
From Eq.~(\ref{eq:equiv}) it follows that
\begin{equation}\label{eq:sol-3R1}
  \mbox{$^3\!R_{\een\parallel}$}=-\dfrac{1}{p_4}\dot{\varepsilon}^\gi_\een-
     \dfrac{p_1}{p_4}\varepsilon_\een-\dfrac{p_2}{p_4}n_\een-
     \dfrac{p_3}{p_4}\vartheta_\een.
\end{equation}
In this way we have expressed the
quantity~$\mbox{$^3\!R_{\een\parallel}$}$ as a linear combination of
the quantities~$\dot{\varepsilon}^\gi_\een$, $\varepsilon_\een$,
$n_\een$ and~$\vartheta_\een$. Upon replacing
$\mbox{$^3\!R_{\een\parallel}$}$ given by~(\ref{eq:sol-3R1}), in
Eqs.~(\ref{subeq:nieuw}), we arrive at the system of equations
\begin{subequations}
\label{subeq:tweede}
\begin{eqnarray}
 \dot{\varepsilon}_\een+q_1\dot{\varepsilon}_\een^\gi+
   \beta_{11}\varepsilon_\een+\beta_{12}n_\een+
   \beta_{13}\vartheta_\een & = & 0, \label{tweede1} \\
 \dot{n}_\een+q_2\dot{\varepsilon}_\een^\gi+
    \beta_{21}\varepsilon_\een+\beta_{22}n_\een+
    \beta_{23}\vartheta_\een & = & 0, \label{tweede2} \\
 \dot{\vartheta}_\een+q_3\dot{\varepsilon}_\een^\gi+
   \beta_{31}\varepsilon_\een+\beta_{32}n_\een+
   \beta_{33}\vartheta_\een & = & 0, \label{tweede3} \\
 \mbox{$^3\!\dot{R}_{\een\parallel}$}+
   q_4\dot{\varepsilon}^\gi_\een+\beta_{41}\varepsilon_\een+\beta_{42}n_\een+
   \beta_{43}\vartheta_\een & = & 0, \label{tweede4} \\
 \varepsilon^\gi_\een+
   q_5\dot{\varepsilon}^\gi_\een+\beta_{51}\varepsilon_\een+\beta_{52}n_\een+
   \beta_{53}\vartheta_\een & = & 0, \label{tweede5}
\end{eqnarray}
\end{subequations}
where the coefficients~$q_i(t)$ and~$\beta_{ij}(t)$ are given by
\begin{equation}\label{eq:betaij}
  q_i=-\dfrac{\alpha_{i4}}{p_4}, \quad
   \beta_{ij}=\alpha_{ij}+q_i p_j.
\end{equation}
We now have achieved that the
quantity~$\mbox{$^3\!R_{\een\parallel}$}$ occurs only in
Eq.~(\ref{tweede4}). Since we are not interested in the
non-physical quantity~$\mbox{$^3\!R_{\een\parallel}$}$, we do not
need this equation any more.

\subparagraph{Step 2.} We now proceed in the same way as in step~1:
eliminating this time the quantity~$\vartheta_\een$ from the system
of equations~(\ref{subeq:tweede}). Differentiating
Eq.~(\ref{tweede5}) with respect to time and eliminating the time
derivatives~$\dot{\varepsilon}_\een$, $\dot{n}_\een$
and~$\dot{\vartheta}_\een$ with the help of
Eqs.~(\ref{tweede1})--(\ref{tweede3}), we arrive at
\begin{equation}
\label{eq:ddot-egi}
  q_5\ddot{\varepsilon}^\gi_\een+r\dot{\varepsilon}^\gi_\een+
     s_1\varepsilon_\een+s_2n_\een+s_3\vartheta_\een=0,
\end{equation}
where the coefficients~$r(t)$ and~$s_i(t)$ are given by
\begin{subequations}
\label{eq:coef-qi}
\begin{eqnarray}
  s_i & = & \dot{\beta}_{5i}-\beta_{51}\beta_{1i}-\beta_{52}\beta_{2i}-
       \beta_{53}\beta_{3i}, \\
  r & = & 1+\dot{q}_5-\beta_{51}q_1-\beta_{52}q_2-\beta_{53}q_3.
\end{eqnarray}
\end{subequations}
From Eq.~(\ref{eq:ddot-egi}) it follows that
\begin{equation}\label{eq:sol-theta1}
  \vartheta\een=-\dfrac{q_5}{s_3}\ddot{\varepsilon}^\gi_\een-
     \dfrac{r}{s_3}\dot{\varepsilon}^\gi_\een-
     \dfrac{s_1}{s_3}\varepsilon_\een-\dfrac{s_2}{s_3}n_\een.
\end{equation}
In this way we have expressed the quantity~$\vartheta_\een$ as a
linear combination of the quantities~$\ddot{\varepsilon}^\gi_\een$,
$\dot{\varepsilon}^\gi_\een$, $\varepsilon_\een$ and~$n_\een$. Upon
replacing~$\vartheta_\een$ given by~(\ref{eq:sol-theta1}) in
Eqs.~(\ref{subeq:tweede}), we arrive at the system of equations
\begin{subequations}
\label{subeq:derde}
\begin{eqnarray}
&&\dot{\varepsilon}_\een-\beta_{13}\dfrac{q_5}{s_3}\ddot{\varepsilon}^\gi_\een+
   \left(q_1-\beta_{13}\dfrac{r}{s_3}\right)\dot{\varepsilon}^\gi_\een+
   \left(\beta_{11}-\beta_{13}\dfrac{s_1}{s_3}\right)\varepsilon_\een
  +\left(\beta_{12}-\beta_{13}\dfrac{s_2}{s_3}\right)n_\een=0, \label{derde1} \\
&&\dot{n}_\een-\beta_{23}\dfrac{q_5}{s_3}\ddot{\varepsilon}^\gi_\een+
   \left(q_2-\beta_{23}\dfrac{r}{s_3}\right)\dot{\varepsilon}^\gi_\een+
   \left(\beta_{21}-\beta_{23}\dfrac{s_1}{s_3}\right)\varepsilon_\een
  +\left(\beta_{22}-\beta_{23}\dfrac{s_2}{s_3}\right)n_\een=0, \label{derde2} \\
&&\dot{\vartheta}_\een-\beta_{33}\dfrac{q_5}{s_3}\ddot{\varepsilon}^\gi_\een+
   \left(q_3-\beta_{33}\dfrac{r}{s_3}\right)\dot{\varepsilon}^\gi_\een+
   \left(\beta_{31}-\beta_{33}\dfrac{s_1}{s_3}\right)\varepsilon_\een
  +\left(\beta_{32}-\beta_{33}\dfrac{s_2}{s_3}\right)n_\een=0, \label{derde3} \\
&&\mbox{$^3\!\dot{R}_{\een\parallel}$}-\beta_{43}\dfrac{q_5}{s_3}\ddot{\varepsilon}^\gi_\een+
   \left(q_4-\beta_{43}\dfrac{r}{s_3}\right)\dot{\varepsilon}^\gi_\een+
   \left(\beta_{41}-\beta_{43}\dfrac{s_1}{s_3}\right)\varepsilon_\een
  +\left(\beta_{42}-\beta_{43}\dfrac{s_2}{s_3}\right)n_\een=0, \label{derde4} \\
&&\varepsilon^\gi_\een-\beta_{53}\dfrac{q_5}{s_3}\ddot{\varepsilon}^\gi_\een+
   \left(q_5-\beta_{53}\dfrac{r}{s_3}\right)\dot{\varepsilon}^\gi_\een+
   \left(\beta_{51}-\beta_{53}\dfrac{s_1}{s_3}\right)\varepsilon_\een
  +\left(\beta_{52}-\beta_{53}\dfrac{s_2}{s_3}\right)n_\een=0. \label{derde5}
\end{eqnarray}
\end{subequations}
We have achieved now that the quantities~$\vartheta_\een$
and~$\mbox{$^3\!R_{\een\parallel}$}$ occur only in
Eqs.~(\ref{derde3}) and~(\ref{derde4}), so that these equations
will not be needed anymore. We are left, in principle, with
Eqs.~(\ref{derde1}), (\ref{derde2}) and~(\ref{derde5}) for the
three unknown quantities~$\varepsilon_\een$, $n_\een$
and~$\varepsilon^\gi_\een$, but we first proceed with all five
equations.

\subparagraph{Step 3.} At first sight, the next steps would be to
eliminate, successively, the quantities~$\varepsilon_\een$
and~$n_\een$ from Eq.~(\ref{derde5}) with the help of
Eqs.~(\ref{derde1}) and~(\ref{derde2}). We then would end up with
a fourth order differential equation for the unknown
quantity~$\varepsilon^\gi_\een$.

However, it is possible to extract a second order equation for the
gauge-invariant energy density from the
equations~(\ref{subeq:derde}). This will now be shown.
Eq.~(\ref{derde5}) can be rewritten
\begin{equation}\label{eq:eindelijk}
  \ddot{\varepsilon}^\gi_\een+a_1\dot{\varepsilon}^\gi_\een+
    a_2\varepsilon^\gi_\een=
    a_3\left(n_\een+\dfrac{\beta_{51}s_3-\beta_{53}s_1}
    {\beta_{52}s_3-\beta_{53}s_2} \varepsilon_\een\right),
\end{equation}
where the coefficients $a_1(t)$, $a_2(t)$ and~$a_3(t)$ are given
by
\begin{subequations}
\label{subeq:vierde}
\begin{eqnarray}
&& a_1 = -\dfrac{s_3}{\beta_{53}}+\dfrac{r}{q_5}, \\
&& a_2 = -\dfrac{s_3}{\beta_{53}q_5}, \\
&& a_3 = \dfrac{\beta_{52}s_3}{\beta_{53}q_5}-\dfrac{s_2}{q_5}.
\end{eqnarray}
\end{subequations}
These are precisely the coefficients
(\ref{eq:alpha-1})--(\ref{eq:alpha-3}) of the main text.
Furthermore, we find
\begin{equation}\label{eq:check}
  \dfrac{\beta_{51}s_3-\beta_{53}s_1}{\beta_{52}s_3-\beta_{53}s_2}=
      -\dfrac{n_\nul}{\varepsilon_\nul(1+w)}.
\end{equation}
Hence,
\begin{equation}\label{eq:hulp3}
  n_\een+\dfrac{\beta_{51}s_3-\beta_{53}s_1}{\beta_{52}s_3-\beta_{53}s_2}
    \varepsilon_\een=
    n_\een-\dfrac{n_\nul}{\varepsilon_\nul(1+w)}\varepsilon_\een.
\end{equation}
With the help of this expression and Eq.~(\ref{eq:lin-gi}) we can
rewrite Eq.~(\ref{eq:eindelijk}) in the form~(\ref{eq:vondst2}).

The derivation of the expressions~(\ref{subeq:coeff})
from~(\ref{subeq:vierde}) and the proof of the
equality~(\ref{eq:check}) is straightforward, but extremely
complicated. We used \textsc{Maple~V}~\cite{MapleV} to perform
this algebraic task.

\subsection{Evolution equations for the contrast functions}
\label{app:contrast}

In this section we derive Eqs.~(\ref{subeq:final}). We start off
with Eq.~(\ref{fir-ord}). From Eq.~(\ref{eq:entropy-gi}) and the
definitions~(\ref{eq:contrast}) it follows that
\begin{equation}\label{eq:sgi-contrast}
  \sigma^\gi_\een=n_\nul\left(\delta_n-\dfrac{\delta_\varepsilon}{1+w}\right).
\end{equation}
Differentiating this equation with respect to~$ct$ yields
\begin{equation}\label{eq:diff-sgi}
  a_4 \sigma^\gi_\een=\dot{n}_\nul\left(\delta_n-\dfrac{\delta_\varepsilon}{1+w}\right)+
  n_\nul\dfrac{1}{c}\dfrac{\dif}{\dif t}
  \left(\delta_n-\dfrac{\delta_\varepsilon}{1+w}\right),
\end{equation}
where we have used Eq.~(\ref{eq:vondst1}). Using
Eqs.~(\ref{FRW2a}), (\ref{eq:alpha-4}) and~(\ref{eq:sgi-contrast})
to eliminate~$\sigma^\gi_\een$, we arrive at Eq.~(\ref{fir-ord})
of the main text.

Finally, we derive Eq.~(\ref{sec-ord}). Upon substituting the
expressions
\begin{subequations}
\label{subeq:afgeleiden}
\begin{eqnarray}
&& \varepsilon^\gi_\een = \varepsilon_\nul\delta_\varepsilon, \\
&&
\dot{\varepsilon}^\gi_\een=\dot{\varepsilon}_\nul\delta_\varepsilon+
      \varepsilon_\nul\dot{\delta}_\varepsilon, \\
&&
\ddot{\varepsilon}^\gi_\een=\ddot{\varepsilon}_\nul\delta_\varepsilon+
      2\dot{\varepsilon}_\nul\dot{\delta}_\varepsilon+
      \varepsilon_\nul\ddot{\delta}_\varepsilon,
\end{eqnarray}
\end{subequations}
into Eq.~(\ref{eq:vondst2}), and dividing by~$\varepsilon_\nul$,
we find
\begin{subequations}
\label{subeq:tilde-alpha}
\begin{eqnarray}
&& b_1=2\dfrac{\dot{\varepsilon}_\nul}{\varepsilon_\nul}+a_1, \\
&& b_2=\dfrac{\ddot{\varepsilon}_\nul}{\varepsilon_\nul}+
      a_1\dfrac{\dot{\varepsilon}_\nul}{\varepsilon_\nul}+a_2, \\
&& b_3=a_3\dfrac{n_\nul}{\varepsilon_\nul}.
\end{eqnarray}
\end{subequations}
where we have also used Eq.~(\ref{eq:sgi-contrast}). These are the
coefficients~(\ref{subeq:coeff-contrast}) of the main text.

\section{Derivation of the standard equation for density perturbations}
\label{app:standard-equation}

In this appendix equations (\ref{eq:delta-rad-peacock-ls}) and
(\ref{eq:delta-standard}) of the main text are derived, for a flat
\textsc{flrw} universe, $\mbox{$^3\!R_\nul$}=0$, with vanishing
cosmological constant, $\Lambda=0$, using the background equations
(\ref{subeq:einstein-flrw})--(\ref{FRW3}) and the linearized
Einstein equations and conservation laws for scalar perturbations
(\ref{subeq:pertub-gi}).

From (\ref{eq:td-w}) it follows that $w$ is constant if and only if
$w=\beta^2$ for all times. Using (\ref{eq:begam3}) it is found for
constant $w$ that $p_n=0$ and $p_\varepsilon=w$, i.e., the pressure
does not depend on the particle number density. Consequently, in the
derivation of equations (\ref{eq:delta-rad-peacock-ls}) and
(\ref{eq:delta-standard}) the equations (\ref{FRW2a}) for
$n_\nul(t)$ and (\ref{FRW4agi}) for $n_\een(t,\vec{x})$ are not
needed. In this case, the equation of state is given by
\begin{equation}\label{eq:state-simple}
    p = w \varepsilon.
\end{equation}
To derive the standard equations (\ref{eq:delta-rad-peacock-ls}) and
(\ref{eq:delta-standard}), it is required that
$u^i_{\een\parallel}=0$, implying that
\begin{equation}\label{eq:theta1-is-0}
    \vartheta_\een(t,\vec{x})=0.
\end{equation}
Since $\nabla^2p_\een=0$ for large-scale perturbations, equation
(\ref{FRW5gi}) is identically satisfied. After substituting
$\varepsilon_\een=\varepsilon_\nul\delta$ into equation
(\ref{FRW4gi}) and eliminating $\dot{\varepsilon}_\nul$ with the
help of equation (\ref{FRW2}), it is found that
\begin{equation}\label{eq:delta-eerste}
    \dot{\delta}+\frac{1+w}{2H}\left(\kappa\varepsilon_\nul\delta+
        \tfrac{1}{2}\,\mbox{$^3\!R_{\een\parallel}$}\right)=0.
\end{equation}
Differentiating equation (\ref{eq:delta-eerste}) with respect to
$x^0=ct$ and using equations (\ref{FRW1}), (\ref{FRW2}),
(\ref{FRW3}) and (\ref{FRW6gi}), yields
\begin{equation}\label{eq:delta-tweede}
    \ddot{\delta}+\tfrac{3}{2}(1+w)H\dot{\delta}-
    \tfrac{3}{4}(1+w)^2\kappa\varepsilon_\nul\delta-
    \tfrac{1}{8}(1+w)(1-3w)\,\mbox{$^3\!R_{\een\parallel}$}=0.
\end{equation}
Eliminating $\mbox{$^3\!R_{\een\parallel}$}$ from equation
(\ref{eq:delta-tweede}) with the help of equation
(\ref{eq:delta-eerste}), yields the standard equation for
large-scale perturbations in a flat \textsc{flrw} universe:
\begin{equation}\label{eq:stand-w}
    \ddot{\delta}+2H\dot{\delta}-
         \tfrac{1}{2}\kappa\varepsilon_\nul\delta(1+w)(1+3w)=0.
\end{equation}
This equation has been derived by Weinberg \cite{c8}, Eq.~(15.10.57)
and Peebles~\cite{c11}, Eq.~(86.11). For $w=\tfrac{1}{3}$ (the
radiation-dominated era) equation (\ref{eq:delta-rad-peacock-ls}) is
found, whereas for $w=0$ (the matter-dominated era) equation
(\ref{eq:delta-standard}) applies.

Using that the general solution of the background equations
(\ref{subeq:einstein-flrw})--(\ref{FRW3}) for
$\mbox{$^3\!R_\nul$}=0$, $\Lambda=0$ and constant~$w$ is given by
\begin{subequations}
\begin{eqnarray}
  H(t) &=& \dfrac{2}{3(1+w)}(ct)^{-1}=H(t_0)\left(\dfrac{t}{t_0}\right)^{-1}, \\
  \varepsilon_\nul(t) &=& \dfrac{4}{3\kappa(1+w)^2}(ct)^{-2}
  =\varepsilon_\nul(t_0)\left(\dfrac{t}{t_0}\right)^{-2},
\end{eqnarray}
\end{subequations}
we find for the general solution of Eq.~(\ref{eq:stand-w})
\begin{equation}\label{eq:gen-sol-stand-w}
    \delta(\tau,\vec{q}) = E_1(\vec{q})\tau^{(2+6w)/(3+3w)} -
       3(1+w)H(t_0)\psi(\vec{q}) \tau^{-1},
\end{equation}
where $E_1(\vec{q})$ is an arbitrary function (the integration
`constant') and $\psi(\vec{q})$ is the gauge function [see
(\ref{eq:gauge-rad}) with $w=\tfrac{1}{3}$ or
(\ref{eq:gauge-matter}) with $w=0$]. The
solution~(\ref{eq:gen-sol-stand-w}) is exactly equal to the result
found by Peebles~ \cite{c11}, \S\,86, Eq.~(86.12).

Finally, we note that in the derivation of
(\ref{eq:delta-rad-peacock-ls}) Peebles, \S 86 uses
Eq.~(\ref{eq:theta1-is-0}) (in his notation: $\theta=0$). In this
case, Peebles' approach yields a physical mode $\delta\propto\tau$
and a gauge mode $\delta\propto\tau^{-1}$. For $\vartheta_\een\neq0$
Peebles finds a physical mode $\delta\propto\tau^{1/2}$. However, in
the approach presented in this article both physical modes
$\delta\propto\tau$ and $\delta\propto\tau^{1/2}$ [see
Eq.~(\ref{delta-H-rad})] follow from \emph{one} second-order
differential equation (\ref{eq:delta-rad}), \emph{without} taking an
explicit value for $\vartheta_\een$.

\section{Symbols and their meaning}\label{notations}

\begin{longtable*}{lll}
\caption{Symbols and their meaning of all quantities, except for those occurring in the appendices.} \\
\hline\hline \textbf{Symbol} & \textbf{Meaning} &
\textbf{Reference Equation} \\ \hline
\endfirsthead
\caption[]{(continued)} \\ \hline \hline
\endhead
\hline
\endfoot
$\vec{\nabla}$& $(\partial_1,\partial_2,\partial_3)$ & --- \\
$\nabla^2 f$ & $\vec{\nabla}\cdot(\vec{\nabla}f)=\delta^{ij}f_{,i,j}$ & (\ref{poisson}) \\
$(\vec{\tilde{\nabla}}f)^i$ & $\tilde{g}^{ij}f_{|j}$ & --- \\
$\vec{\tilde{\nabla}}\cdot\vec{v}$ & $v^k{}_{|k}$ & (\ref{longitudinal}), (\ref{transversal}) \\
$(\vec{\tilde{\nabla}}\wedge\vec{u}_\een)_i$ &
          $\epsilon_i{}^{jk}u_{\een j|k}=\epsilon_i{}^{jk}u_{\een j,k}$ & (\ref{rotation}) \\
$\tilde{\nabla}^2 f $ &
      $\vec{\tilde{\nabla}}\cdot(\vec{\tilde{\nabla}}f=\tilde{g}^{ij}f_{|i|j})
      $ & (\ref{Laplace})\\
$\partial_{0}$& derivative with respect to $x^{0}=ct$& ---\\
$\partial_{i}$& derivative with respect to $x^{i}$& ---\\
hat: $\hat{}$&   computed with respect to
$\hat x$& (\ref{subeq:split-e-n})\\
dot: $\dot{}$ & derivative with respect to $x^0=ct$& (\ref{def-gammas})\\
tilde: $\tilde{}$ & computed with respect to three-metric $\tilde{g}_{ij}$& (\ref{m2}), (\ref{con3FRW}) \\
superindex: ${}^\gi$ &   gauge-invariant& (\ref{subeq:gi-en})\\
superindex: ${}^{|k}$ & contravariant derivative with respect to
      $x^k$: $\zeta^{|k}=g^{kj}_\nul\zeta_{|j}$ & (\ref{decomp-hij-par}) \\
subindex: ${}_\nul$ & background quantity& (\ref{subeq:ent-split})\\
subindex: ${}_\een$& perturbation of first order & (\ref{subeq:ent-split}) \\
subindex: ${}_\twee$& perturbation of second order&  (\ref{subeq:exp-scalar})--(\ref{subeq:exp-vec-tens}) \\
subindex: ${}_{;\lambda}$& covariant derivative with respect to $x^\lambda$& (\ref{lie1})\\
subindex: ${}_{|k}$ & covariant derivative with respect to $x^k$ & (\ref{threecov})\\
subindex: ${}_{,\mu}$ & derivative with respect to $x^\mu$& (\ref{concoef1}) \\
subindex: ${}_{\parallel}$ & longitudinal part of a vector or tensor& (\ref{decomp-symh})\\
subindex: ${}_{\perp}$ & perpendicular part of vector or  tensor & (\ref{decomp-symh}) \\
subindex: ${}_{\ast}$ & transverse and traceless part of a tensor & (\ref{decomp-symh}) \\
$\beta$ & $c^{-1}$ times speed of sound: $\sqrt{\dot{p}_{(0)}/\dot{\varepsilon}_{(0)}}$& (\ref{begam2})\\
$\Gamma^\lambda{}_{\mu\nu}$ & connection coefficients & (\ref{concoef1})\\
$\gamma$ & arbitrary function & (\ref{Rij-paral}) \\
$\delta_\varepsilon$ & energy contrast function (gauge-invariant) & (\ref{eq:contrast})\\
$\delta^\mu{}_\nu$ & Kronecker delta & (\ref{u0})\\
$\delta_n$ & particle number contrast function (gauge-invariant) & (\ref{eq:contrast})\\
$\delta_\mathrm{T}$ &  temperature contrast function (gauge-invariant) & (\ref{eq:delta-T})\\
$\varepsilon$& energy density& (\ref{eps1})\\
$\epsilon_i{}^{jk}$ & Levi-Civita tensor,
    $\epsilon_1{}^{23}=+1$ & (\ref{rotation})  \\
$\zeta,\phi$ & potentials due to
      relativistic density perturbations & (\ref{decomp-hij-par}), (\ref{RnabEE})  \\
$\eta$ & bookkeeping parameter equal to 1& (\ref{subeq:exp-scalar})--(\ref{subeq:exp-vec-tens})\\
$\theta$ &  expansion scalar in four-space& (\ref{exp1}) \\
$\vartheta$ & expansion scalar in three-space & (\ref{driediv})\\
$\kappa$ &  $8\pi G/c^4$& (\ref{kappa}) \\
$\varkappa_{ij}$ & time derivative metric coefficients & (\ref{def-gammas})\\
$\Lambda$& cosmological constant & (\ref{ein-verg}) \\
$\lambda$ & wavelength & (\ref{pw12}) \\
$\mu$ & thermodynamic potential & (\ref{eq:sec-law-thermo})\\
$\mu_\mathrm{m}$ & reduced wave length (matter) & (\ref{eq:const-mu})\\
$\mu_\mathrm{r}$ & reduced wave length (radiation) & (\ref{xi})\\
$\xi^\mu$& first order space-time translation& (\ref{func}) \\
$\pi$ & arbitrary function & (\ref{Rij-paral}) \\
$\varrho_\een$  & $\varepsilon^\gi_\een/c^2$ & --- \\
$\sigma$ & arbitrary scalar &  (\ref{sigma}) \\
$\sigma^\gi_\een$ &  abbreviation for
      $n_\een^\gi-n_\nul\varepsilon^\gi_\een/[\varepsilon_\nul(1+w)]$ &
      (\ref{eq:hulp-entropy})--(\ref{eq:entropy-gi}), (\ref{eq:sigmagi})\\
$\tau$ & dimensionless time &  (\ref{tau}) \\
$\phi, \zeta$ & potentials due to relativistic density perturbations & (\ref{decomp-hij-par}),  (\ref{RnabEE}) \\
$\varphi$ & Newtonian potential  & (\ref{poisson}), (\ref{eq:ident})\\
$\chi^i$  & arbitrary three-vector & (\ref{xi-syn})  \\
$\psi$& first order time translation (gauge function) & (\ref{subeq:split-e-n}),  (\ref{func}), (\ref{defpsi})\\
$\omega$& arbitrary scalar &  (\ref{omegahat3})\\
$A^{\alpha\ldots\beta}{}_{\mu\ldots\nu}$& arbitrary tensor&  (\ref{lie1})\\
$A_{\mu\nu}$& arbitrary rank two tensor&  (\ref{lieder})\\
$a$ & scale factor or radius of universe& (\ref{m2}) \\
$a_1,a_2,a_3,a_4$& coefficients in perturbation equations& (\ref{subeq:coeff})\\
$b_1,b_2,b_3$& coefficients in perturbation equations& (\ref{subeq:coeff-contrast})\\
$C$ & arbitrary first order constant & (\ref{eq:psi-K})\\
$c$  &   speed of light & ---\\
$\dif s^2$& line element in four-space& (\ref{line-element}) \\
$E$ & energy within volume $V$ & (\ref{eq:sec-law-thermo}) \\
$G$ & Newton's gravitation constant & --- \\
$g_{\mu\nu}$& metric tensor& (\ref{killing})\\
$\hat g_{\mu\nu}$& metric with respect to $\hat x^\mu$& (\ref{killing})\\
$\tilde{g}_{ij}$& time-independent metric of three-space & (\ref{m2})\\
$H$ &  $c^{-1}\mathcal{H}$ & (\ref{Hubble}) \\
$\mathcal{H}$& Hubble function: $\mathcal{H}=(\dif a/\dif t)/a$ & --- \\
$h_{ij}$& minus first order perturbation of metric & (\ref{hij})\\
$J_{\pm\nu}$ & Bessel function of the first kind &
(\ref{eq:peacock-sol}), (\ref{eq:matter-non-physical})  \\
$K$& constant relating Ricci and metric tensors& (\ref{Riccimaxsym})\\
$k$ & $k=-1$ (open), $0$ (flat), $+1$ (closed) & (\ref{gFLRW})\\
$k_\mathrm{B}$ & Boltzmann's constant & (\ref{state-mat}) \\
$\mathcal{L}_\xi$& Lie derivative with respect to $\xi^\mu$& (\ref{lie1}) \\
$m$ & rest mass of particle of cosmological fluid & (\ref{eq:rest-energy})\\
$m_\mathrm{H}$ & proton rest mass & (\ref{state-mat}) \\
$N$ & number of particles within volume $V$ & (\ref{eq:sec-law-thermo})\\
$N^\mu$& particle density four-flow& (\ref{eq:current})\\
$n$ & particle number density &  (\ref{en1})\\
$p$ & pressure & (\ref{toestand}) \\
$p_\varepsilon$, $p_n$ & partial derivatives of pressure & (\ref{perttoes1})\\
$p_{nn}$, $p_{\varepsilon n}$ & partial derivatives of pressure& (\ref{pne}) \\
$q=|\vec{q}|$ & magnitude of wave vector: $2\pi/\lambda$ & (\ref{pw12})\\
$\vec{q}$ & wave vector & (\ref{pw12}) \\
$\mbox{$^3\!R$}$ & Ricci scalar in three-space& (\ref{drieR})\\
$R_{\mu\nu}$& Ricci tensor in four-space & (\ref{Ricci1})\\
$\mbox{$^3\!R_{ij}$}$ & Ricci tensor in three-space & (\ref{Ricci-drie})\\
$S$ & entropy within a volume $V$ & (\ref{eq:sec-law-thermo}) \\
$s$ & entropy per particle $s:=S/N$ & (\ref{eq:sec-law-2})\\
$T$ & absolute temperature & (\ref{eq:sec-law-thermo})\\
$T^{\mu\nu}$& energy momentum tensor& (\ref{Tmunu})\\
$t$ &  cosmological time & ---\\
$t_0$ & initial cosmological time & --- \\
$t_\mathrm{p}$ & present cosmological time & (\ref{poisson}) \\
$U^\mu$& cosmological four-velocity $U^\mu U_\mu=c^2$   & (\ref{subeq:ent}) \\
$u^\mu$ & $c^{-1}U^\mu$ & --- \\
$\vec{U}$ & spatial velocity & (\ref{eq:nrl-limit}) \\
$\vec{u}$ & $c^{-1}\vec{U}$ & (\ref{decomp-u}) \\
$V^\mu$& arbitrary four-vector& (\ref{lievec})\\
$w$ & pressure divided by energy & (\ref{begam1}) \\
$x$& space-time point $x^\mu=(ct,\mbox{\boldmath{$x$}})$& ---\\
$\hat{x}^\mu$& locally transformed coordinates&  (\ref{func})\\
\vec{x}&  spatial point $\vec{x}=(x^1,x^2,x^3)$ &  --- \\
$Y_{\pm\nu}$ & Bessel function of the second kind
&(\ref{eq:peacock-sol})
\end{longtable*}

\begin{acknowledgments}
One of the authors (\textsc{pgm}) is indebted to the director of
the Institute of Theoretical Physics of the University of
Amsterdam, prof.\ dr.\ ir.\ F.\ A.\ Bais, for moral support and
travel expenses.
\end{acknowledgments}


\bibstyle{plain}

\end{document}